\documentclass[preprint]{aastex}
\usepackage{geometry}                
\usepackage{graphicx}
\usepackage{amssymb}
\usepackage{multirow}
\usepackage{rotating}
\usepackage{graphicx}
\usepackage{bm}

\begin{document}

\title{Standardizing Type Ia Supernova Absolute Magnitudes Using Gaussian Process Data Regression}

\author
{
    A.~G.~Kim,\altaffilmark{1}
    R.~C. Thomas,\altaffilmark{2}
    G.~Aldering,\altaffilmark{1}
    P.~Antilogus,\altaffilmark{3}
    C.~Aragon,\altaffilmark{1}
    S.~Bailey,\altaffilmark{1}
    C.~Baltay,\altaffilmark{4}
    S.~Bongard,\altaffilmark{3}
    C.~Buton,\altaffilmark{5}
    A.~Canto,\altaffilmark{3}
    F.~Cellier-Holzem,\altaffilmark{3}
    M.~Childress,\altaffilmark{1,6}
    N.~Chotard,\altaffilmark{7,8}
    Y.~Copin,\altaffilmark{9}
    H.~K. Fakhouri,\altaffilmark{1,6}
    E.~Gangler,\altaffilmark{9}
    J.~Guy,\altaffilmark{3} 
    M.~Kerschhaggl,\altaffilmark{5}
    M.~Kowalski,\altaffilmark{5}
    J.~Nordin,\altaffilmark{1}
    P.~Nugent,\altaffilmark{2}
    K.~Paech,\altaffilmark{5}
    R.~Pain,\altaffilmark{3}
    E.~Pecontal,\altaffilmark{10}
    R.~Pereira,\altaffilmark{9}
    S.~Perlmutter,\altaffilmark{1,6}
    D.~Rabinowitz,\altaffilmark{4}
    M.~Rigault,\altaffilmark{9}  
    K.~Runge,\altaffilmark{1}
    C.~Saunders,\altaffilmark{1,6}
    R.~Scalzo,\altaffilmark{11}
    G.~Smadja,\altaffilmark{9}
    C.~Tao,\altaffilmark{7, 12}
    B.~A.~Weaver,\altaffilmark{13}
    C.~Wu\altaffilmark{3,8}
}

\altaffiltext{1}
{
    Physics Division, Lawrence Berkeley National Laboratory, 
    1 Cyclotron Road, Berkeley, CA, 94720
}

\altaffiltext{2}
{
    Computational Cosmology Center, Computational Research Division, Lawrence Berkeley National Laboratory, 
    1 Cyclotron Road MS 50B-4206, Berkeley, CA, 94720
}

\altaffiltext{3}
{
    Laboratoire de Physique Nucl\'eaire et des Hautes \'Energies,
    Universit\'e Pierre et Marie Curie Paris 6, Universit\'e Denis Diderot Paris 7, CNRS-IN2P3, 
    4 place Jussieu, 75252 Paris Cedex 05, France
}
\altaffiltext{4}
{
    Department of Physics, Yale University, 
    New Haven, CT, 06250-8121
}
\altaffiltext{5}
{
    Physikalisches Institut, Universit\"at Bonn,
    Nu\ss allee 12, 53115 Bonn, Germany
}
\altaffiltext{6}
{
    Department of Physics, University of California Berkeley,
    366 LeConte Hall MC 7300, Berkeley, CA, 94720-7300
}

\altaffiltext{7}
{
    Tsinghua Center for Astrophysics, Tsinghua University, Beijing 100084, China 
}

\altaffiltext{8}
{
    National Astronomical Observatories, Chinese Academy of Sciences, Beijing 100012, China
}

\altaffiltext{9}
{
    Universit\'e de Lyon, F-69622, Lyon, France ; Universit\'e de Lyon 1, Villeurbanne ; 
    CNRS/IN2P3, Institut de Physique Nucl\'eaire de Lyon.
}

\altaffiltext{10}
{
    Centre de Recherche Astronomique de Lyon, Universit\'e Lyon 1,
    9 Avenue Charles Andr\'e, 69561 Saint Genis Laval Cedex, France
}
\altaffiltext{11}
{
    Research School of Astronomy and Astrophysics,
    The Australian National University,
    Mount Stromlo Observatory,
    Cotter Road, Weston Creek ACT 2611 Australia
}
\altaffiltext{12}
{
    Centre de Physique des Particules de Marseille , 163, avenue de Luminy - Case 902 - 13288 Marseille Cedex 09, France
}

\altaffiltext{13}
{
    Center for Cosmology and Particle Physics,
    New York University,
    4 Washington Place, New York, NY 10003, USA
}

\begin{abstract}
We present a novel class of models for Type Ia supernova time-evolving spectral energy distributions (SED) and absolute magnitudes: they are
each modeled as stochastic functions
described by Gaussian processes.  The values of the SED and absolute magnitudes are defined through
well-defined regression prescriptions, so that data directly inform the models. 
As a proof of concept, we
implement a
model for synthetic photometry built from
the spectrophotometric time series from the Nearby Supernova Factory.
Absolute magnitudes at peak $B$ brightness are calibrated to 0.13 mag in the  $g$-band
and to as low as 0.09 mag in
the $z=0.25$ blueshifted $i$-band, where the dispersion
includes contributions from measurement uncertainties and peculiar velocities.
The methodology can be applied to spectrophotometric
time series of supernovae that span a range of redshifts to simultaneously standardize
supernovae together with fitting cosmological parameters. 
\end{abstract}


\section{Introduction}
Type I supernovae were found to be good standard candles with a low dispersion in visible-wavelength peak luminosities
\citep{1968AJ.....73.1021K}.
As supernovae were further subclassified, Type Ia supernovae (SNe Ia) were found to be even better calibrated candles by accounting for temporal \citep{1993ApJ...413L.105P}
and color \citep{1998A&A...331..815T} information transmitted
in their light curves.  
Modern distance determinators  fit supernova measurements to a time-evolving spectral energy distribution (SED)
model with free parameters that adjust the light-curve shapes and colors.
An individual supernova's peak absolute magnitude 
is derived directly from the fit \citep[e.g.][MLCS2k2, BayeSN]{2007ApJ...659..122J,2011ApJ...731..120M} or from a secondary relation
based on the fit parameters \citep[e.g.][SALT2, SIFTO, SNooPy]{2007A&A...466...11G, 2008ApJ...681..482C, 2011AJ....141...19B}.

There are challenges in empirically modeling the family of SN~Ia SEDs and relating supernova observables with absolute magnitude.
Temporal coverage and sampling, and SN-frame photometric coverage vary from supernova to supernova; seldom are there data
of different supernovae taken at common phase and band that can be directly compared.  Data are fit to SED models 
to transform these heterogeneous observations into parameters that have common meaning across all supernovae.  Constructing
these models requires physical insight, diligent scrutiny of data, and/or extreme flexibility in order to ensure capturing the parameters
that describe SN~Ia heterogeneity.  Once those parameters are identified, selecting the functional form to model their relationship with absolute magnitude
is non-trivial: the behavior is known to be non-linear  and there is the difficulty of choosing a function that
leaves no residual bias over the full parameter space while not overfitting the realized data set.  We therefore seek an alternative
to choosing the explicit functions for
the time-evolving SED and the parameter--absolute magnitude relationship.

This article introduces an alternative approach: the evolving SED for a supernova plus foreground dimming are modeled as a Gaussian process:
a function whose value at each input is known to lie within a Normal distribution but whose precise value is unknown
and whose values at different inputs may be correlated \citep{GPML}.
The simple form
of the Normal distribution
leads to analytic expressions for the likelihood of the data for the Gaussian process model,
and for the predicted mean and covariance of the fluxes at any wavelength and epoch.
Modeled as a Gaussian process, the description of the underlying SED of a single supernova
is influenced by each datum and thus has more degrees of freedom than
either of the SEDs underlying MLCS2k2 and SALT2. 
Here supernova absolute magnitudes are modeled as a function
of the flux values at different epochs and wavelengths regressed using the first Gaussian process.
In one case a linear model is used.  In a second
Gaussian processes are again used to model supernova absolute magnitudes as a non-linear function.
  Previous applications of Gaussian processes in the analysis of supernova data are
given in \citet{2010PhRvL.105x1302H,2012arXiv1204.2272S}.

Our approach contrasts with the usage of Gaussian processes in \citet{2011ApJ...731..120M}.
We model the covariance between single-band absolute magnitudes of different supernovae with a
parameterized function that depends on  differences between light-curve shapes and colors.
The absolute magnitude 
of a supernova is regressed from the absolute magnitudes
of other supernovae.  The distance is then inferred from the observed and absolute magnitude
in that band; independent information potentially available in the distances
from other bands is not included.
\citet{2011ApJ...731..120M} focus on supernova distance and its covariance with a suite of light-curve parameters.
There independent elements of the covariance matrix are fit in training.
The distance of a supernova is then inferred directly from its light curve parameters and the trained covariance matrix.

The approach presented in this article
can be used in the analysis of existing supernova data and can inform the design
of future experiments.   To date, the Nearby Supernova Factory \citep[SNfactory,][]{2002SPIE.4836...61A} 
has obtained and processed spectrophotometric time series
for a sample of 132 supernovae within the linear Hubble flow. 
These data can serve as a template sample to search for indicators of diversity in luminosity,
and contribute to
the low-redshift anchor when combined with available high-redshift SN Ia data and future photometric surveys such as the
Dark Energy Survey\footnote{\url{http://www.darkenergysurvey.org/}},
LSST\footnote{\url{http://www.lsst.org/lsst/}} or high-redshift spectrophotometric surveys as espoused by the Interim Science Working Group
of the Joint Dark
Energy Mission  \citep{JDEM-ISWG}.

In \S\ref{band:sec}, we present in detail how Gaussian processes can
determine SN Ia distances from broad-band photometry based on training with a low-redshift spectrophotometric
sample.  The framework for analyzing spectrophotometric data to calibrate
absolute magnitudes and/or simultaneously fit for cosmological parameters is sketched in
\S\ref{spec:sec}.  We finish with conclusions in \S\ref{conclusions:sec}.

\section{Case Illustration: Inferring Luminosity from Broad-Band Photometry}
\label{band:sec}
The methodology  to
determine absolute magnitudes
from broad-band light curves is applied to synthetic photometry generated from low-redshift SNfactory flux-calibrated spectroscopy
(spectrophotometry).
Two steps are involved.  First,
the photometry in all bands of all supernovae is modeled as a Gaussian process. Based on that model,
interpolated multi-band light curves are constructed for each supernova.  Then, the light-curve shapes
and colors derived from the interpolations
are treated as parameters that are related to absolute magnitudes using a second Gaussian process.
Based on that model, the absolute magnitude of a supernova with given light-curve shapes and colors
is predicted.

The motivation is to calibrate absolute magnitudes for supernovae at redshift $z$ observed
in the Dark Energy Survey (DES) $griz$ photometric system.  We work in the supernova restframe,
treating observations as if taken in the appropriate blueshifted DES filter set.  Focusing the analysis for
specific redshifts is  made possible for the first time by our
extensive set of spectrophotometric data and eliminates K-correction uncertainties
\citep{1996PASP..108..190K}. 
In this article we consider four redshifts and thus four filter sets corresponding to the restframe coverage of $z=0.00, 0.25, 0.50, 0.75$ supernovae.
The absolute magnitudes in each of the blueshifted $griz$ bands
at the epoch of $B$-band peak brightness are calibrated.

The analysis procedure for each redshift of interest is summarized as follows:
\begin{enumerate}
\item Synthetic photometry of SNfactory data is generated.  (\S\ref{data:sec})
\item The synthetic light curves from all supernovae are modeled as a Gaussian process.  Best-fit parameters for the model are determined.
(\S\ref{training:sec}).
\item SN light curves for a grid of epochs and bands and their associated covariance matrix  are constructed by interpolating between a
supernova's photometry  using the regression feature of Gaussian processes.   Subtracting the interpolated absolute magnitude at $B$-peak
for a fiducial band gives
light-curve shapes and colors.  For computational convenience the shapes and colors are re-expressed in terms of their principal components.
(\S\ref{lcparam:sec})
\item The absolute magnitude at $B$-peak as a function of the light-curve shapes and colors is modeled either as a second Gaussian process or a linear function and
trained.  (\S\ref{absmag:sec})
\item The trained procedure is applied to predict absolute magnitudes for an independent validation set.
This is done four times, running with completely independent validation sets
and corresponding complementary training sets.  We report 
the average weighted rms (wrms) and standard deviation of the four-fold validation runs. (\S\ref{validation:sec})
\end{enumerate}

\subsection{Computing}
Certain steps of Gaussian process regression involve the inversion or
decomposition of matrices having dimension equal to the number of data
points $n$.  To handle large data sets (thousands of points and up), we
implemented a framework of Gaussian process algorithms especially for
use with high-performance parallel architectures.  However, we note that
there are alternative strategies for mitigating the $\mathcal{O}({n^3})$
matrix inversion cost, such as selective pruning of data and sparse
linear algebra methods, but exploring them is beyond the scope of the
present paper.  Our implementation uses the well-tested ScaLAPACK
parallel distributed linear algebra library \citep{scalapack}.

The code base has the flexibility to operate on multiple platforms.
Development was performed using one or two processors of a 2.53 GHz Intel Core 2 Duo MacBook Pro.
The full analysis was run using 144 processors of Hopper, NERSC's Cray XE6.
A single training/validation analysis took 5 wall-clock minutes to complete.

High-performance computing facilitated code development and training: we were able to
quickly cycle through tests using different Gaussian process models and
optimize parameters in likelihood fits.  Once training of the model is completed, determining the
absolute magnitudes of the validation set is not computationally intensive.

\subsection{Data}
\label{data:sec}
The input to our analysis is the portion of the spectrophotometric data set obtained by
the SNfactory between 2004 and 2009 with the SuperNova Integral Field
Spectrograph \citep[SNIFS,][]{2004SPIE.5249..146L}
that had been fully processed as of early 2011.  SNIFS is a fully integrated
instrument optimized for automated observation of point sources on a
structured background over the full ground-based optical window at
moderate spectral resolution ($R \sim 500$).  It consists of a
high-throughput wide-band pure-lenslet integral field spectrograph
\citep[IFS, ``\`a la TIGER;''][]{1995A&AS..113..347B,2000ASPC..195..173B,2001MNRAS.326...23B}, a
multi-filter photometric channel to image the field in the vicinity of
the IFS for atmospheric transmission monitoring simultaneous with
spectroscopy, and an acquisition/guiding channel.  The IFS possesses a
fully-filled $6\farcs 4 \times 6\farcs 4$ spectroscopic field of view
subdivided into a grid of $15 \times 15$ spatial elements, a
dual-channel spectrograph covering 3200--5200~\AA\ and 5100--10000~\AA\
simultaneously, and an internal calibration unit (continuum and arc
lamps).  SNIFS is mounted on the south bent Cassegrain port of the
University of Hawaii 2.2~m telescope on Mauna Kea, and is operated
remotely.  Observations are reduced using the SNfactory's dedicated data
reduction pipeline, similar to that presented in \S4 of \citet{2001MNRAS.326...23B}.
A discussion of the software pipeline is presented in
\citet{2006ApJ...650..510A} and is updated in \citet{2010ApJ...713.1073S}.  A detailed
description of host-galaxy subtraction is given in \citet{2011MNRAS.418..258B}.

The instrument characteristics and observing strategy enable the
construction of spectrophotometric time-series of supernovae sampled every  2 to 3
days up until $\sim 25$ days after maximum, and then 5 to 7 days out to $\sim 45$ days
after maximum.  The median signal-to-noise within 2.5-days of peak brightness is 10.2 per 2.4 \AA\ bin.
In our sample, 83\% of the objects have $z>0.03$ where peculiar velocities of order 300~km~s$^{-1}$ are expected to contribute less than 0.07~mag to the absolute magnitude dispersion (in fact 98\% have $z>0.01$). At the other extreme, 90\% have $z<0.08$ so are insensitive to present uncertainties in $\Omega_M$ or the dark energy equation of state, $w$.

The spectral time-series used are corrected for Milky Way dust
extinction \citep{1989ApJ...345..245C,1998ApJ...500..525S} but no attempt is made to correct for the
effects of circumstellar or host galaxy dust extinction.  Reference observations when the
supernova has faded away facilitate host-galaxy subtraction \citep{2011MNRAS.418..258B}.
Synthetic
photometry light curves are fit using SALT2 \citep{2007A&A...466...11G}
to obtain the date of $B$-band maximum in order to assign a phase to
each spectrum.  For our analysis, each spectral time series is
blue-shifted to rest-frame and the fluxes normalized to represent that
observed at a common distance for all supernovae assuming a flat universe with $\Omega_M=0.28$, $h=0.7$--- thus, ratios
in flux between supernovae can be interpreted as ratios in
luminosity or equivalently differences in absolute magnitude.

Synthetic photometry and uncertainties are generated for each of the blueshifted DES filter sets when there is sufficient wavelength
coverage, including the few cases when data from only one arm of the SNIFS spectrograph is available.  
Spectra from the two arms generally provide coverage of DES $gri$ for $z=0.00$, $griz$ for $z=0.25$, and $riz$ for $z=0.50, 0.75$;
these redshifts provide different restframe wavelength ranges with which to calibrate supernova absolute magnitudes. 
Photon fluxes are calculated and
are converted to magnitudes, as we find that the
the dominant sources of Gaussian dispersion are multiplicative (e.g.\ flux calibration) rather than additive
(e.g.\ detector dark current).
As the magnitudes refer to supernovae normalized to be at  a common distance
we refer to them as absolute magnitudes; the zeropoint of the magnitude system is irrelevant for the purposes of this article.

Application of SALT2 on Johnson-Cousins synthetic photometry
 of the full input data set
calibrates absolute magnitudes to a weighted rms of 0.147 mag, similar to the 0.16
mag found in earlier data productions
\citep{2009A&A...500L..17B,2011A&A...529L...4C}.

\subsection{Modeling Supernova Absolute Magnitudes}
\subsubsection{Modeling Supernova Light Curves}
\label{training:sec}
We begin by training a parameterized model for the multi-band light curves that make up our data set.
For a single blueshifted filter set, the multi-band light curves underlying supernovae are modeled as arising from a Gaussian process,
\begin{equation}
m_{\left(t,\lambda\right)} \sim GP\left(\bar{m}(t,\lambda;\mathbf{\bar{m}}_0),k_m(t,\lambda,t',\lambda';l_{k_m}, \bm{\sigma}_{k_m} )\right),
\label{xgp:eqn}
\end{equation}
where $m_{\left(t,\lambda\right)}$ is the photometric magnitude at epoch $t$ relative to $B$-band peak in the filter indexed by $\lambda$ .
(The definition of a Gaussian process and associated equations used in the analysis of this article are given in  \S\ref{appendix:sec}.)
Subscripted $(t, \lambda)$ represent points whereas those in unsubscripted parentheses are function variables.
This notation means that for a set of coordinates $\{\left(t,\lambda\right)\}$, the magnitude values are drawn from a Normal distribution
\begin{equation}
m_{\{\left(t,\lambda\right)\}} \sim \mathcal{N}\left(\{\bar{m}(t,\lambda;\mathbf{\bar{m}}_0)\},K \right),
\end{equation}
where the elements of the model covariance matrix are
\begin{equation}
K_{\left(t,\lambda\right),\left(t',\lambda'\right)} = k_m(t,\lambda,t',\lambda';l_{k_m}, \bm{\sigma}_{k_m} ).
\end{equation}
Measurements of a realized supernova are then also described by a Gaussian process,
 \begin{equation}
m^{\star}_{\left(t,\lambda\right)} \sim GP\left(\bar{m}(t,\lambda;\mathbf{\bar{m}}_0),k_m(t,\lambda,t',\lambda';l_{k_m}, \bm{\sigma}_{k_m} )+ n_m(t,\lambda,t',\lambda';\bm{\sigma}_{n_m})\right).
\label{xgp2:eqn}
\end{equation}

Given a set of measurements $m^{\star}_{\left(t,\lambda\right)}$, the likelihood that the data arises from a Gaussian process can be expressed analytically,
as well as the mean and covariance of the  probability distribution function of a finite set of underlying points
$m_{\left(t,\lambda\right)}$ (see \S\ref{appendix:sec}).

For our model, we choose as mean function
$\bar{m}(t,\lambda;\mathbf{\bar{m}}_0)$ the synthetic photometry of band $\lambda$ from the updated templates of
\citet{2007ApJ...663.1187H} offset by
zeropoints $\mathbf{\bar{m}}_0$ for each band.
The mean function could include a parameter for  the date of peak $B$ magnitude
but the spectral time series we work with has been preprocessed and is already expressed in terms of days relative to peak
determined by SALT2.
A zero mean function would lead to expectations of zero flux in temporal regions lacking data;  the Hsiao mean is used to avoid
abrupt deviations toward zero where there are temporal gaps in the light curves, whose locations differ  from supernova to supernova.

The kernel represents the covariance of the underlying flux: we use the square exponential kernel
\begin{equation}
 k_m(t,\lambda,t',\lambda';l_{k_m}, \bm{\sigma}_{k_m} ) = \sigma_{k_m}^2(\lambda)\delta_{\lambda \lambda'} \exp{
 \left[
 -\left(\frac{t-t'}{l_{k_m}}\right)^2 
 \right]
 },
 \end{equation}
 where each band has its own normalization factor  $ \sigma_{k_m}(\lambda)$ but shares a common time-scale $l_{k_m}$.
The $l_{k_m}$ does not represent the time scale of the light curves themselves, but rather those of deviations from the mean model. 
The kernel does not correlate the light curves in different bands.  Although supernova flux across bands is correlated,
for this stage of the analysis we preferred imposing no prior assumptions on color.  We opt for the
square exponential kernel as it is frequently used in the Gaussian process literature; model testing comparing the
likelihoods of other kernels is deferred for future study.

Measurements of $m$ include a nugget term that contains all other sources of variance in the photometry.
Our nugget model is
\begin{equation}
n_m(t,\lambda,t',\lambda';\bm{\sigma}_{n_m})=\left( \sigma_{n_m}^2(\lambda) + \mathrm{var}(m_{\left(t,\lambda\right)}) \right)\delta_{t,t'}\delta_{\lambda,\lambda'} ,
\end{equation}
where within the scope of our analysis there are no two measurements with the same $(t,\lambda)$ so the nugget applies per measurement.
Each band has its own intrinsic dispersion and the SNfactory photon noise has variance $\mathrm{var}(m_{\left(t,\lambda\right)})$.
The measurement covariance is very small given the small overlap of the $griz$ bands, and so is ignored. 

The kernel includes an additional term that accounts for radial peculiar velocities 
\citep{2006PhRvL..96s1302B}:
\begin{equation}
\sigma_{\alpha\beta}^2=\left[\sigma_{z}^{pec}\frac{5}{\ln{(10)}}\frac{(1+z_\alpha)}{\chi H}\right]^2\delta_{\alpha\beta},
\label{peculiar:eqn}
\end{equation}
where the peculiar velocity dispersion, $\sigma_{z}^{pec}$, can be either
a fixed or fit parameter, and $z_\alpha$ is the actual redshift of the supernova;
in this article we fix $\sigma_z^{pec}=300$ km s$^{-1}$.
Here the $\alpha$ and $\beta$ are supernova indices, so the
$\delta_{\alpha\beta}$ reflects the
common realization of peculiar velocity for a single supernova.  At
low redshifts the conformal distance can be conveniently approximated as
$\chi \approx z-0.75\Omega_Mz^2$ for a flat $\Lambda$CDM cosmology.

For each of four training/validation runs, 75\% of the SNfactory supernovae are designated as the training set.
Photometry with $S/N<50$ are culled as often being associated
with poor extractions and a supernova must have at least 16
light-curve points total over all bands to be included.
Cuts are applied after the division into 
training and validation sets, so each of the analyses do not have exactly the same sample sizes: after
the signal-to-noise and number-of-point cuts there remains 80-90 training
supernovae  per run.

All  training-set
supernovae are used to fit the free parameters (and hyperparameters) that give the maximum likelihood of the  Gaussian process model. The hyperparameters
$l_{k_m}$, $\bm{\sigma}_{k_m}$,  and $\bm{\sigma}_{n_m}$ are common to all SNe Ia and are fit simultaneously to the full ensemble
of data, whereas a distinct  $\mathbf{\bar{m}}_0$ is fit for each supernova.
The best-fit hyperparameters for  $l_{k_m}$, $\bm{\sigma}_{k_m}$, and $\bm{\sigma}_{n_m}$ for one of the 4-fold training runs
are shown in Table~\ref{lcgp:tab}; these values do not differ significantly between the
different realizations.  The strength of the model covariance between points in a single light curve range from 0.08--0.14 mag 
with correlation-length scales of around 6  days.
The intrinsic dispersion is 0.05--0.07 mag except in the restframe UV where the dispersion is 0.14 mag.

\begin{table}
\center
\begin{tabular}{c|ccccccccc}
System $z$ &  $l_{k_m}$ & $\sigma_{k_m}(g)$ & $\sigma_{k_m}(r)$& $\sigma_{k_m}(i)$ & $\sigma_{k_m}(z)$& $\sigma_{n_m}(g)$ & $\sigma_{n_m}(r)$& $\sigma_{n_m}(i)$ & $\sigma_{n_m}(z)$\\ \hline
0.00&5.63&  0.09&  0.09&  0.12&   \ldots &  0.05&  0.06&  0.07&  \ldots  \\ \hline
0.25&5.87&  0.13&  0.08&  0.09&  0.13&  0.14&  0.05&  0.06&  0.06 \\ \hline
0.50&6.18&   \ldots &  0.11&  0.08&  0.10&   \ldots &  0.06&  0.05&  0.05 \\ \hline
0.75&6.56&   \ldots &  0.10&  0.11&  0.11&   \ldots &  0.14&  0.06&  0.05 \\ \hline
\end{tabular}
\caption{Best-fit hyperparameters of the Gaussian process that models the supernova multi-band light curves, given in Equation~\ref{xgp:eqn}.
Given are the results of one of the four training sets for each of the blueshifted DES photometric systems.
All values have magnitude units except $l_{k_m}$, given in supernova-frame days.
\label{lcgp:tab}}
\end{table}

The regressed $g$-band date of maximum has a distribution with mean
and standard deviation of $0.058 \pm 0.76$ days. The temporal alignment of our model
is consistent with that of SALT2.

To summarize this first step in the analysis, supernova light curves are described by a parameterized
model for the covariance in their residuals about a parameterized mean function.  Light-curve photometry of
a supernova training set are used to optimize the parameters that best describe their scatter.  

\subsubsection{Supernova Parameterization}
\label{lcparam:sec}
We now turn to the supernova parameters used to describe the light curve of a single event, in analogy to the light-curve shape and color parameters
used in classical light-curve fitters.  The parameters adopted in this article are the (linearly transformed)
values of interpolated light curves.

Since we are ultimately interested in determining the magnitude at peak brightness,
both training and validation sets are pared down to include only those supernovae
having a first measurement at least two days before $B$ peak
to avoid extrapolating the value of $M_{(\lambda_0, t_0)}$.  This leaves 49--57 training supernovae.

\begin{figure}[htbp] 
   \centering
   \includegraphics[width=1.75in,angle=270]{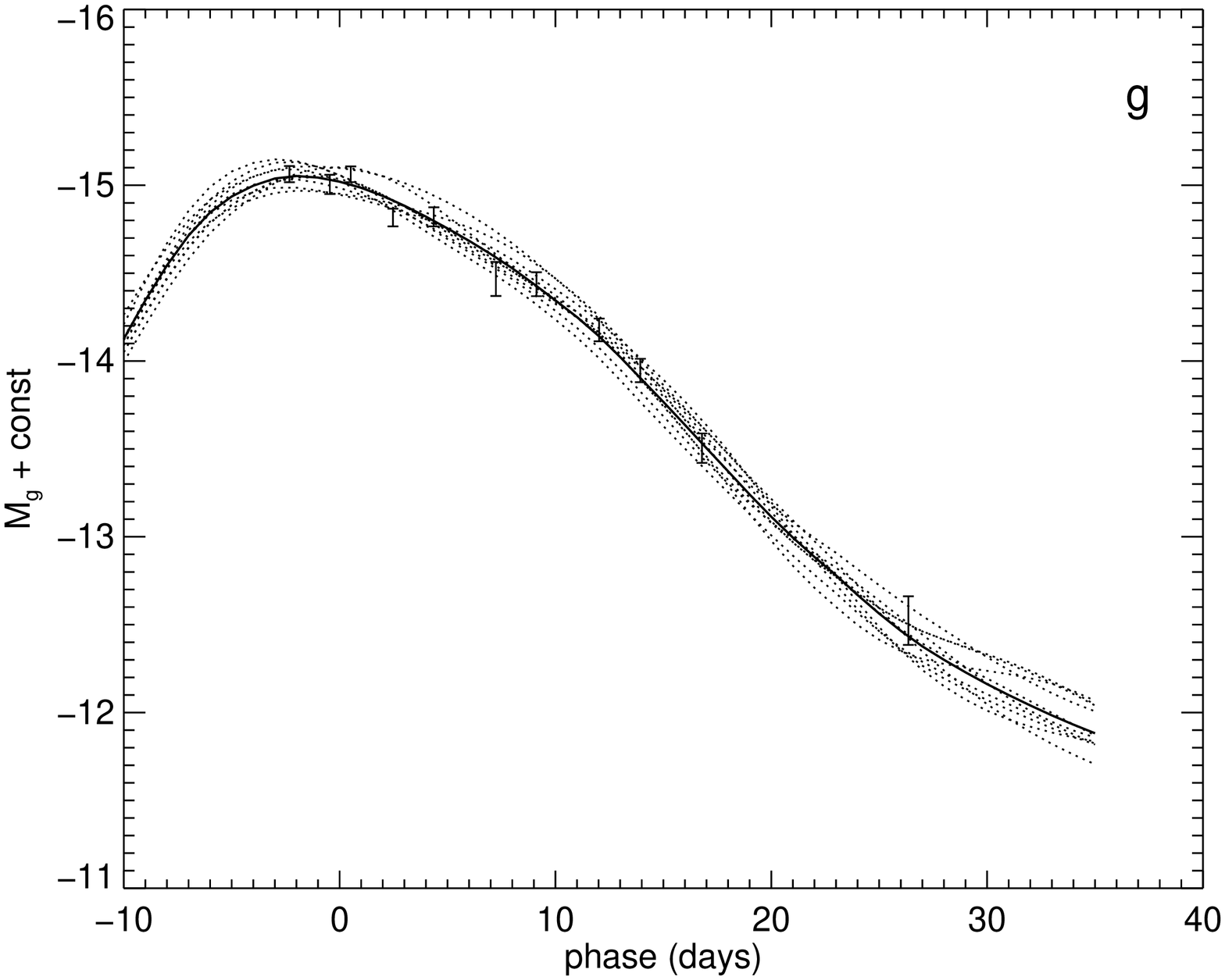} 
   \includegraphics[width=1.75in,angle=270]{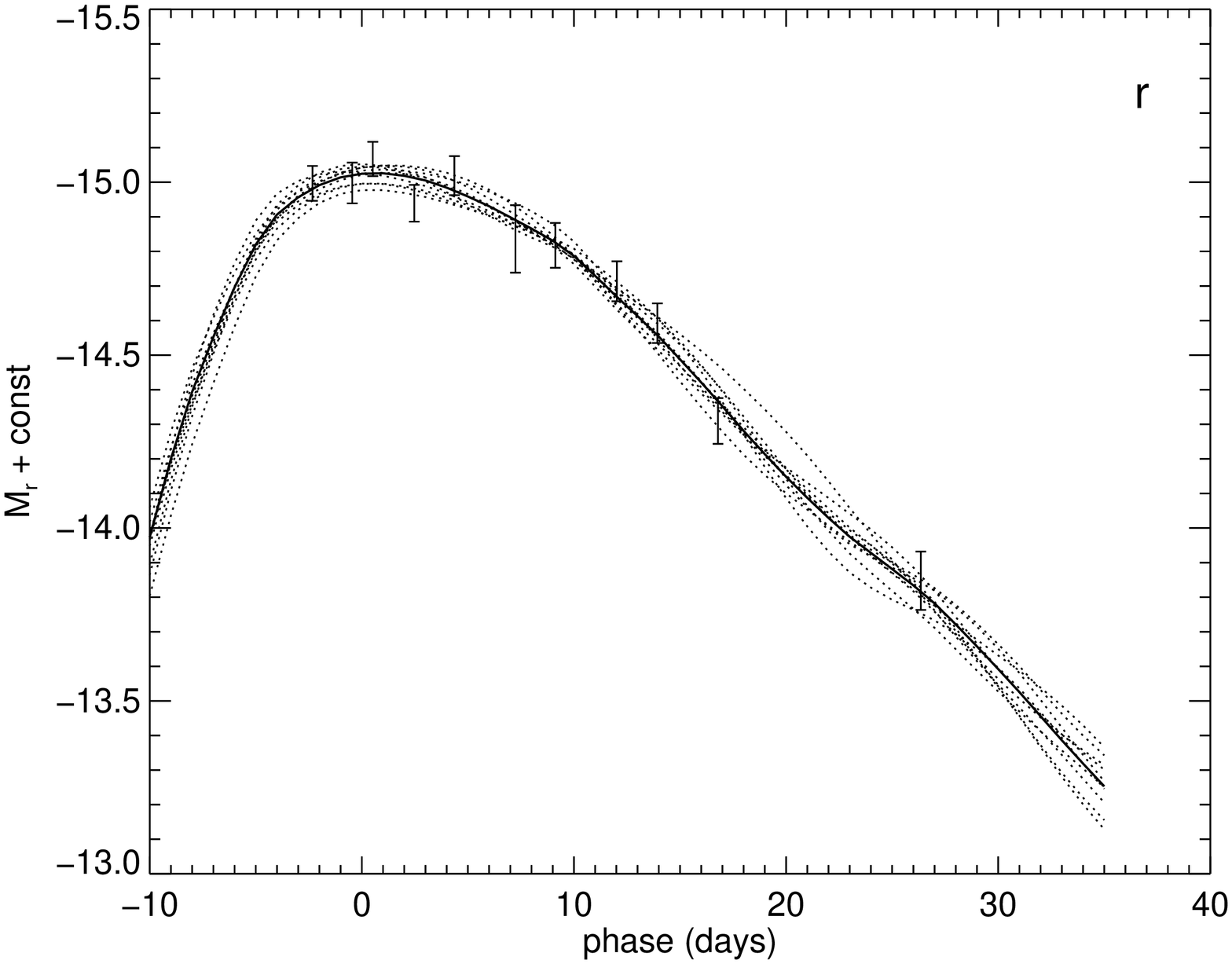} 
  \includegraphics[width=1.75in,angle=270]{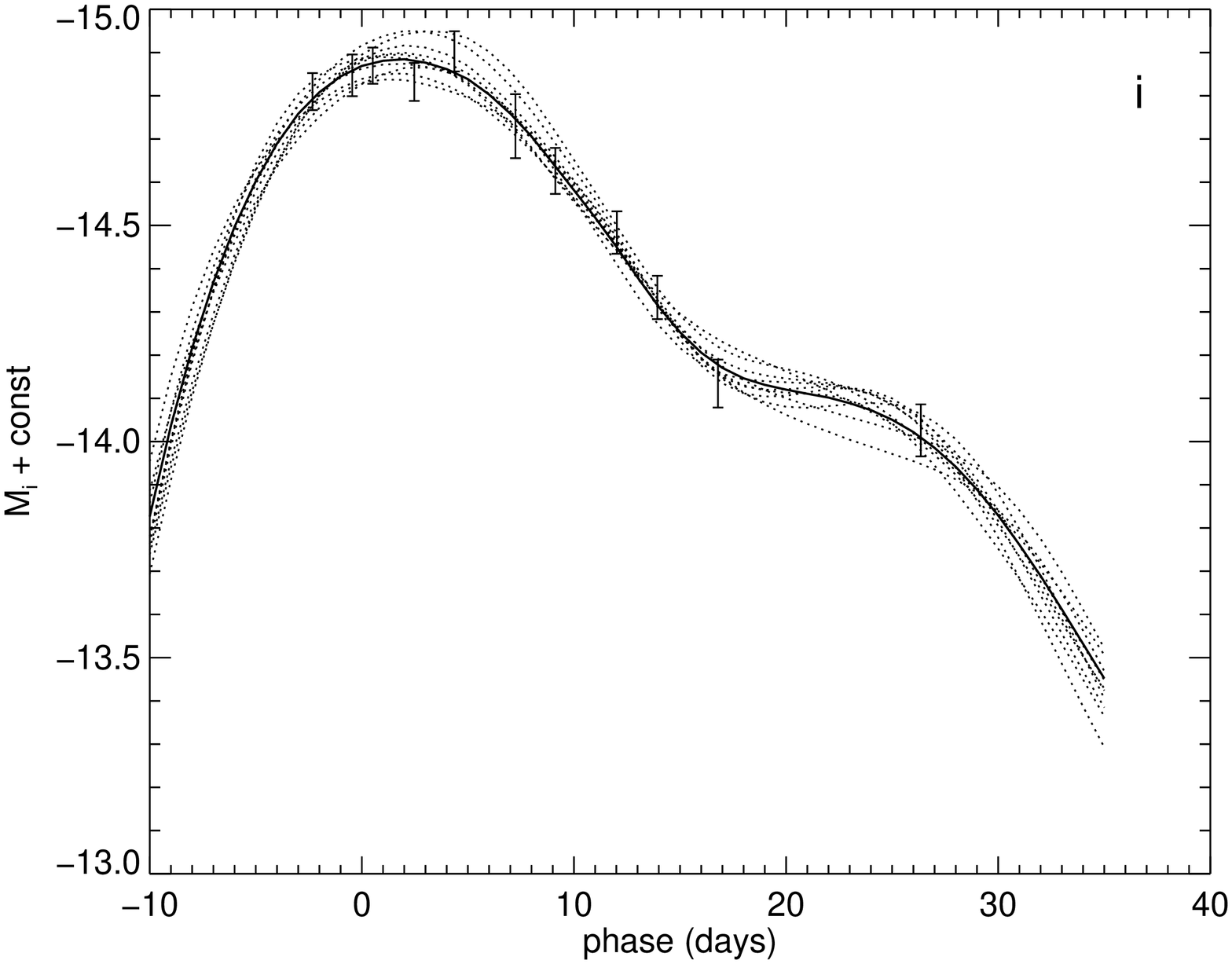} 
  \includegraphics[width=1.75in,angle=270]{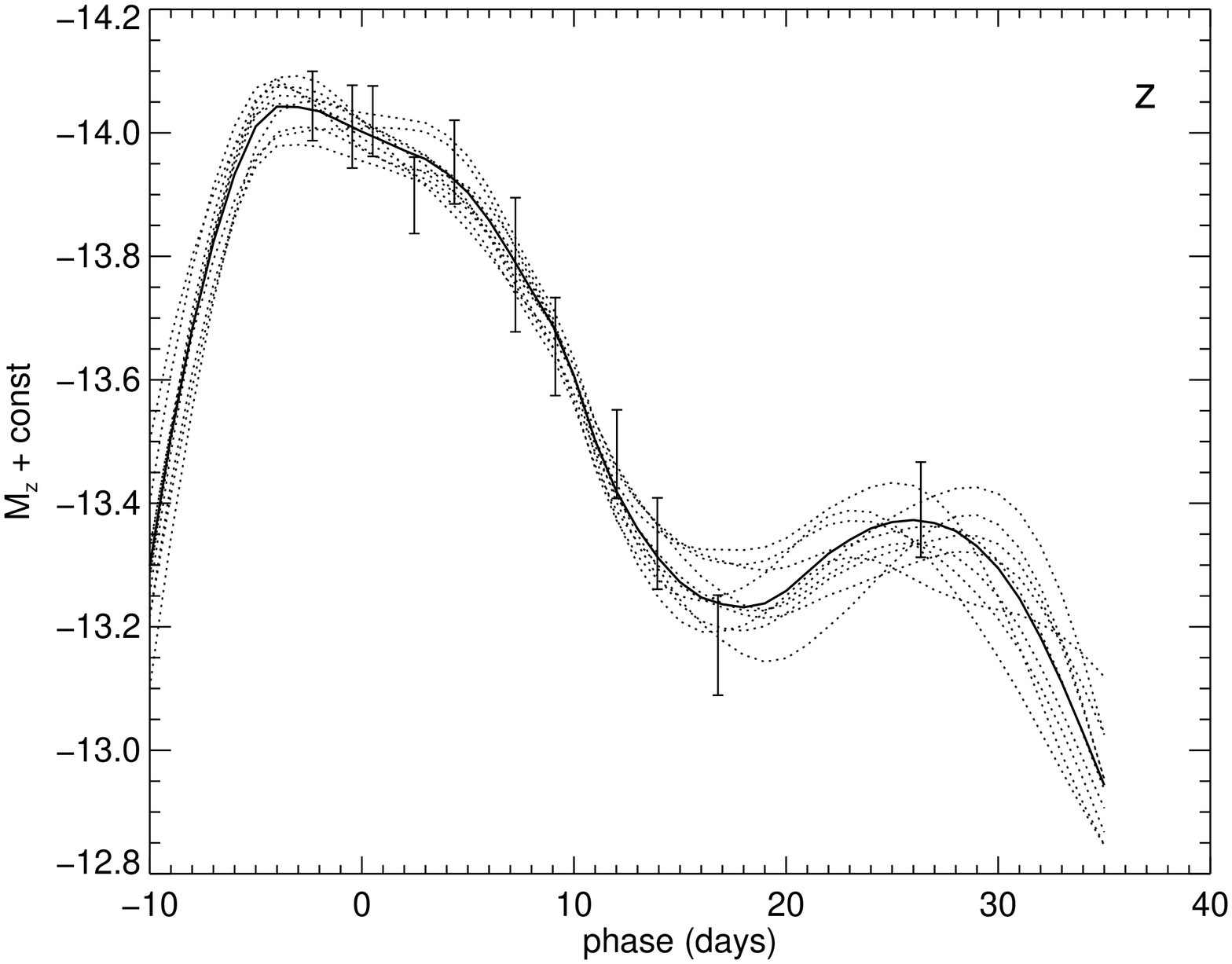} 
  \caption{Synthetic photometry, predicted mean (solid), and ten realized (dashed) $griz$ light curves for a random
  validation
  supernova (not used in the training) in the run with $z=0.25$ filter set.
   \label{lc:fig}}
\end{figure}

\begin{figure}[htbp] 
   \centering
   \includegraphics[width=1.75in,angle=270]{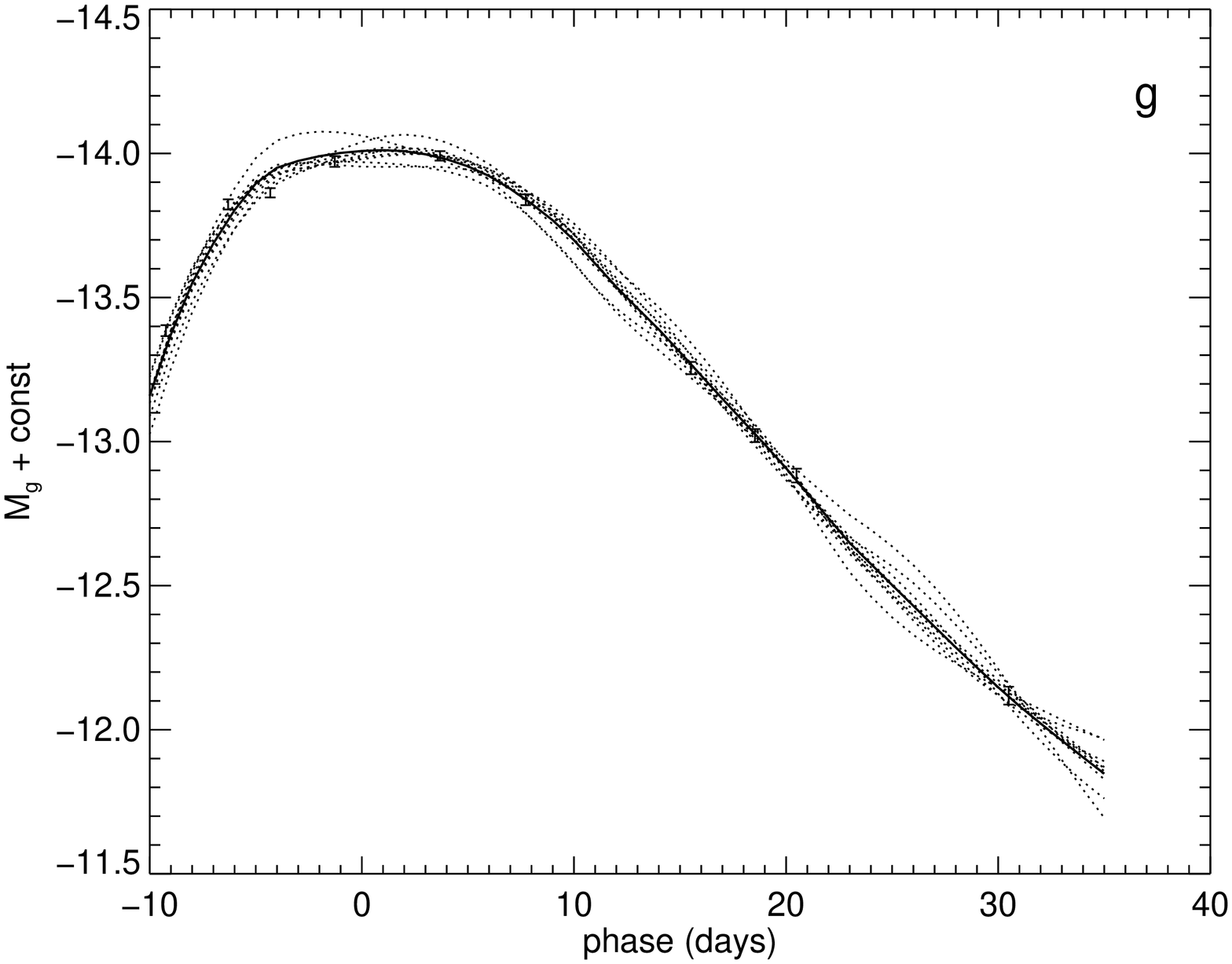} 
   \includegraphics[width=1.75in,angle=270]{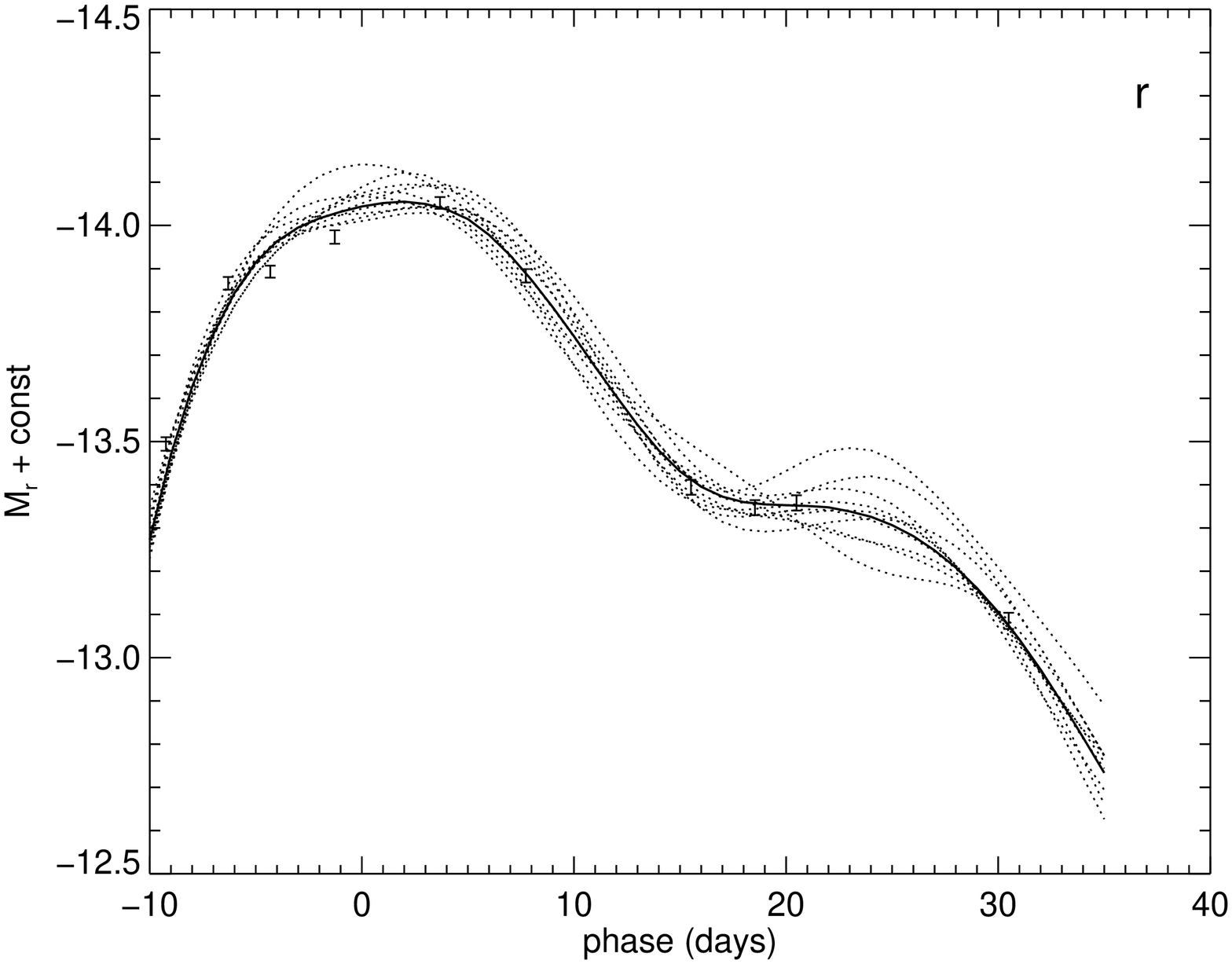} 
  \includegraphics[width=1.75in,angle=270]{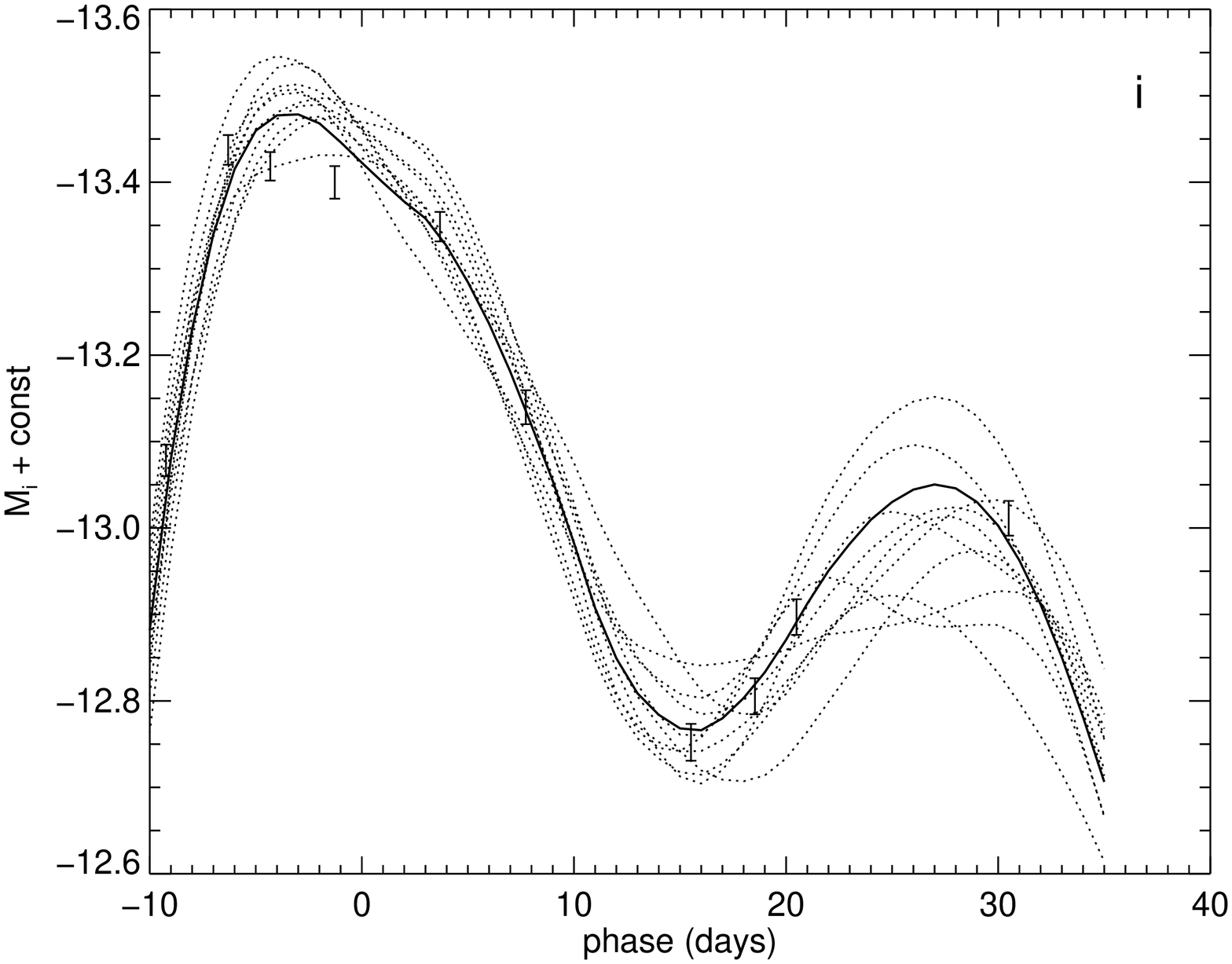} 
  \caption{Light curves for SN~\#19 in the $z=0$ filter set with an arbitrary normalization offset.  The predicted peak magnitudes for this object
  are off by $\sim 0.6$ mag.
  \label{bad1lc:fig}}
\end{figure}

\begin{figure}[htbp] 
   \centering
   \includegraphics[width=1.75in,angle=270]{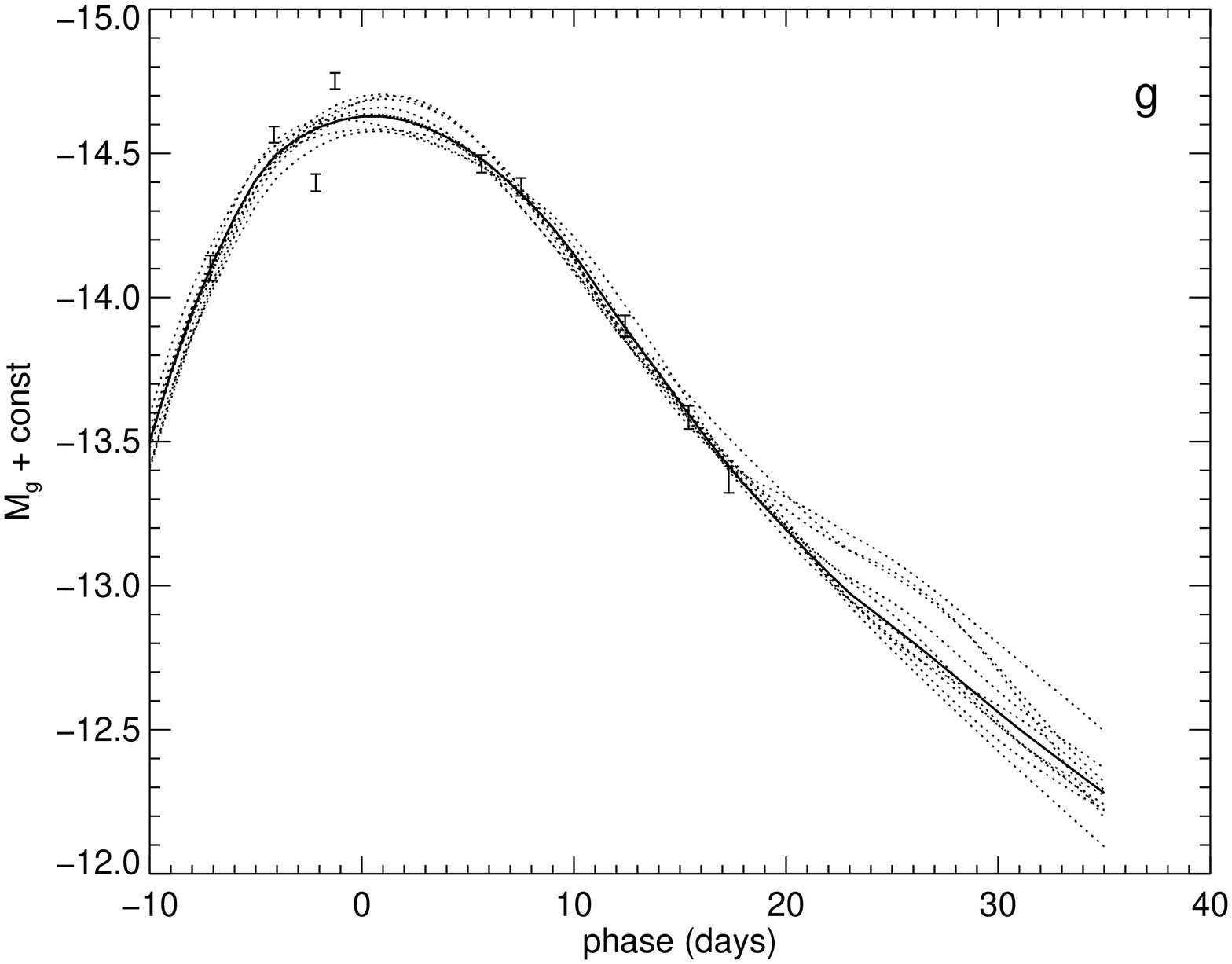} 
   \includegraphics[width=1.75in,angle=270]{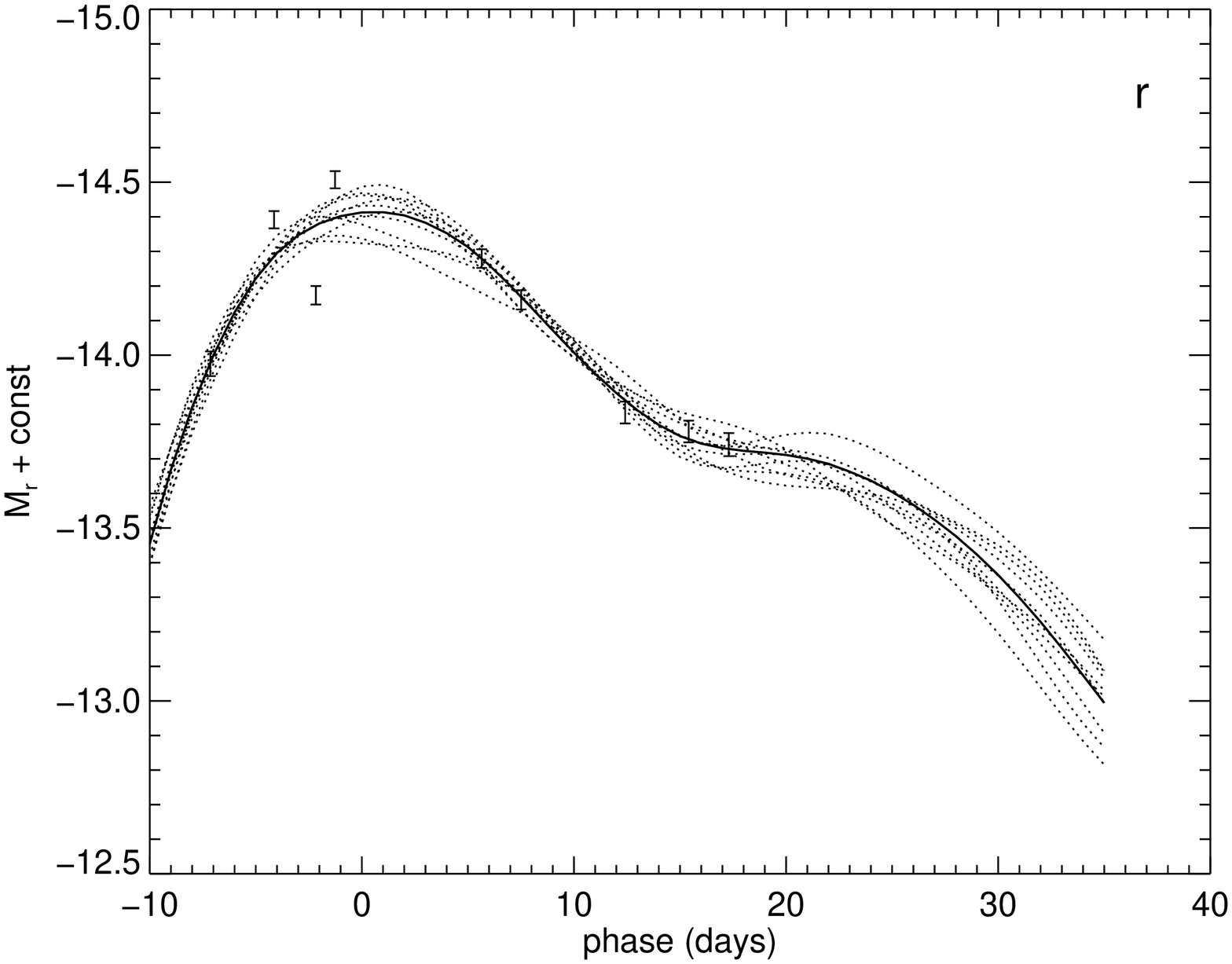} 
   \includegraphics[width=1.75in,angle=270]{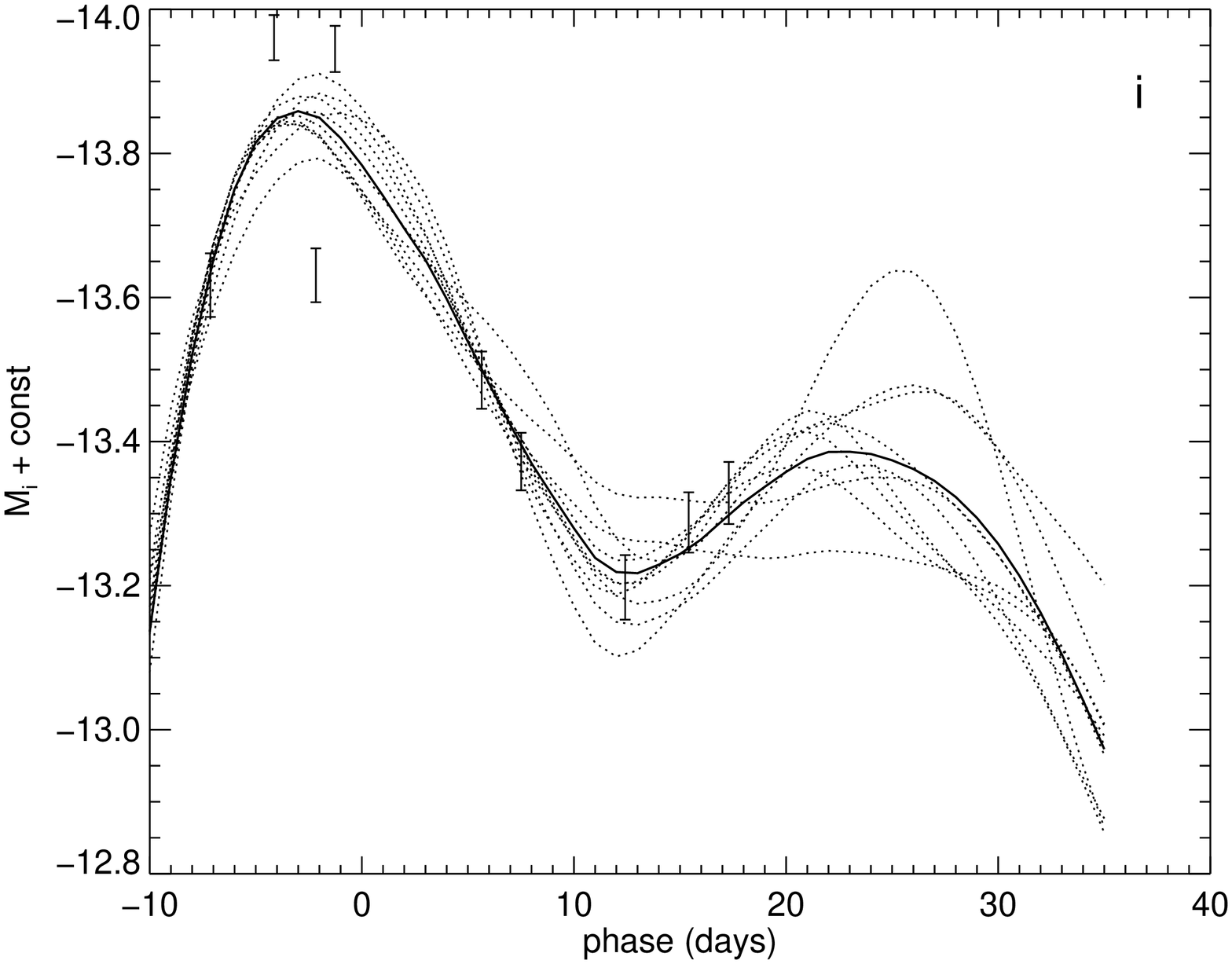} 
  \caption{Light curves in the $z=0$ filter set for SN~\#13, whose predicted peak magnitudes are off by $\sim 0.3$ mag.
   \label{bad2lc:fig}}
\end{figure}

Observed supernovae have disparate temporal sampling; those data must be  transformed
into a set of parameters that can be directly compared between supernovae.  The 
Gaussian process model of
the light curve is used to predict the fluxes on a common set of epochs, daily from $-10$ to 35 rest-frame days relative to $B$ peak, in the blueshifted DES bands
that are spanned by SNfactory data. (The choice of  daily sampling of the light curves was arbitrary and probably finer than necessary.)
The predictions (Eqn.~\ref{gpmean:eqn}) have covariance (Eqn.~\ref{gpcov:eqn}) so 
fifty light-curve grids are realized  per supernova to
 include uncertainties in the training and propagate uncertainties in the inferences.
Predictions and realizations of regressed light curves from sample supernova photometric measurements
are shown in Figures~\ref{lc:fig}, \ref{bad1lc:fig},  and \ref{bad2lc:fig}.  

We want to calibrate the supernova absolute magnitude in some band $\lambda_0$ at the phase of peak $B$ brightness $t_0$:
this is referred to as the ``{\it true}'' absolute magnitude and is estimated by the $M_{(\lambda_0, t_0)}$ predicted by our model.
This information is removed from the predicted light curves by subtracting  $M_{(\lambda_0, t_0)}$ from all other magnitudes.
What remains is a light-curve shape in the selected band $\lambda_0$ and colors for the other bands,
giving 137 ($46\times3-1$ for the case of three bands) or 183 ($46\times4-1$  for four bands) potential parameters to describe a supernova realization.

For practical reasons, we re-express the light-curve shapes and colors
in terms of their principal components ordered by the dimensions that carry the leading population variance by applying
a principal component analysis (PCA) on all 50 realizations of all training supernovae.
The  coordinates of the leading eigenvectors that account for the 95\% of the variance are used as the supernova parameters, and are denoted as $\mathbf{x}$.
This reduces the dimensionality of the problem down to $\sim 15$ in anticipation that the truncated components carry little real signal
that correlates with absolute magnitude.  As a result of this transformation each parameter no longer represents a specific color at a specific phase,
but rather the contribution to the temporal evolution of all colors from a single PCA eigenvector.
Note that due to the correlated noise in the regressed light curves, PCA is not encapsulating
pure supernova diversity efficiently in a minimal number of components.

\begin{figure}[htbp] 
   \centering
   \includegraphics[width=1.75in,angle=270]{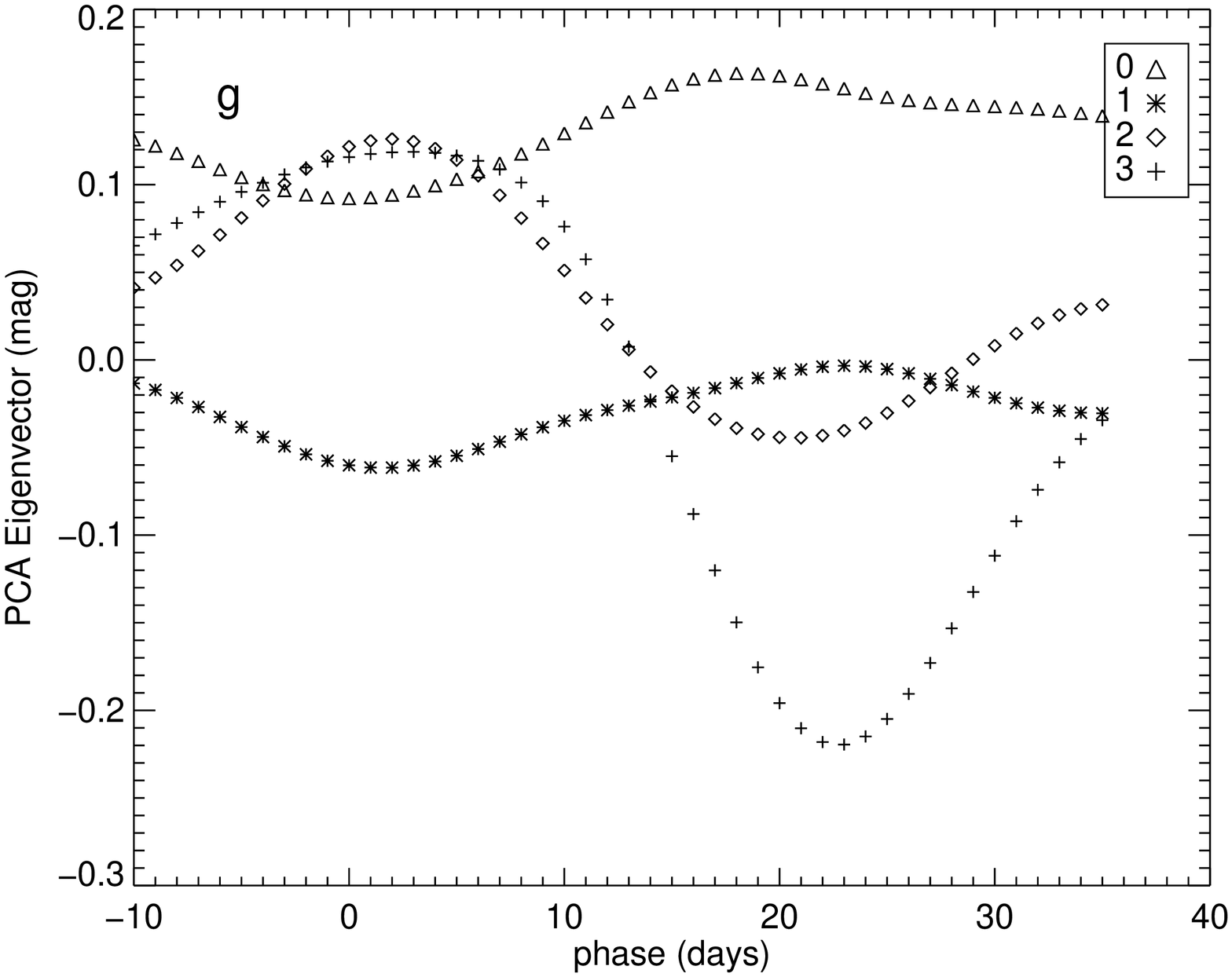} 
   \includegraphics[width=1.75in,angle=270]{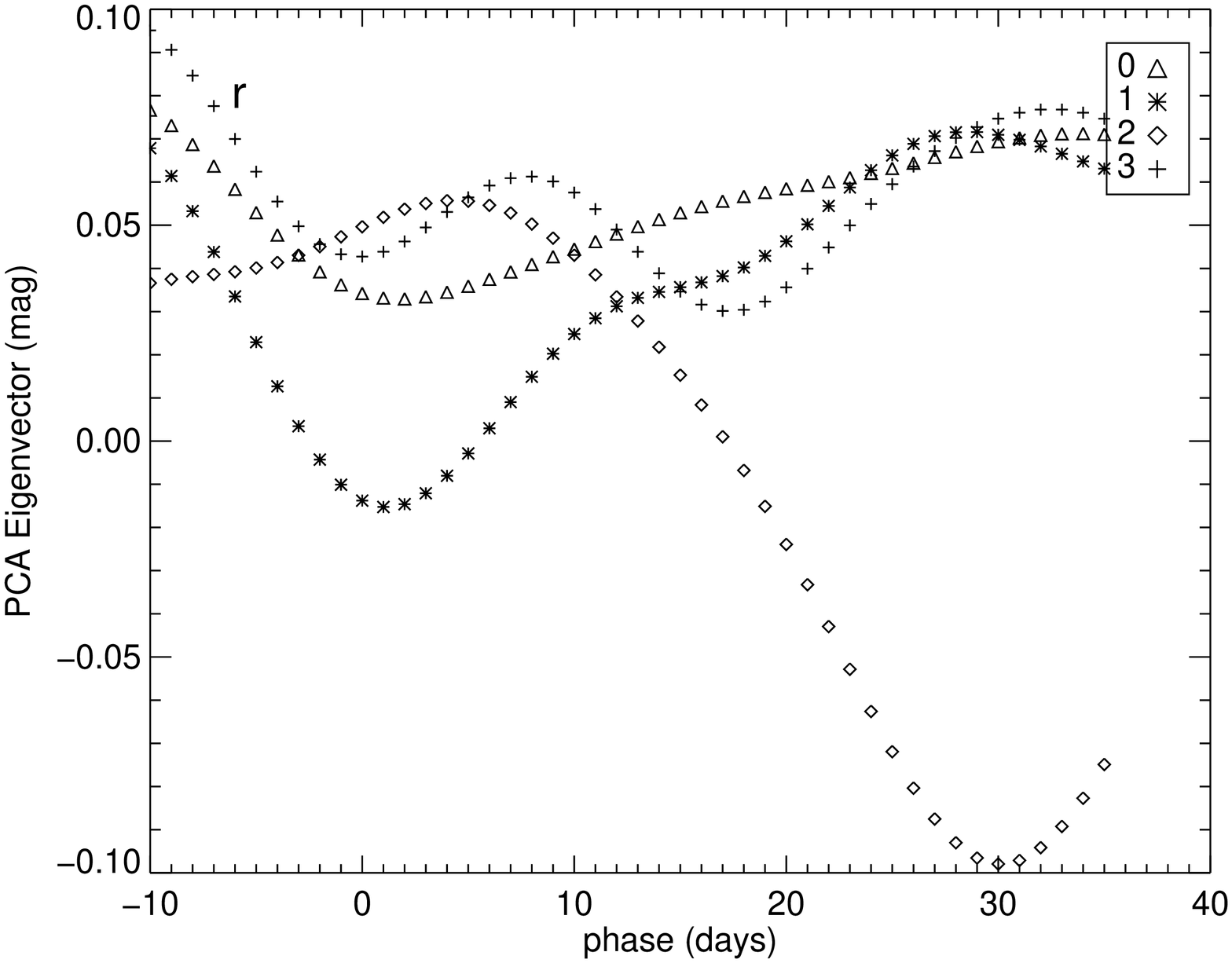} 
  \includegraphics[width=1.75in,angle=270]{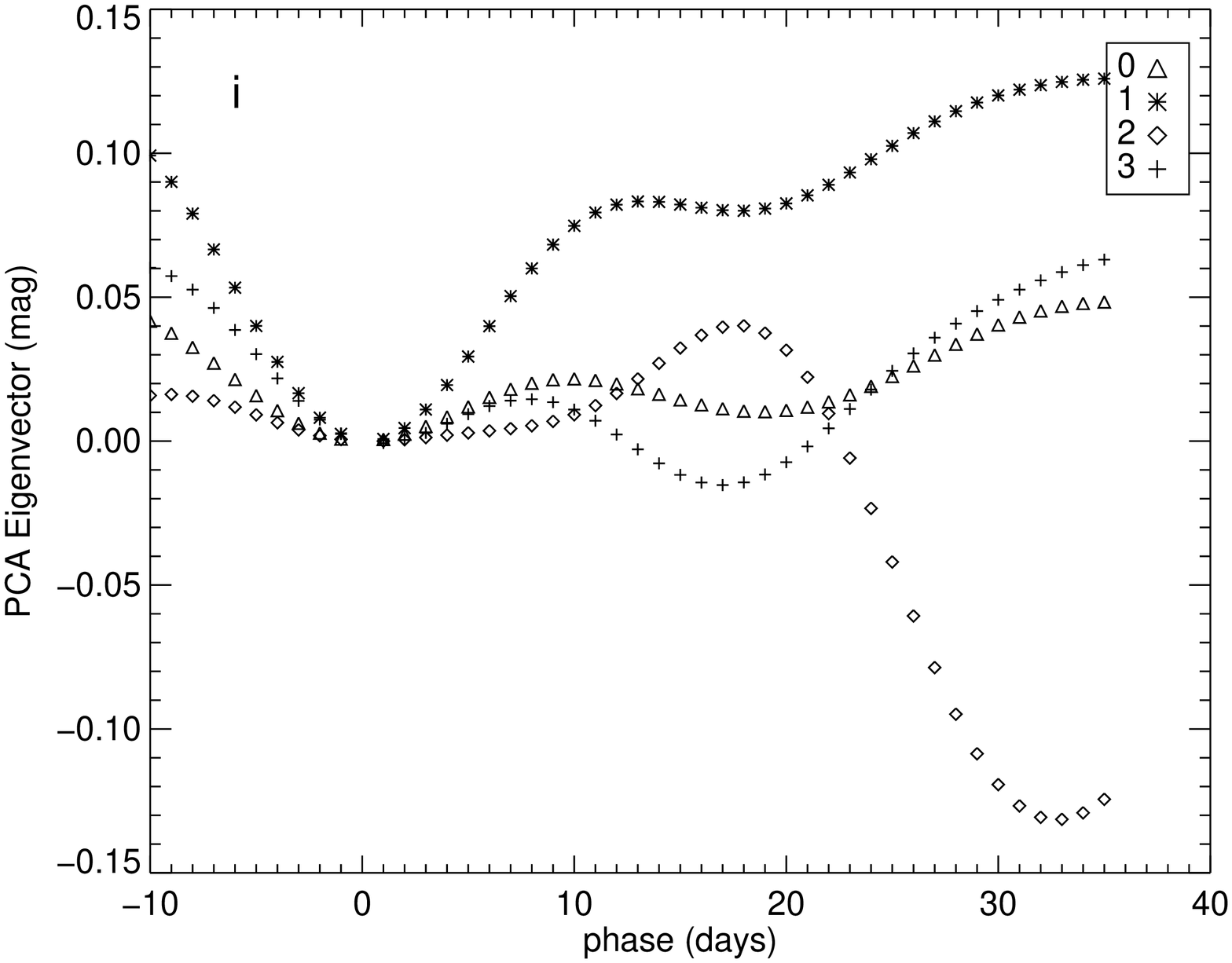} 
  \includegraphics[width=1.75in,angle=270]{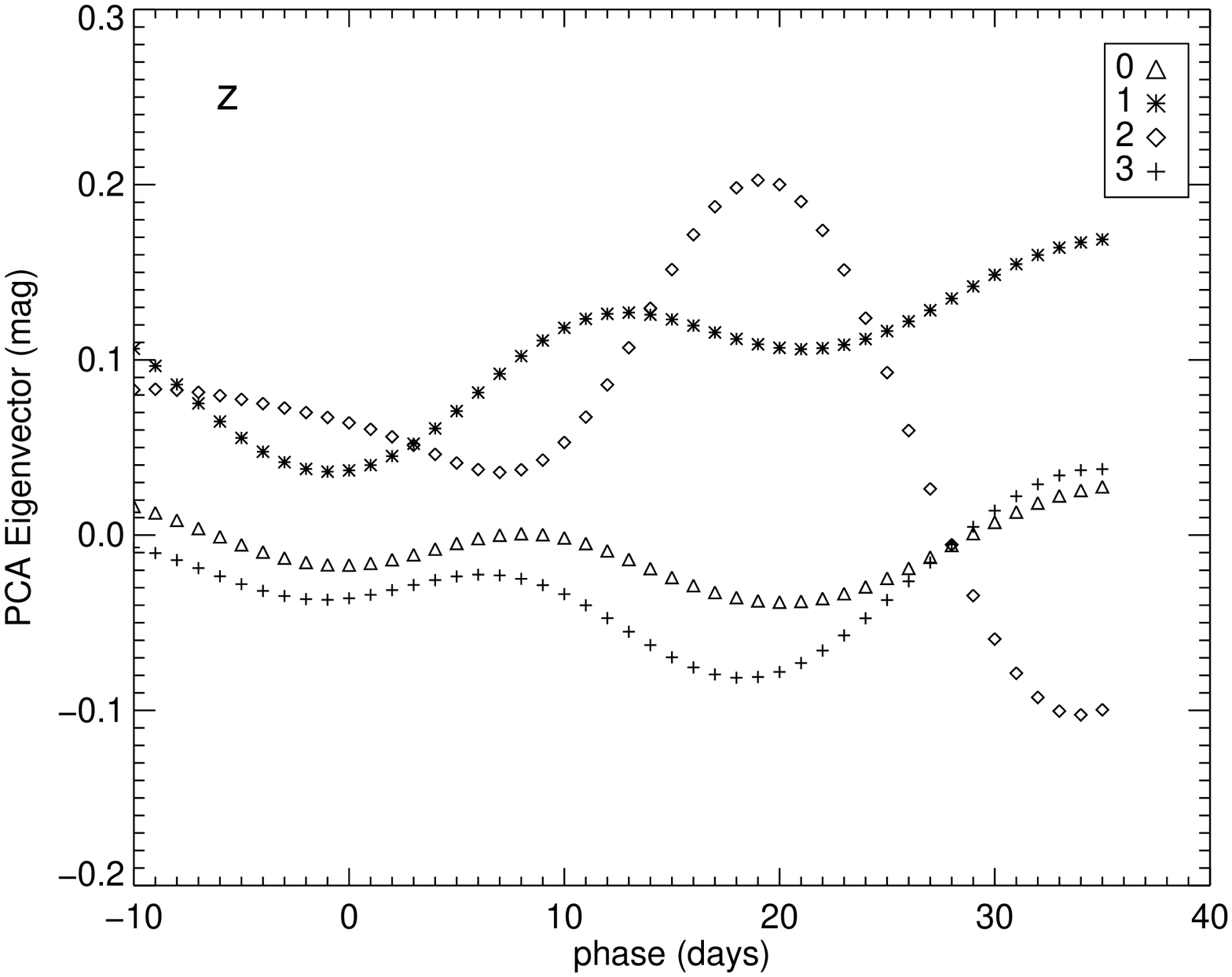} 
  \caption{First four PCA eigenvectors of the $z=0.25$ filter set plotted as a function of corresponding
  phase relative to $B$ peak split into $g$ (top left), $r$ (top right), $i$ (bottom left), and $z$ (bottom right), from
  a run calibrating the $i$ flux (hence the gap in the bottom left panel).  These eigenvectors capture
  54\%, 15\%, 9\%, and 5\% of the population variance.
     \label{pca:fig}}
\end{figure}

\begin{figure}[htbp] 
   \centering
   \includegraphics[width=4in,angle=270]{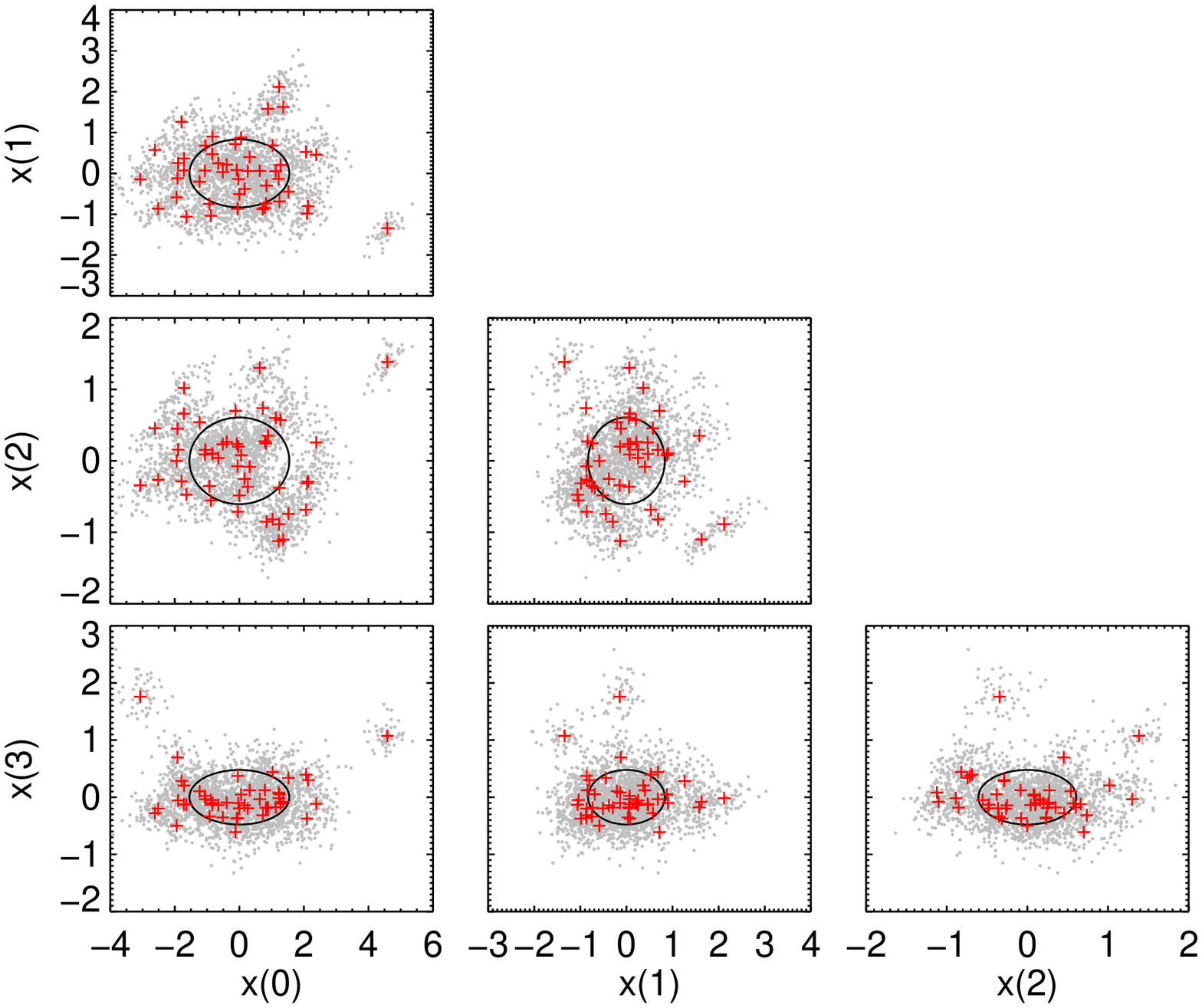} 
  \caption{Distribution of the first four PCA coefficients for  one of the runs calibrating the $i$-band of the $z=0.25$ filter set. 
  The predicted mean of each training-set supernova is shown in red with the 50 associated realizations shown in black.
  The ellipses have semi-major and semi-minor axes with the standard deviations of the distribution $\sigma_{x_i}$. \label{pcacoeff:fig}}
\end{figure}

As a representative example of one of the training sets, we show results from one of the runs calibrating the $i$-band of the $z=0.25$ filter set.
The first several PCA eigenvectors describing light-curve shapes and colors (after subtraction of the mean) are shown in Figure~\ref{pca:fig}.  
These eigenvectors capture
  54\%, 15\%, 9\%, and 5\% of the population variance.
The distribution of PCA coefficient values for the training set is shown in Figure~\ref{pcacoeff:fig}.  For each component $i$, the coefficients
have standard deviation $\sigma_{x_i}$.
In solid (red) are the coefficients for the mean light curves and colors for each training-set supernova.  Surrounding each of those points
is a cloud of gray points of the coefficients from the fifty realizations; the size of the cloud is reflective of the dispersion in the Gaussian process model
as well as the data quality. 
The $\sim 50$ training-set supernovae  do not fill a continuous distribution in this space, there are clear outliers due to realizations of a single supernova.
The distinct blobs containing multiple supernovae seen in the $x(0)$--$x(2)$ plot  are not significant as they are not as prominent in the three other training runs.

\begin{figure}[htbp] 
   \centering
   \includegraphics[width=1.75in,angle=270]{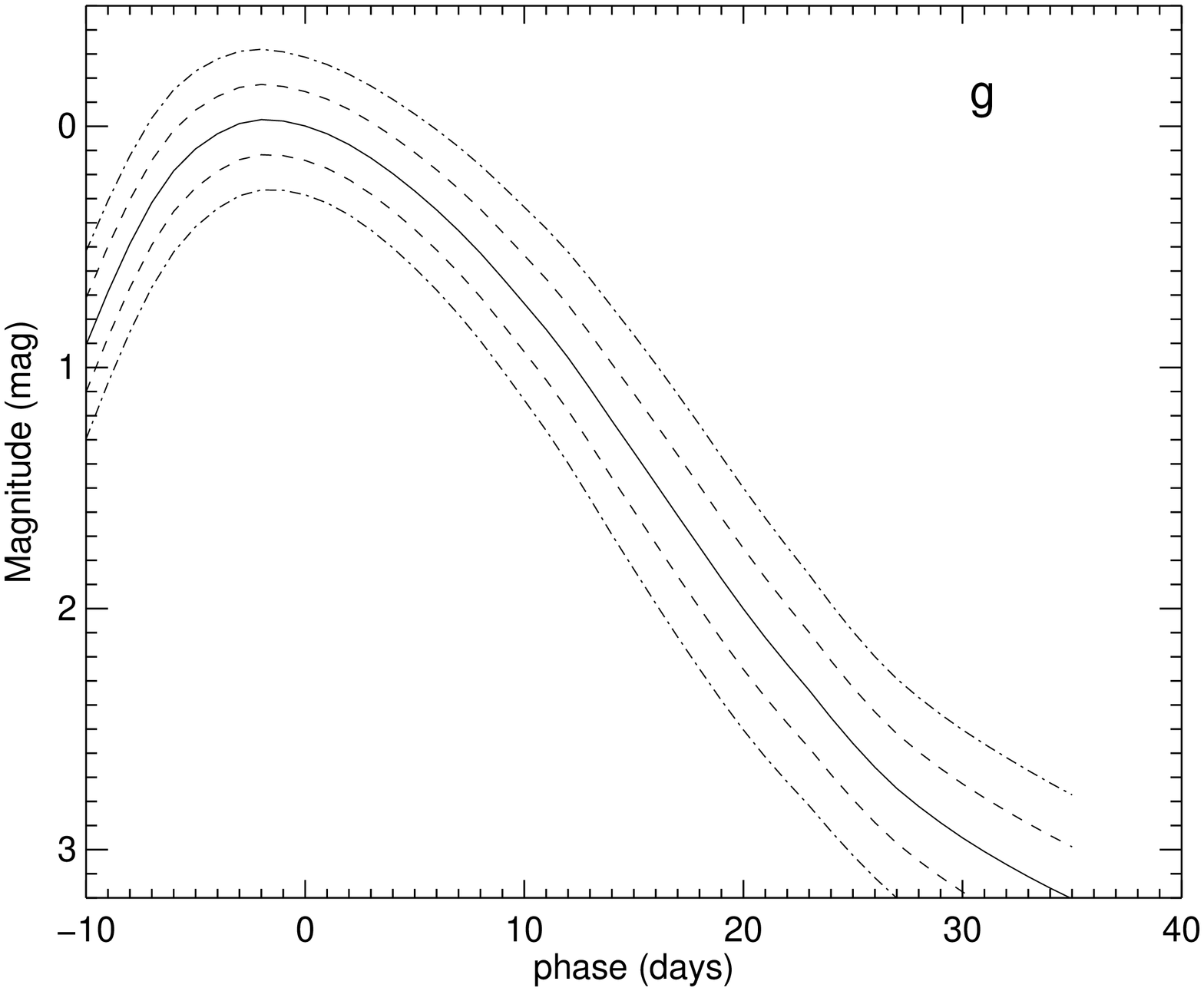} 
   \includegraphics[width=1.75in,angle=270]{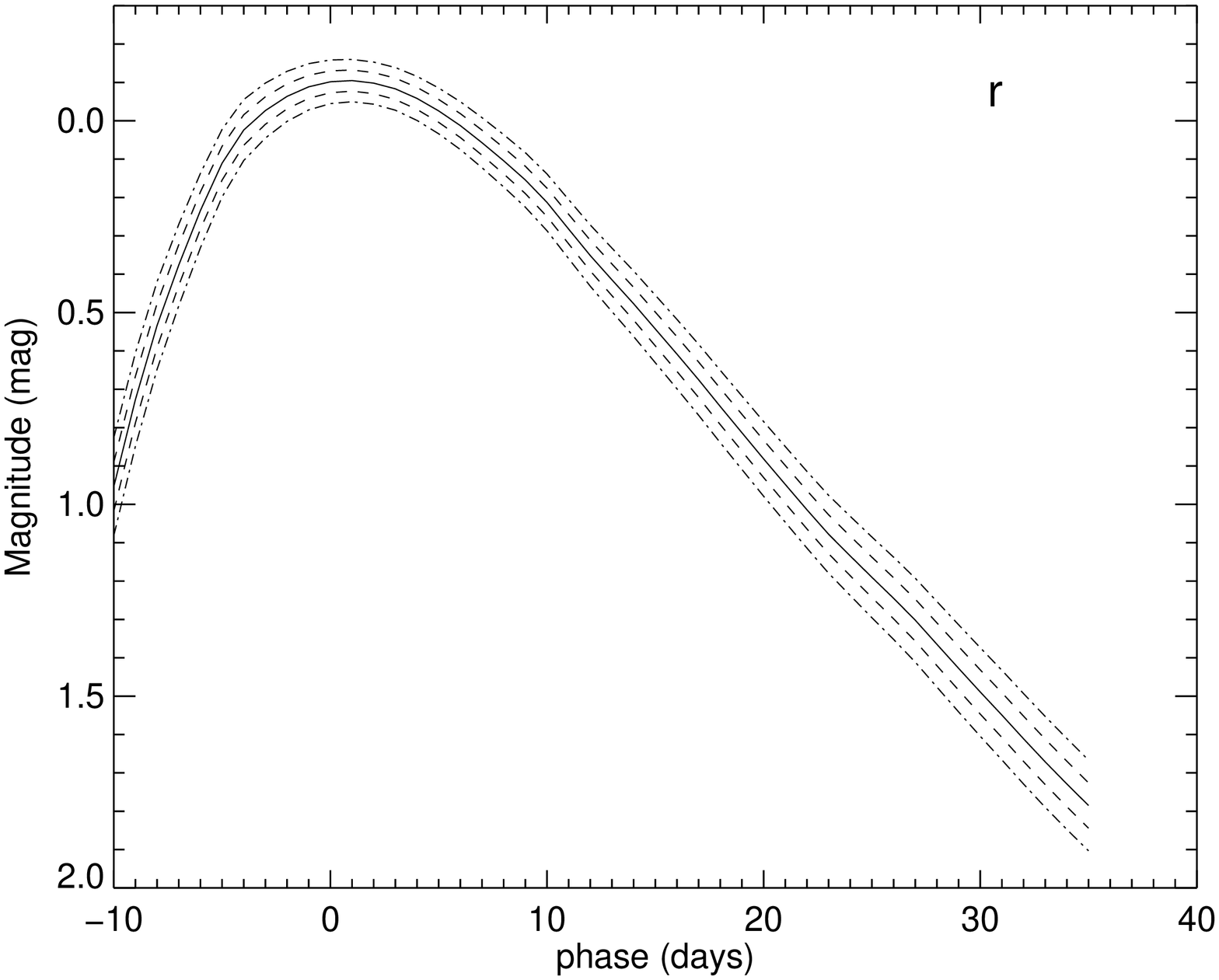} 
  \includegraphics[width=1.75in,angle=270]{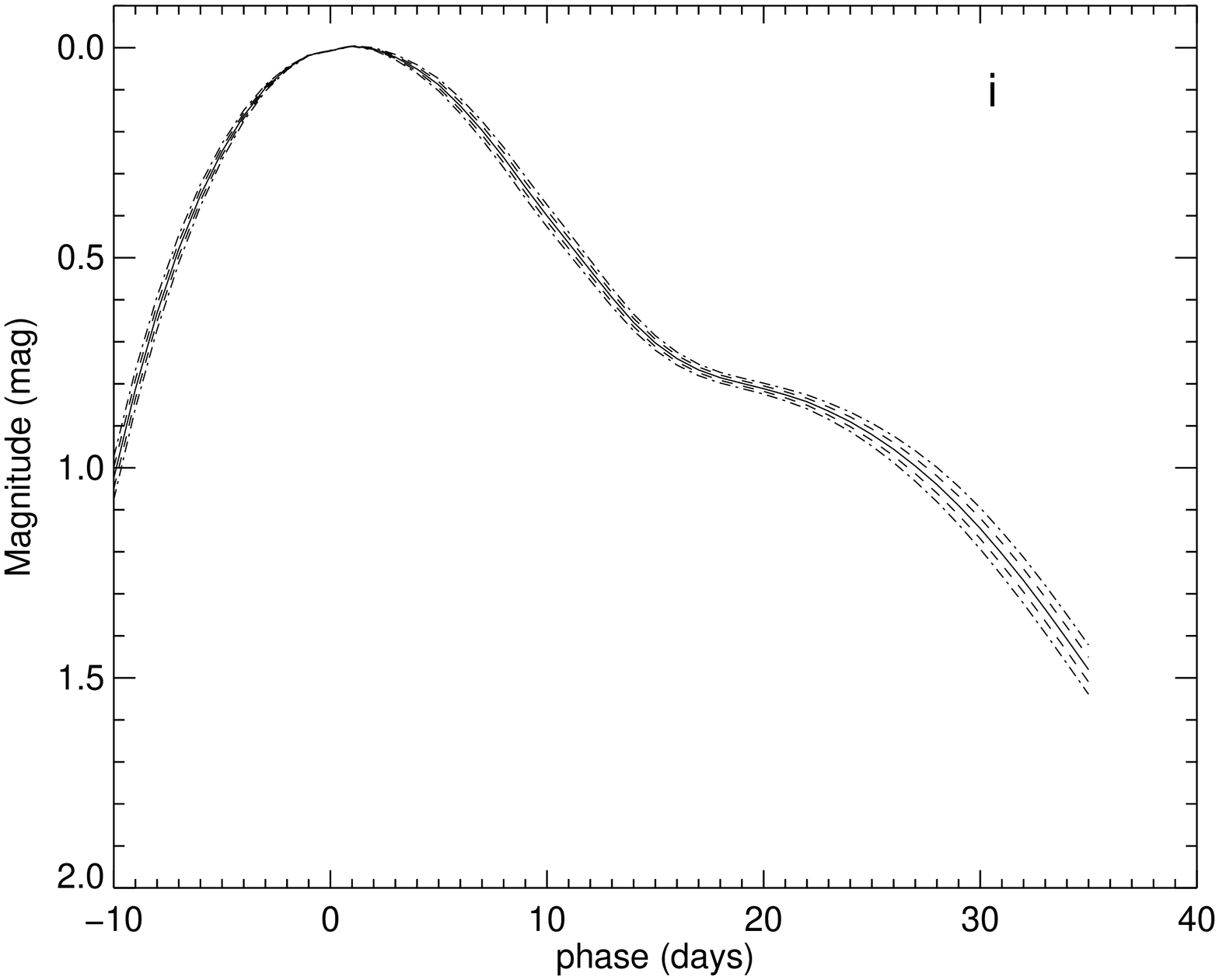} 
  \includegraphics[width=1.75in,angle=270]{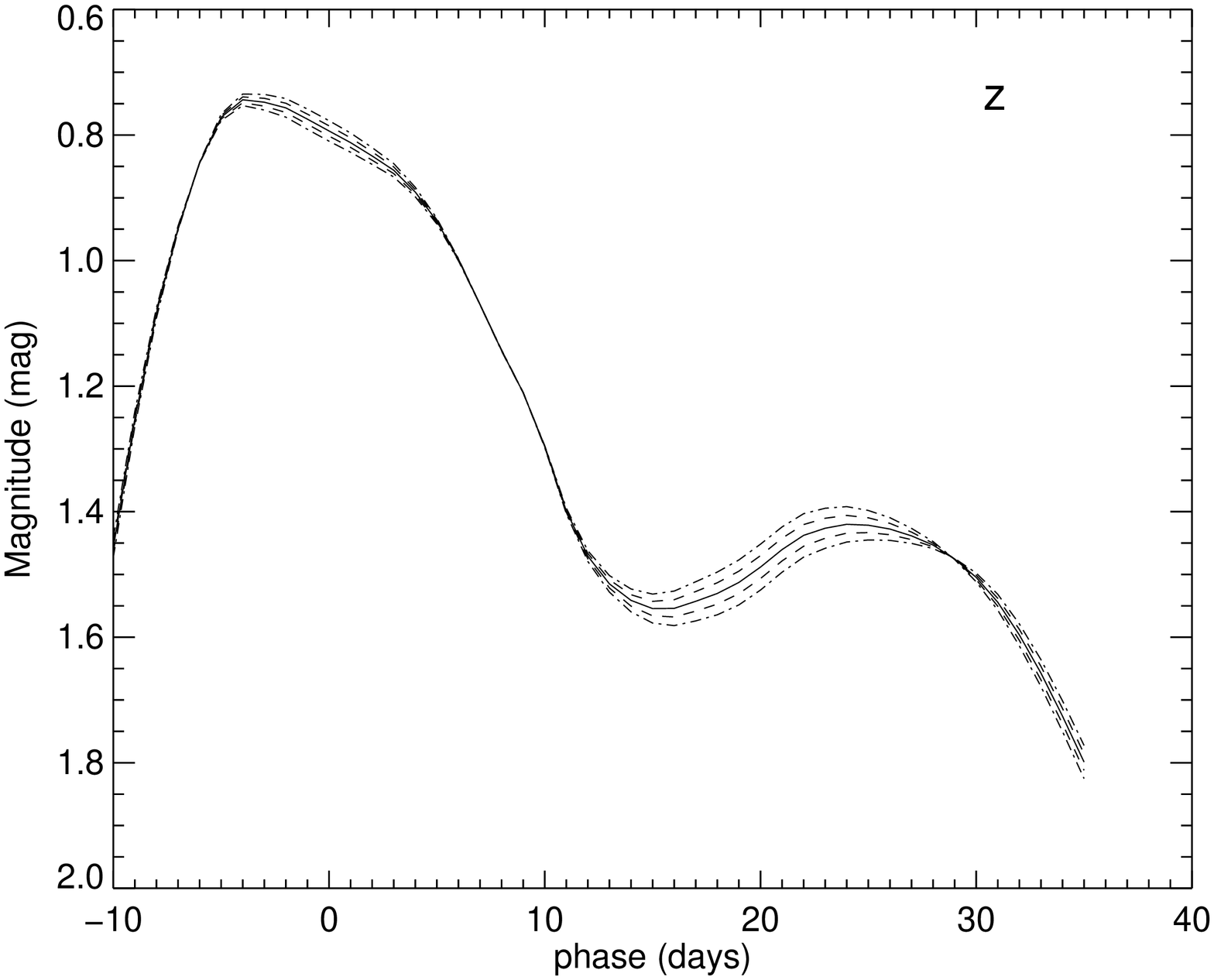} 
  \caption{Mean and perturbed light curves from 1-- and 2--$\sigma_{x_0}$ 
  changes in the $x(0)$ parameter
  for the $z=0.25$ filter set from
  a run calibrating the $i$ flux, plotted as a function of corresponding
  phase relative to $B$ peak. The four bands are plotted: $g$ (top left), $r$ (top right), $i$ (bottom left), and $z$ (bottom right). 
     \label{pca0:fig}}
\end{figure}

\begin{figure}[htbp] 
   \centering
   \includegraphics[width=1.75in,angle=270]{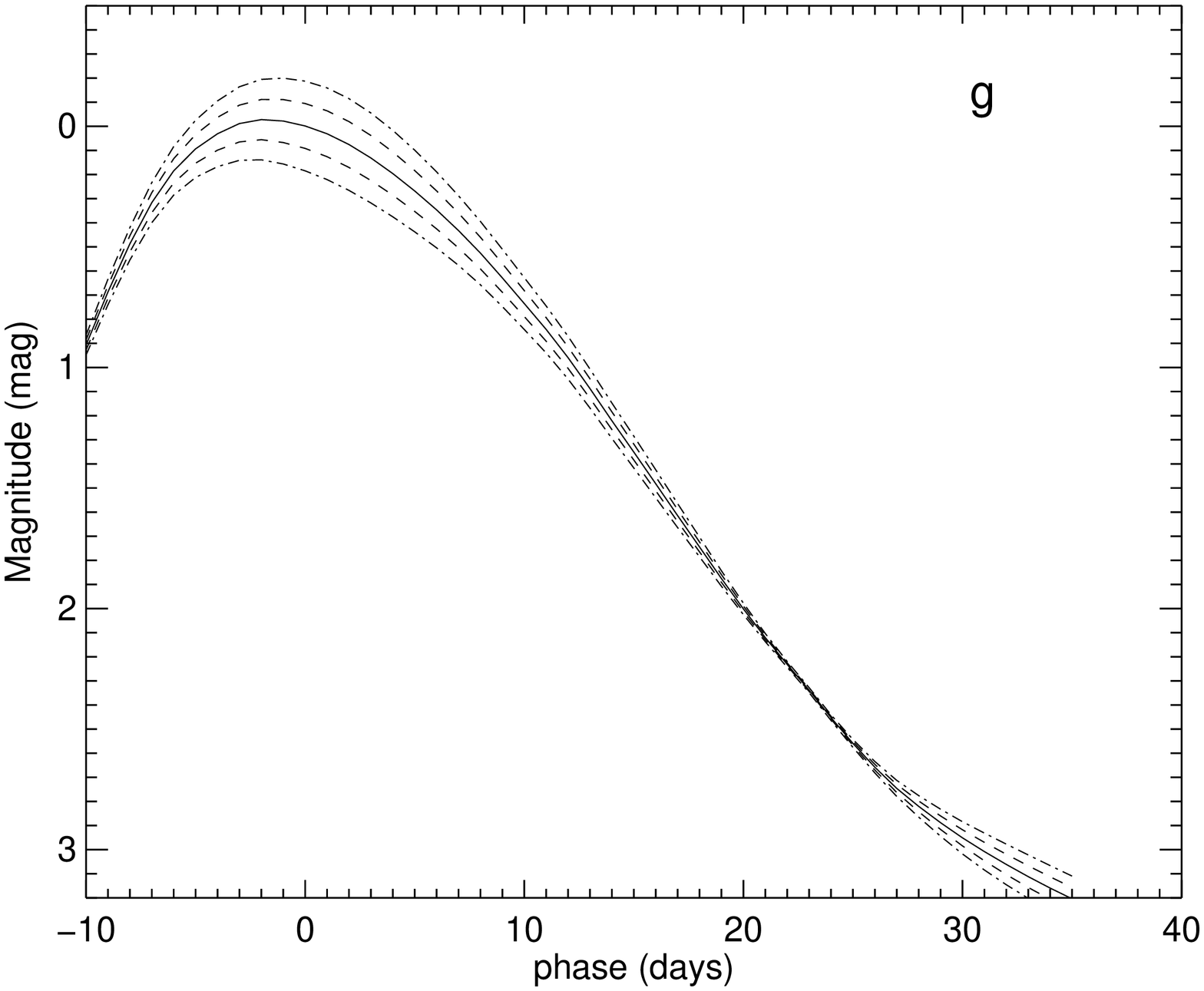} 
   \includegraphics[width=1.75in,angle=270]{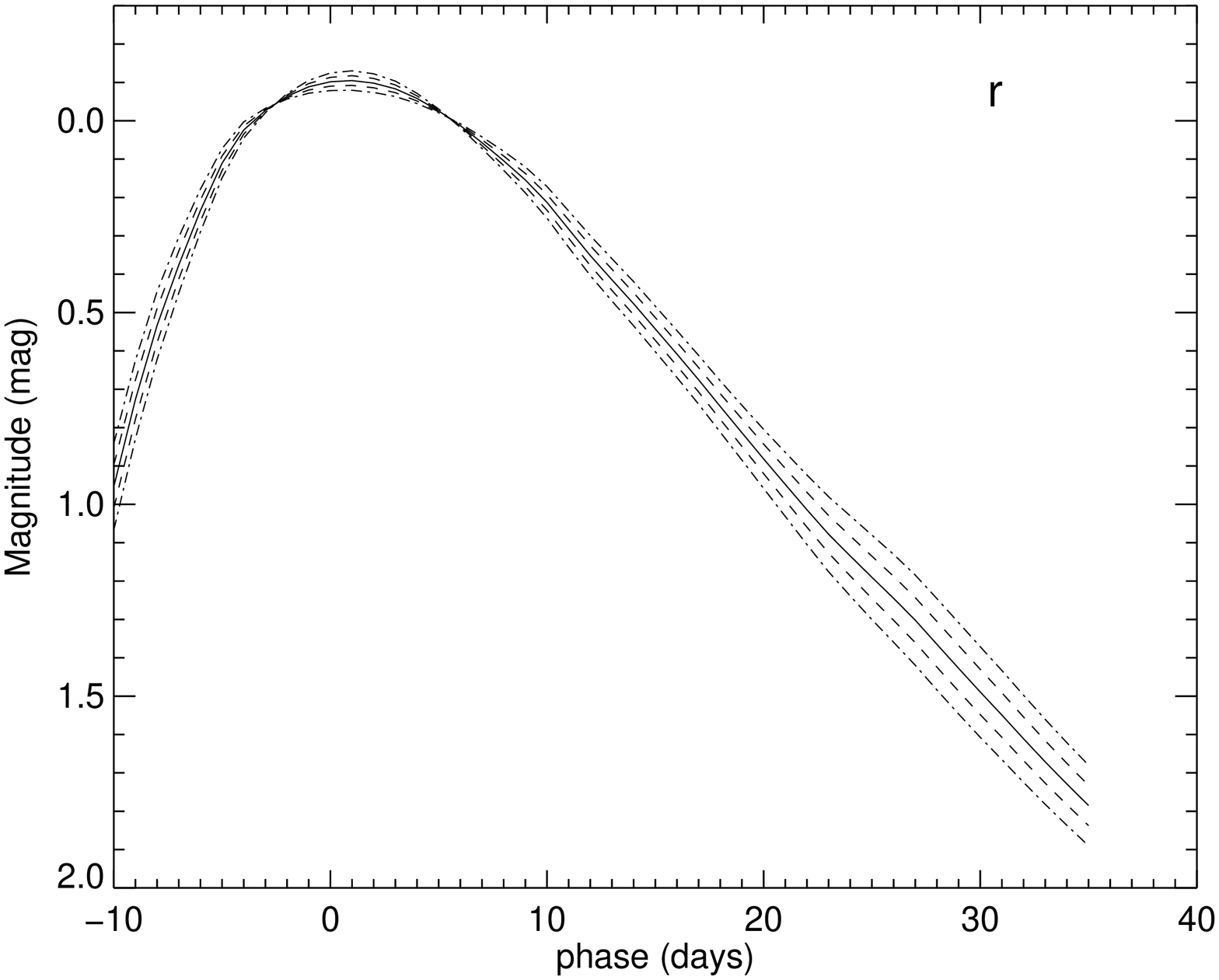} 
  \includegraphics[width=1.75in,angle=270]{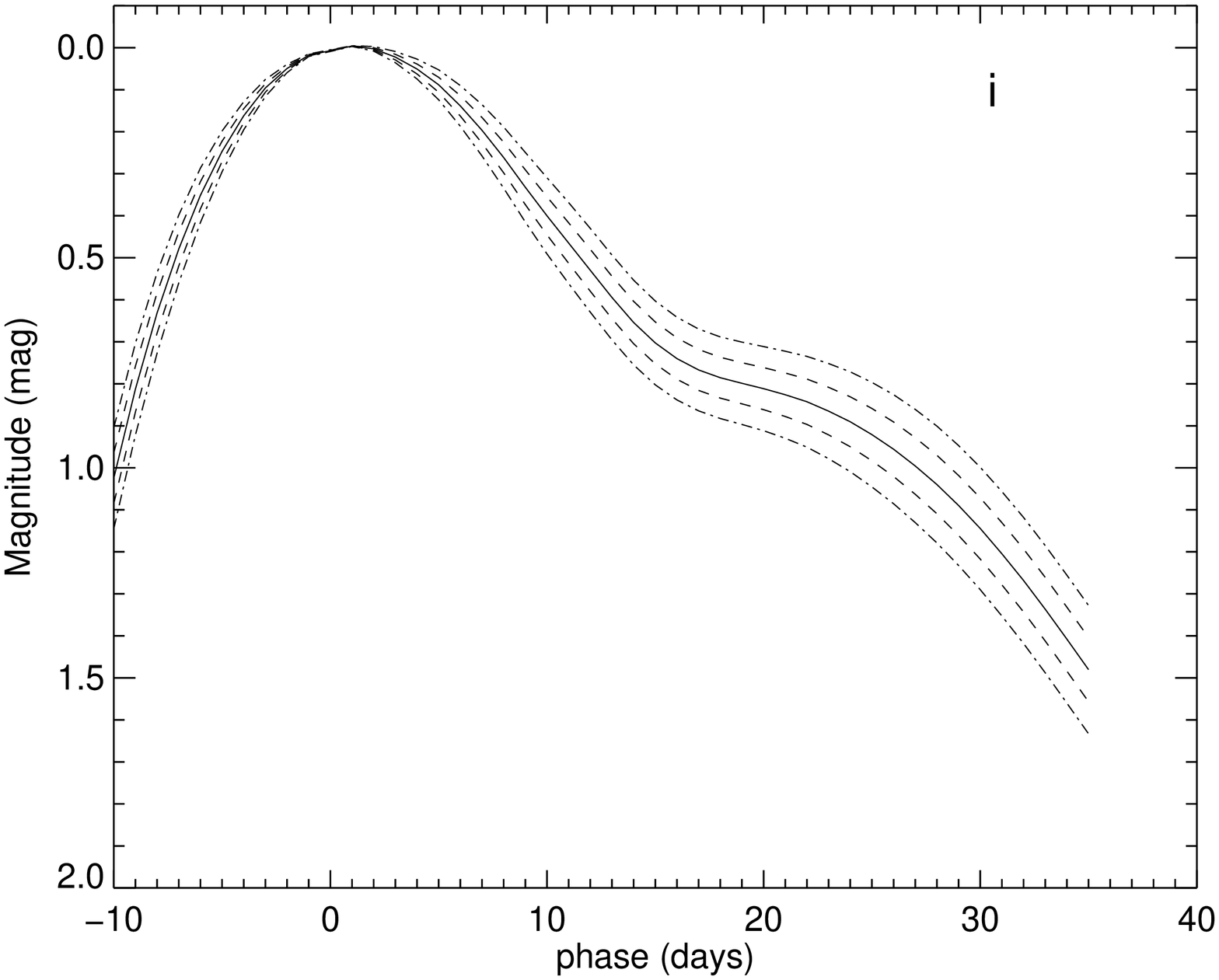} 
  \includegraphics[width=1.75in,angle=270]{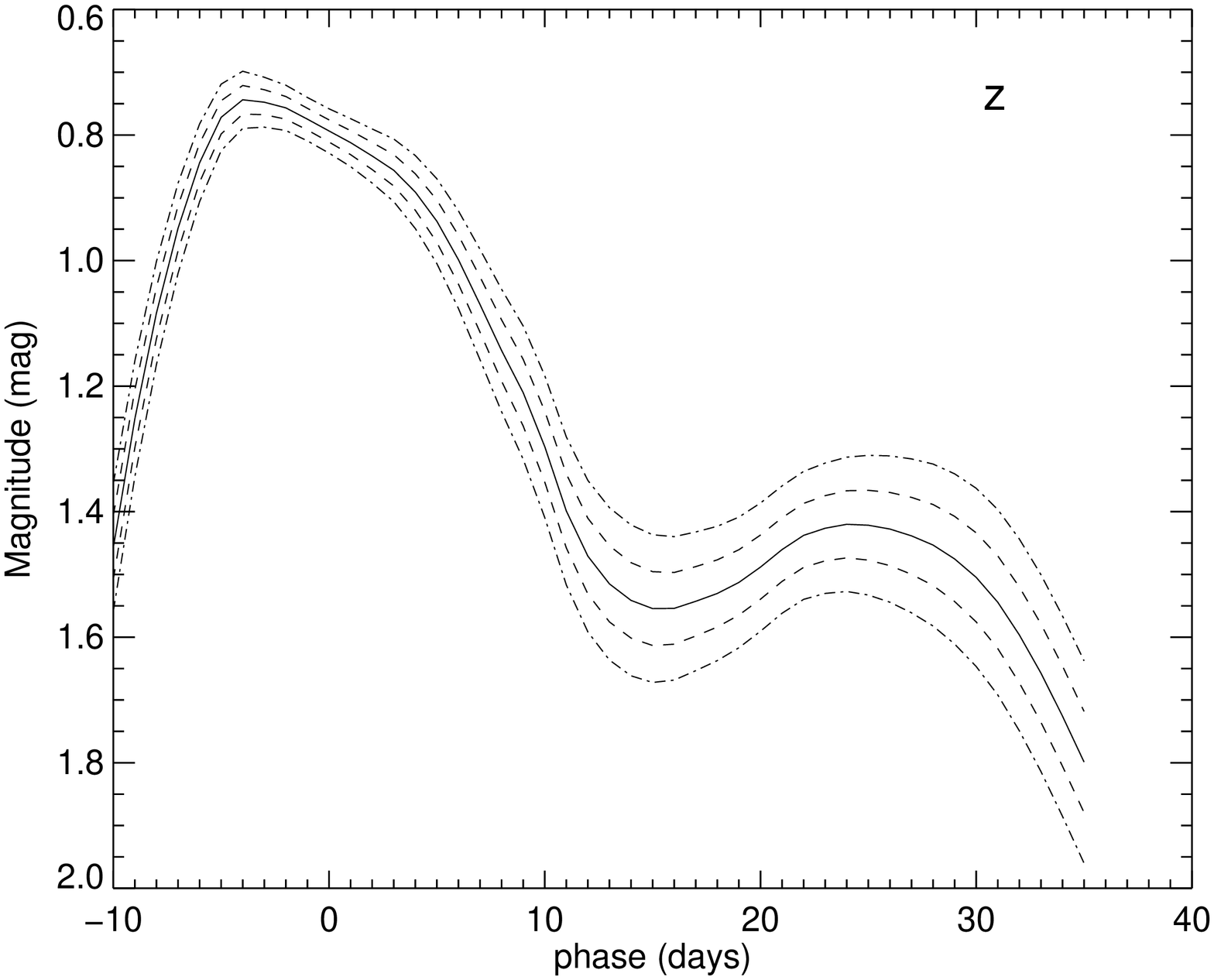} 
  \caption{As in Figure~\ref{pca0:fig} except for perturbations in the $x(1)$ parameter.
     \label{pca1:fig}}
\end{figure}

\begin{figure}[htbp] 
   \centering
   \includegraphics[width=1.75in,angle=270]{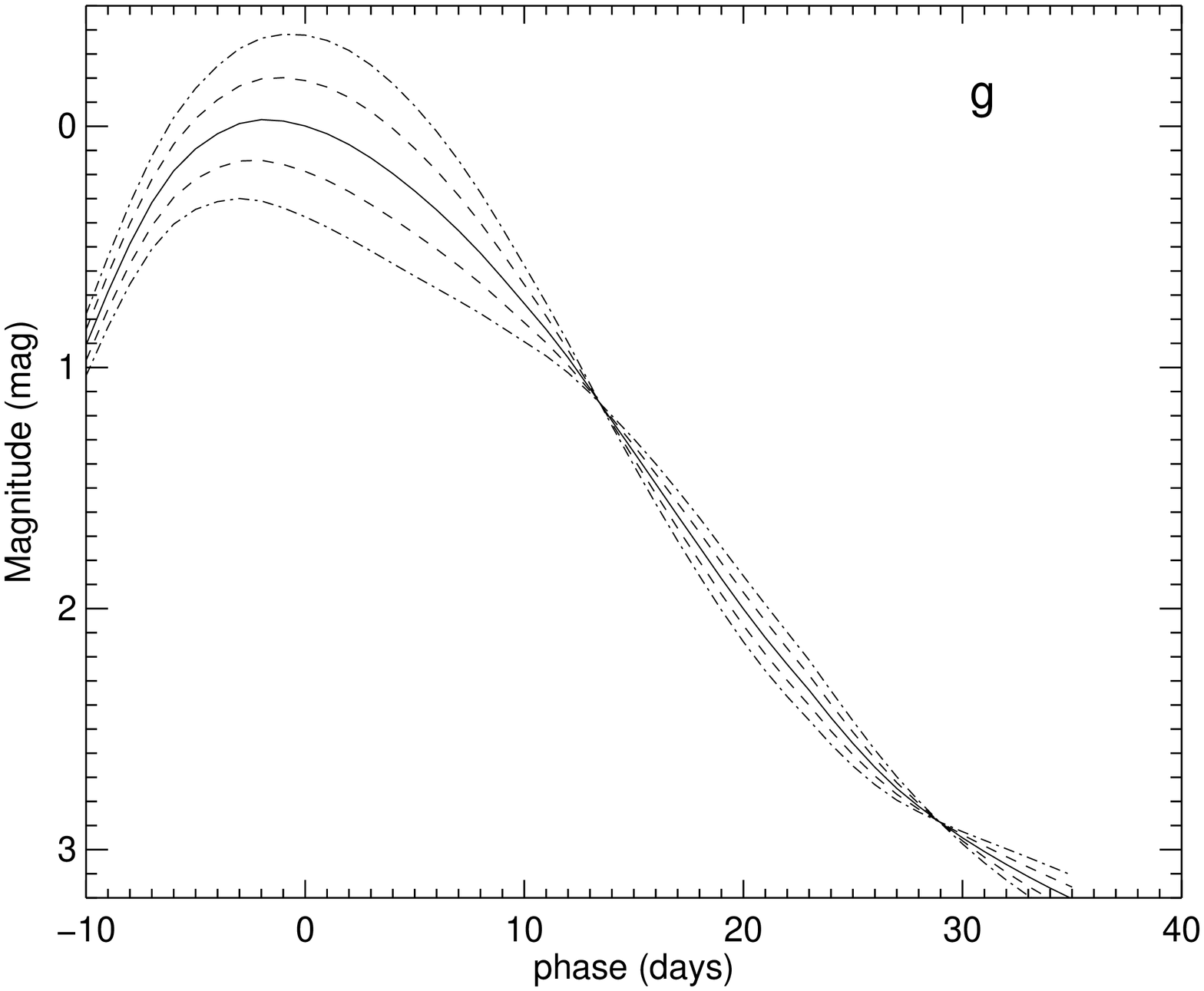} 
   \includegraphics[width=1.75in,angle=270]{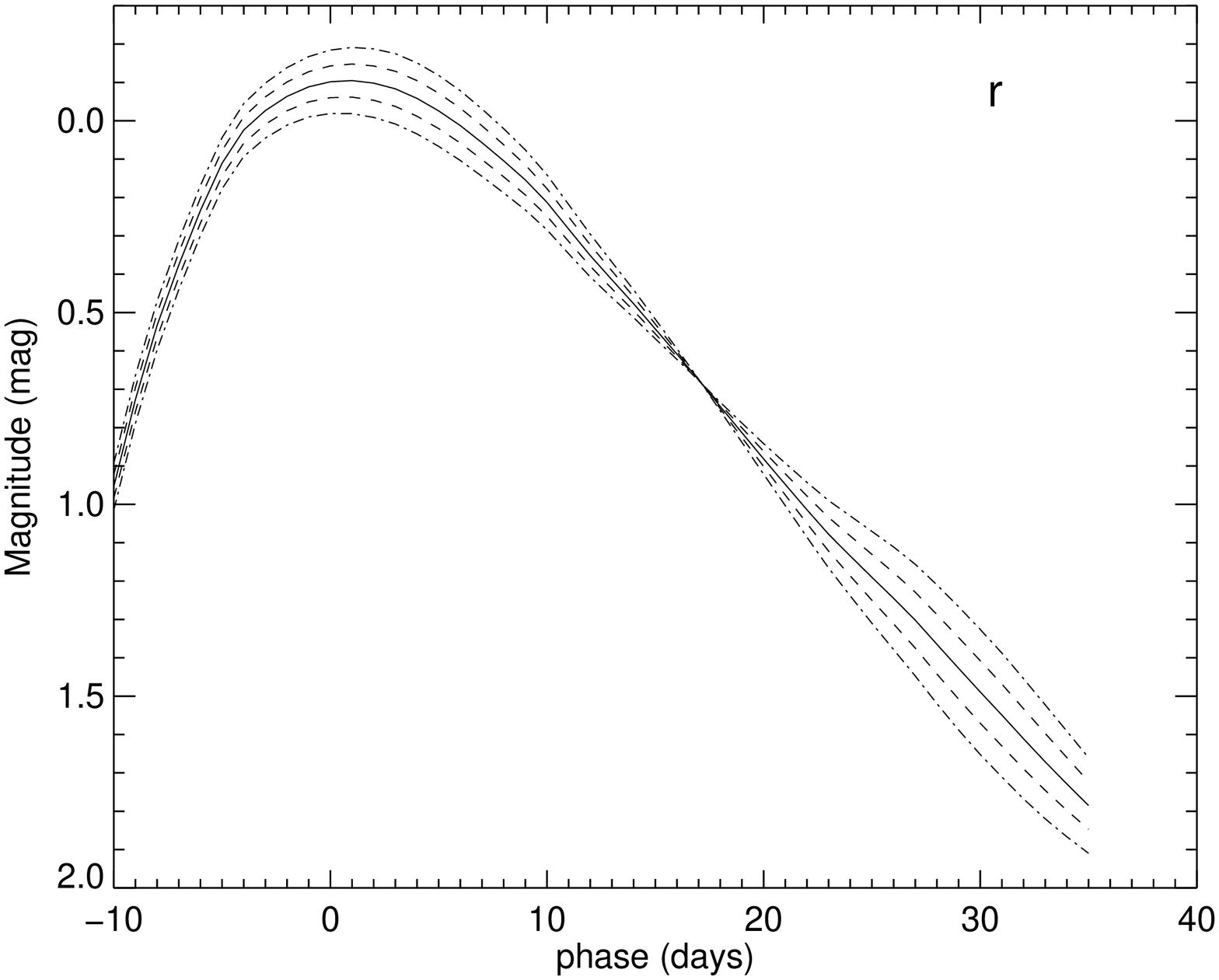} 
  \includegraphics[width=1.75in,angle=270]{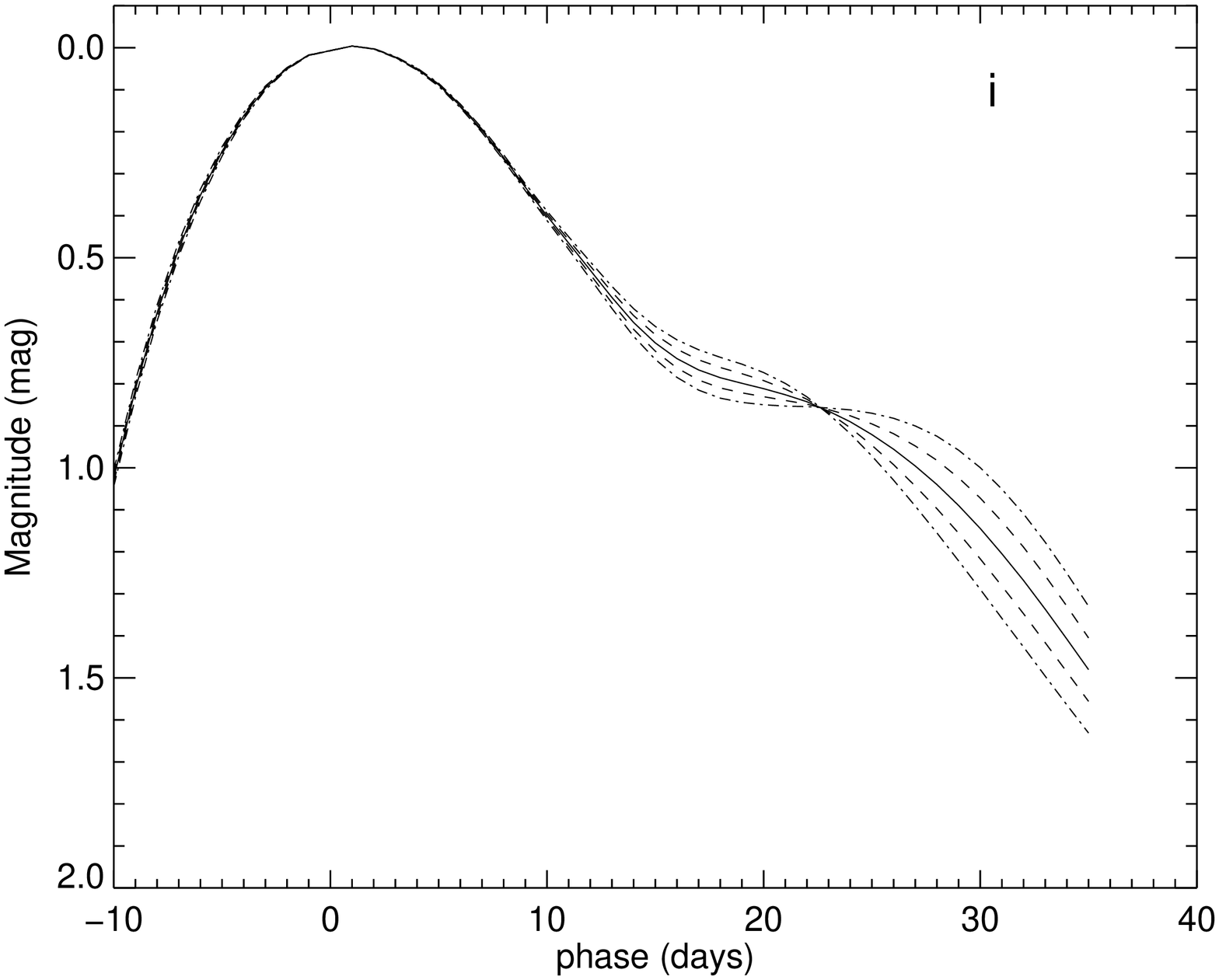} 
  \includegraphics[width=1.75in,angle=270]{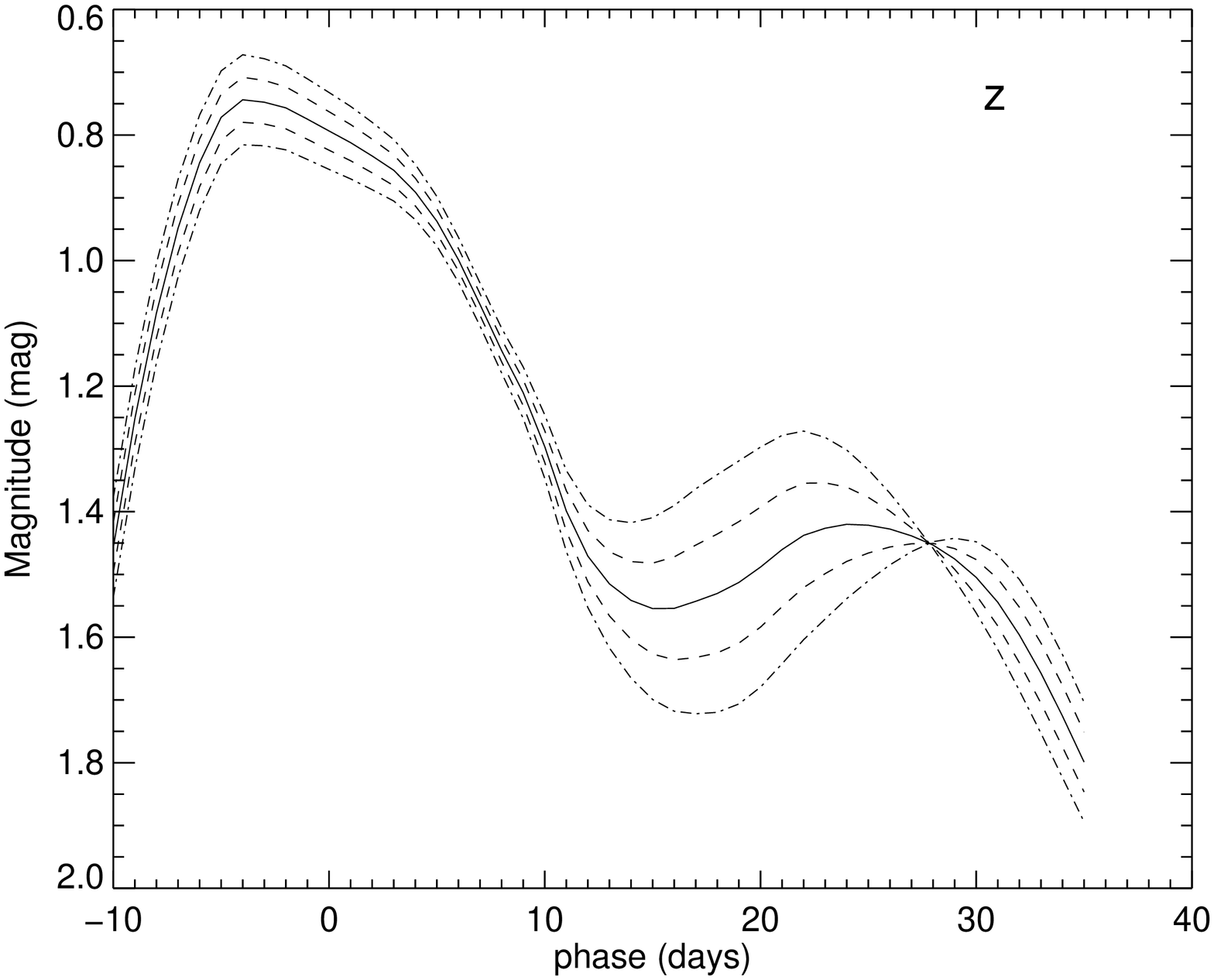} 
  \caption{As in Figure~\ref{pca0:fig} except for perturbations in the $x(2)$ parameter.
     \label{pca2:fig}}
\end{figure}

\begin{figure}[htbp] 
   \centering
   \includegraphics[width=1.75in,angle=270]{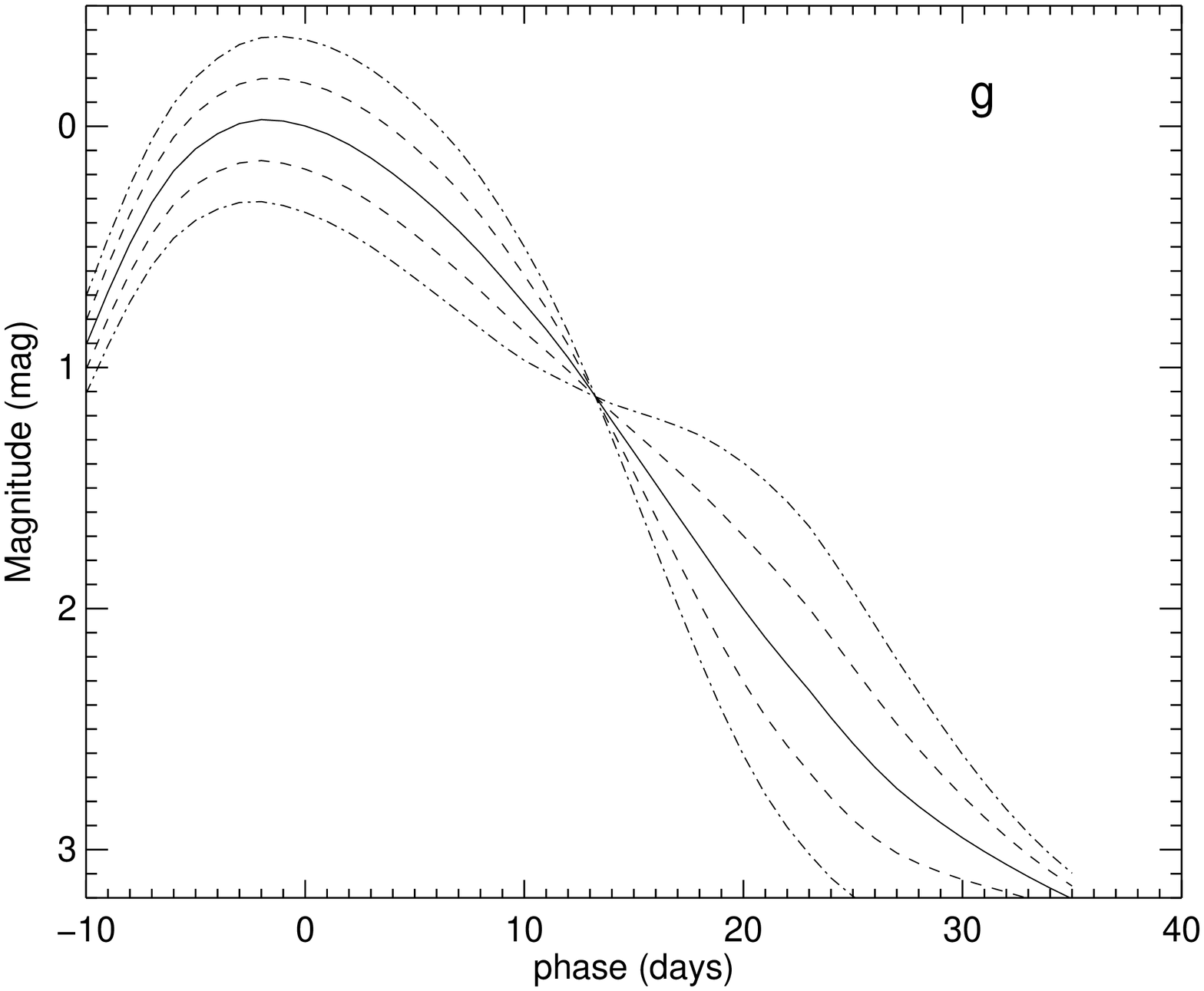} 
   \includegraphics[width=1.75in,angle=270]{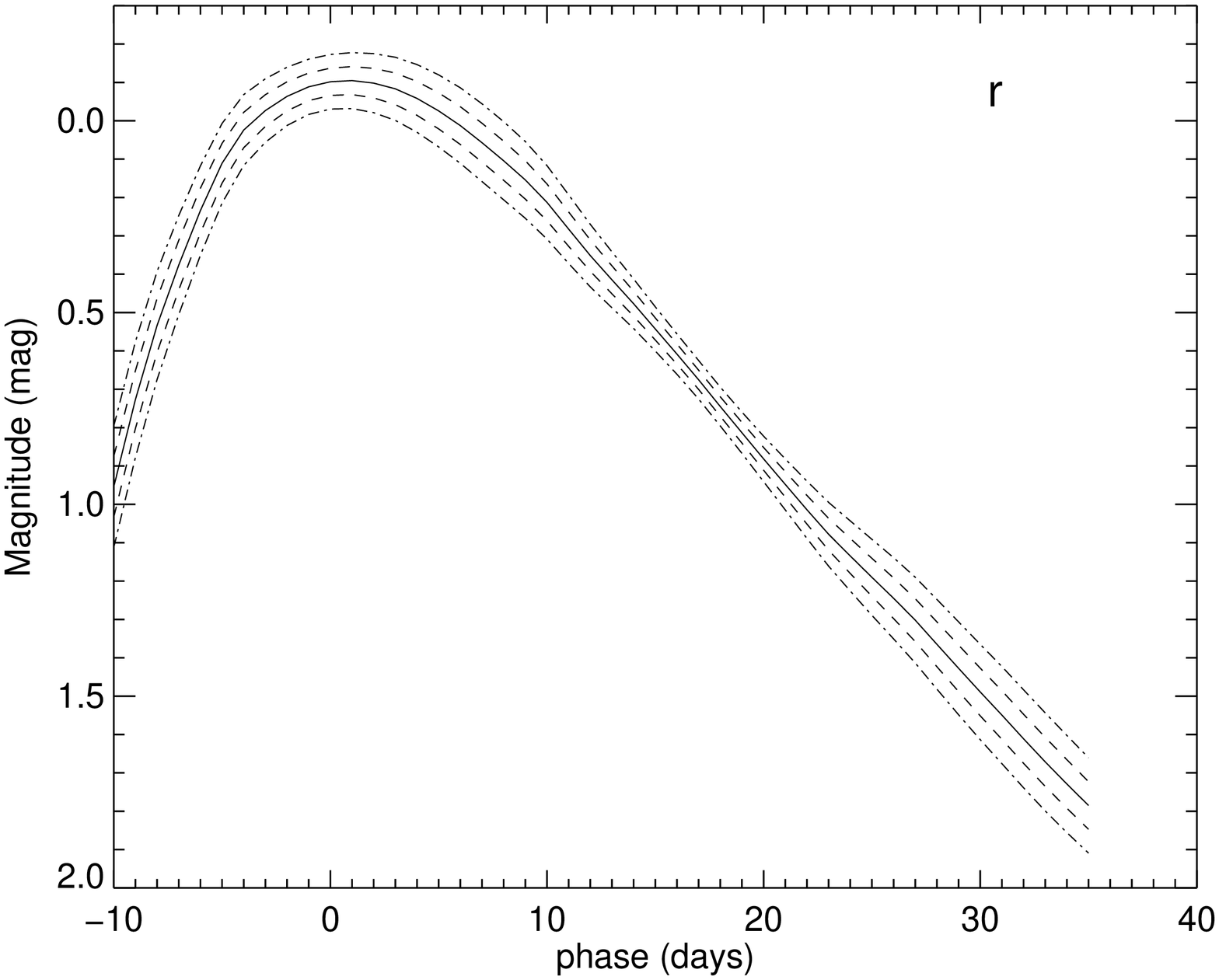} 
  \includegraphics[width=1.75in,angle=270]{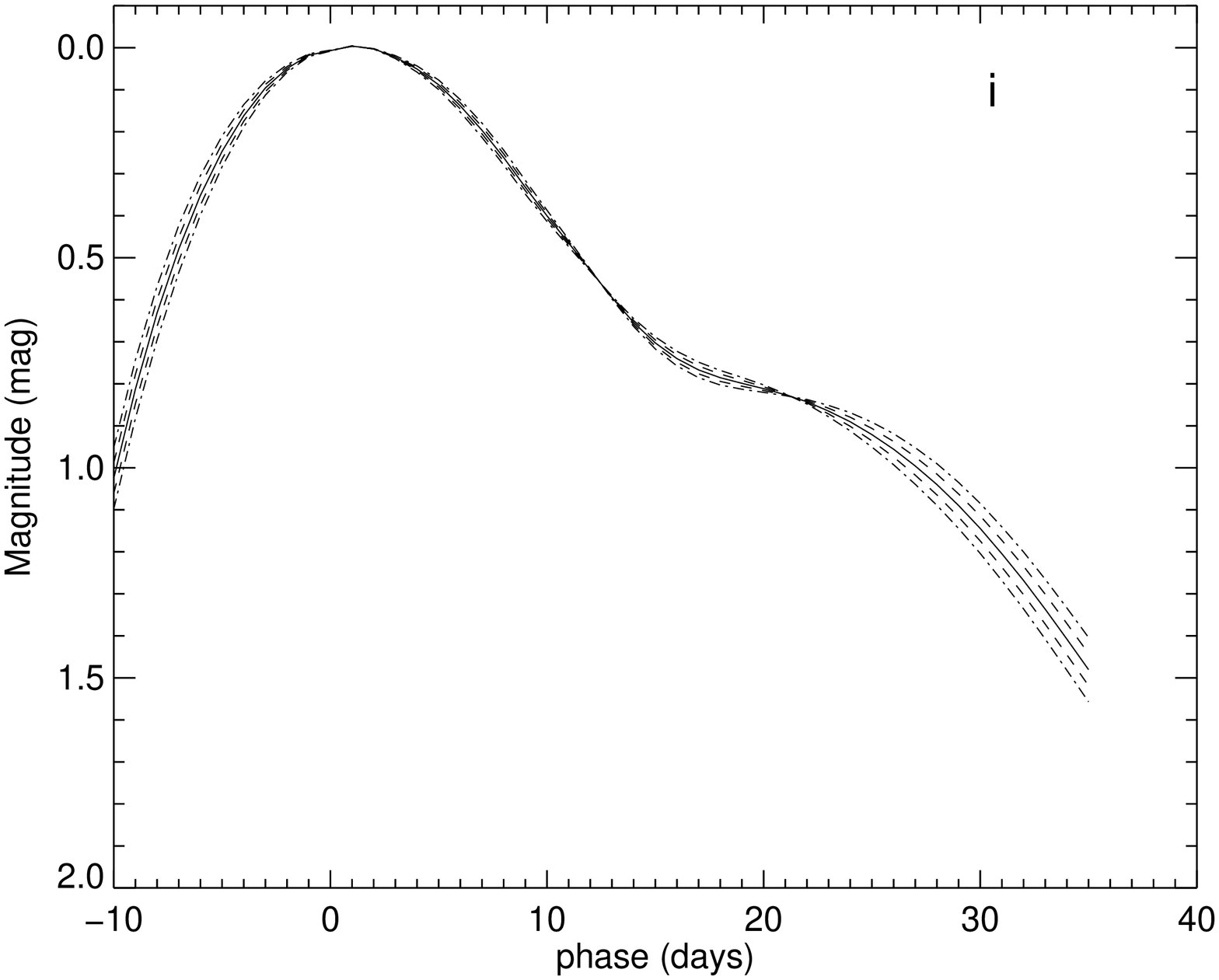} 
  \includegraphics[width=1.75in,angle=270]{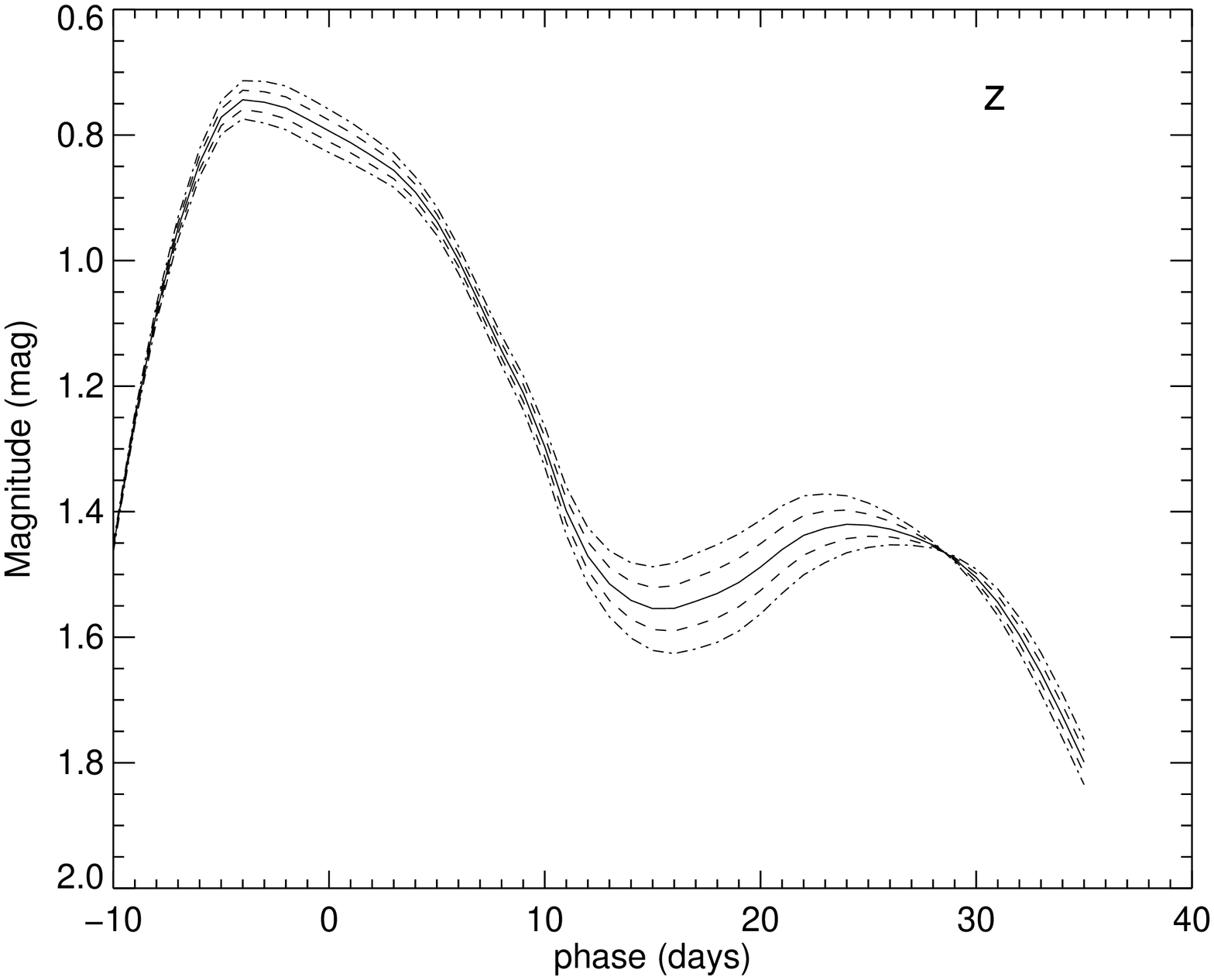} 
  \caption{As in Figure~\ref{pca0:fig} except for perturbations in the $x(3)$ parameter.
     \label{pca3:fig}}
\end{figure}

Mean and perturbed light curves from 1-- and 2--$\sigma_{x_j}$ changes in the $x(j)$ parameter
are shown for $j=0,1,2$ in Figures~\ref{pca0:fig}, \ref{pca1:fig}, \ref{pca2:fig}
and \ref{pca3:fig}.  The $j=0$ eigenvector (Figure~\ref{pca0:fig}) behaves as if it were due to dust: the correction is 
(mostly)  phase-independent
and monotonically decreases from blue to redder bands.
  The $j=1$ eigenvector (Figure~\ref{pca1:fig}) generates light curves that
vary in width, with supernovae with relatively skinny blueshifted-$g$ (observer-frame central wavelength of 380 nm) light curves having relatively
broader light curves in the redder bands.
The $g-X$ colors varies but the other colors do not.
The $j=2$ eigenvector (Figure~\ref{pca2:fig}) has a particularly strong influence in the shape and colors in the near-UV band.  It
has the property that the curves intersect at later phases; the effect is pronounced in the blueshifted $i$ band where the phase
and height of the secondary maximum vary.  Light-curve shapes and all color combinations also 
are influenced by this component.
The $j=3$ eigenvector (Figure~\ref{pca3:fig}) is similar to $j=2$ and
has a particularly exaggerated influence on the shape of the near-UV band.

\begin{figure}[htbp] 
   \centering
   \includegraphics[width=5.25in]{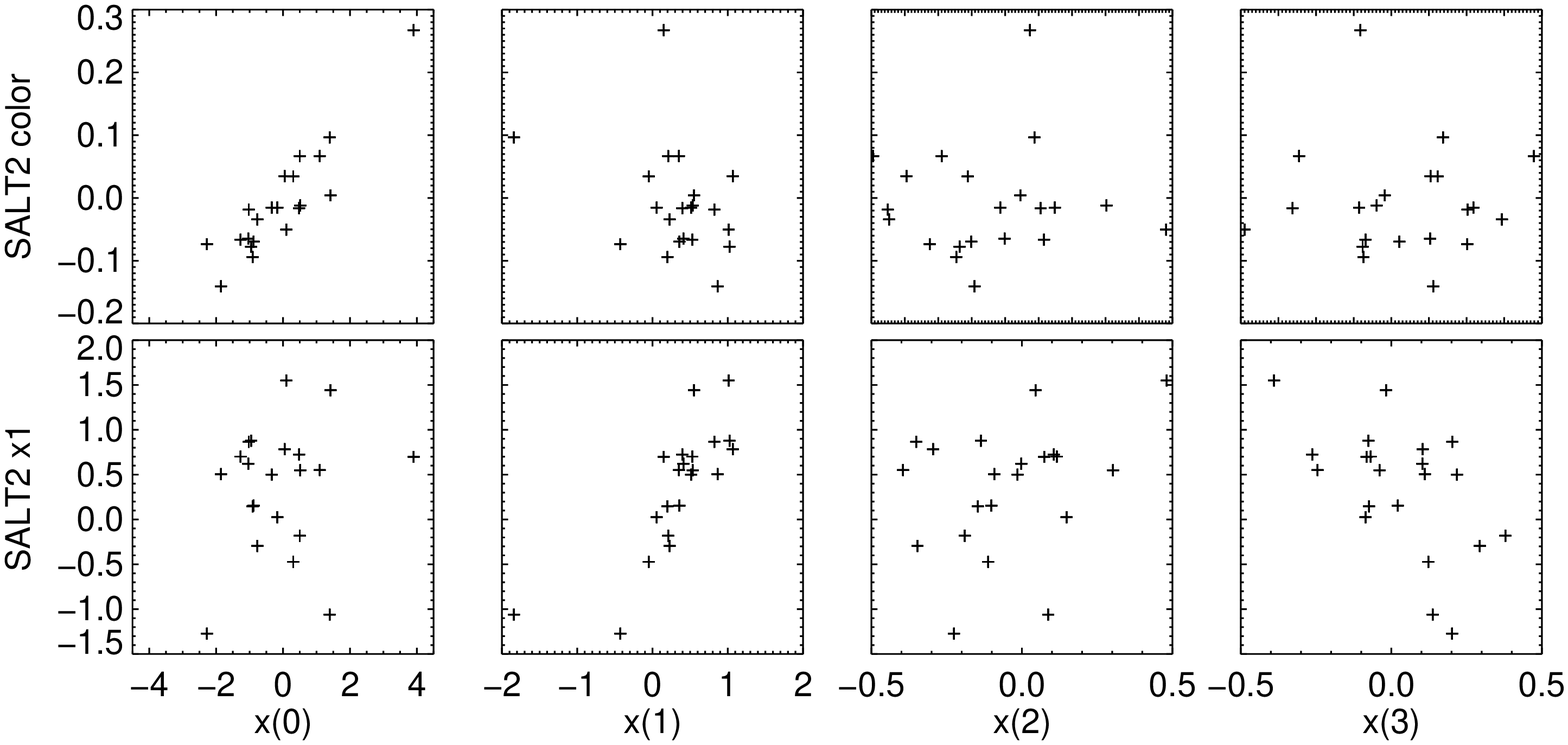} 
  \caption{Scatter plots of the first four PCA coefficients, $x(0)$, $x(1)$, $x(2)$,  and $x(3)$
   with the SALT2 color and $x_1$ (light-curve shape)  parameters
  for one of the runs for the $z=0$, filter set. 
   The correlation coefficients of color with the PCA coefficients are
  $\{-0.91.-0.33,0.02,-0.03\}$ and of $x_1$ with  the PCA coefficients
  $\{-0.18.0.80,0.27,-0.47\}$.  
  \label{pca_salt2:fig}}
\end{figure}

The PCA coefficients are correlated with the SALT2 color and $x_1$ parameters.  Figure~\ref{pca_salt2:fig} contains scatter plots
of the first four principal components with the SALT2 color and $x_1$ parameters for one of the runs for the $z=0$ filter set:
the first PCA component is strongly correlated with color and the second
and fourth are correlated with $x_1$.  

To summarize this second step of the analysis, the mean multi-band flux and covariance of the underlying light curve of an individual supernova
are calculated for a set  of  epochs and filters using the sparsely measured photometry combined with the trained light-curve Gaussian process model.
Each realization of the
light curves is decomposed into one ``true'' absolute magnitude and light-curve shapes and color curves.  Anticipating that absolute-magnitude variations are
encoded in SN Ia heterogeneity,  for the
light-curve shapes and colors we transform to a new coordinate-system that is defined with a PCA on the training set.

\subsubsection{Parameterized Model for Supernova Absolute Magnitudes}
\label{absmag:sec}
\begin{figure}[htbp] 
   \centering
   \includegraphics[width=1.75in,angle=270]{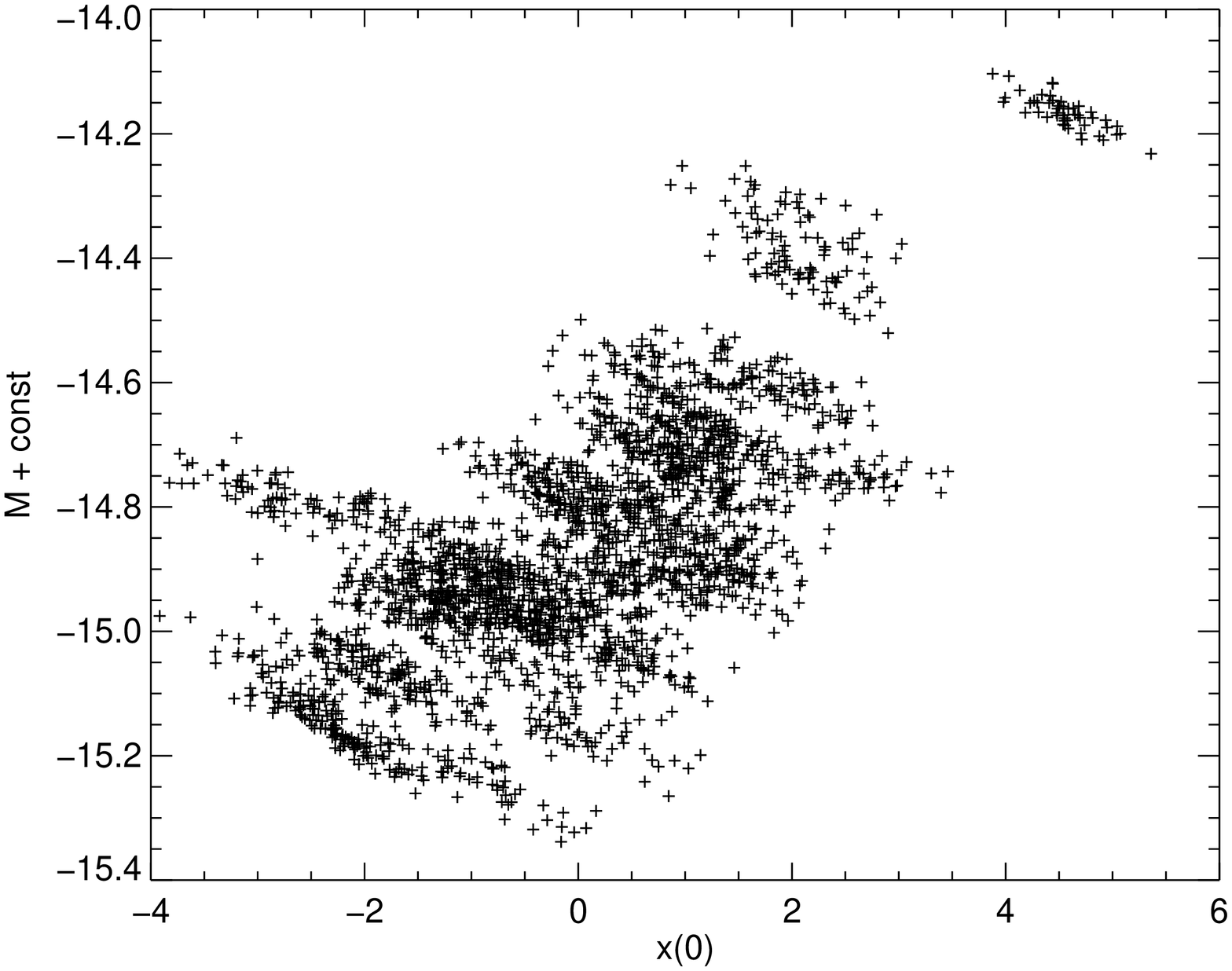} 
   \includegraphics[width=1.75in,angle=270]{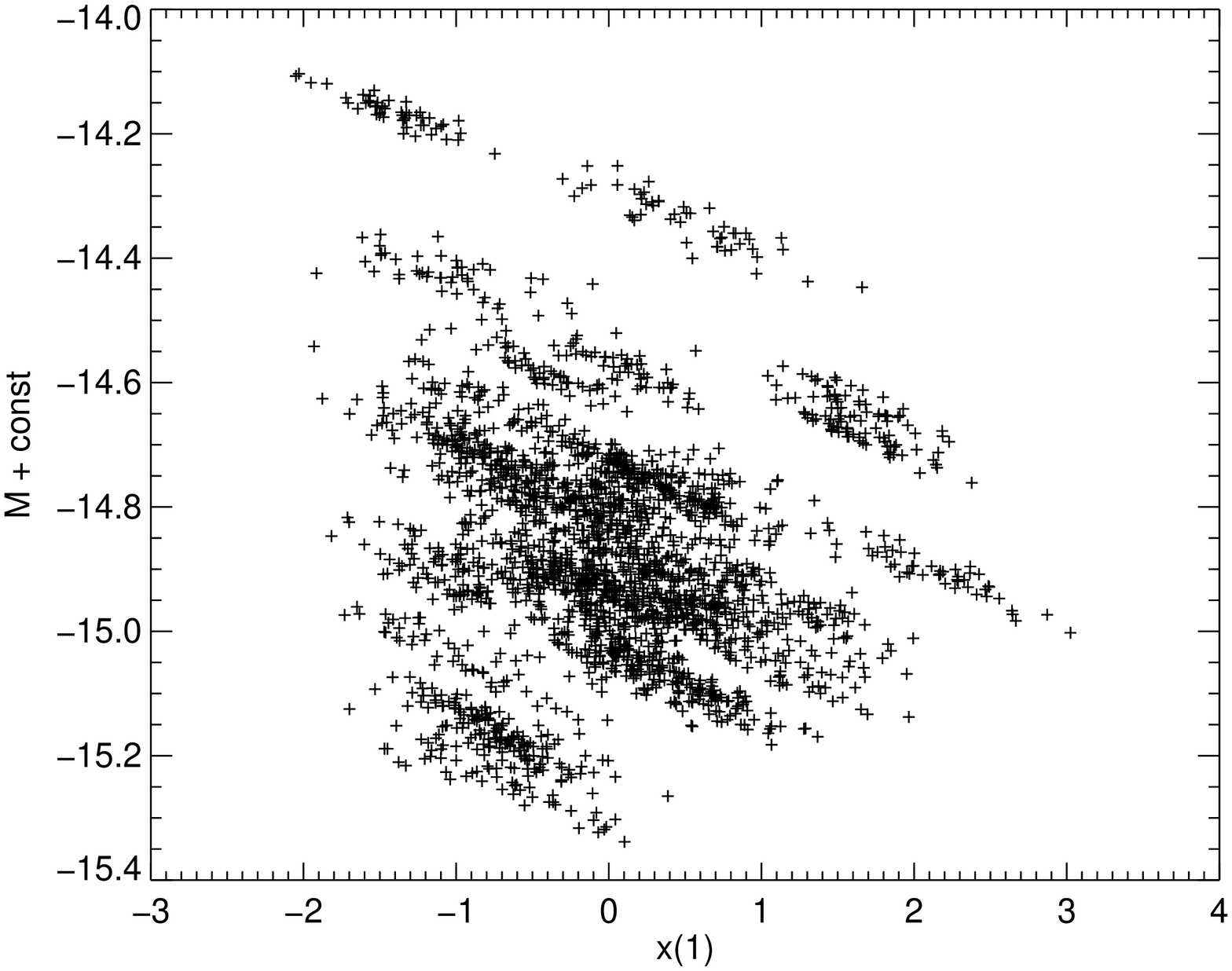} 
  \includegraphics[width=1.75in,angle=270]{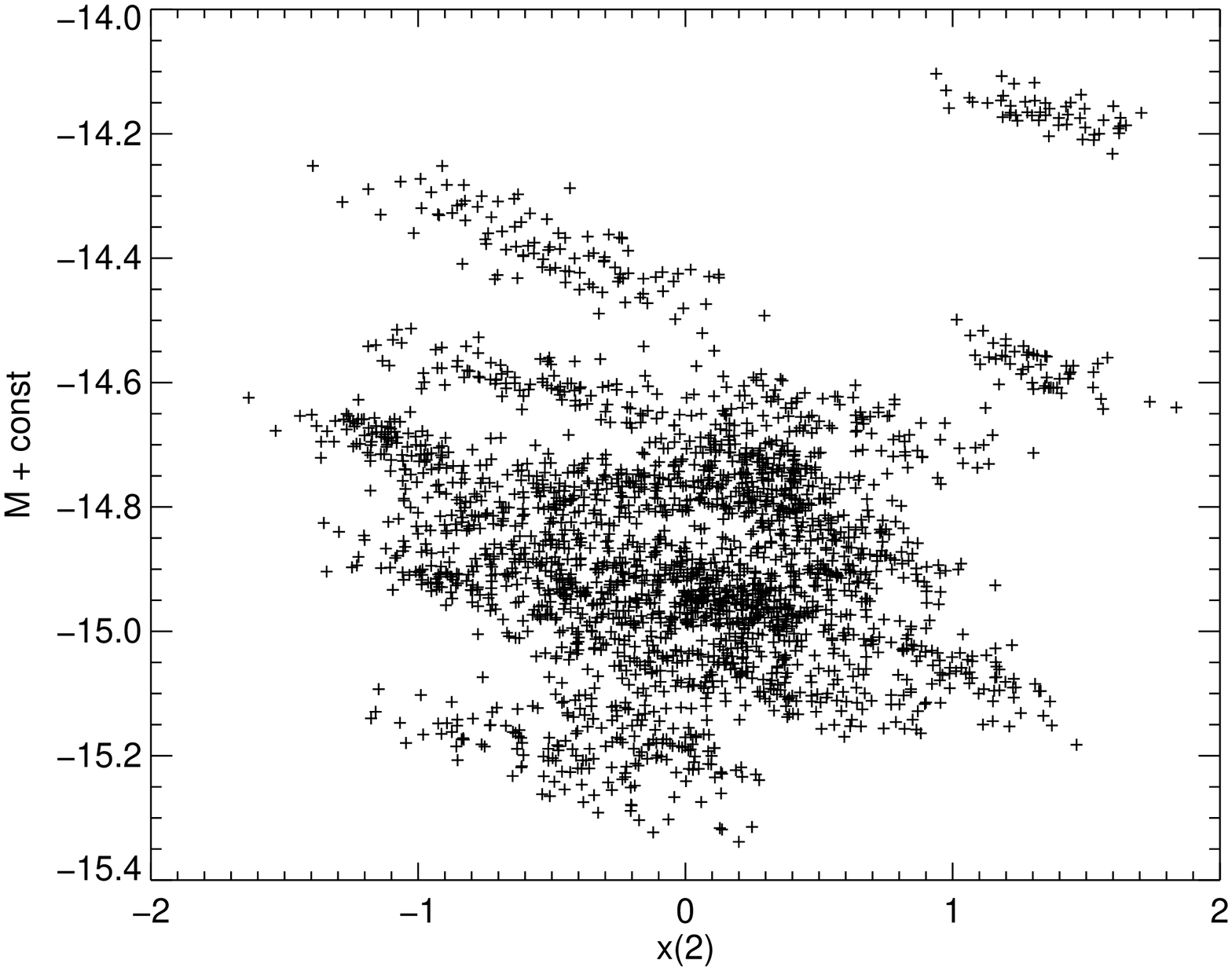} 
  \includegraphics[width=1.75in,angle=270]{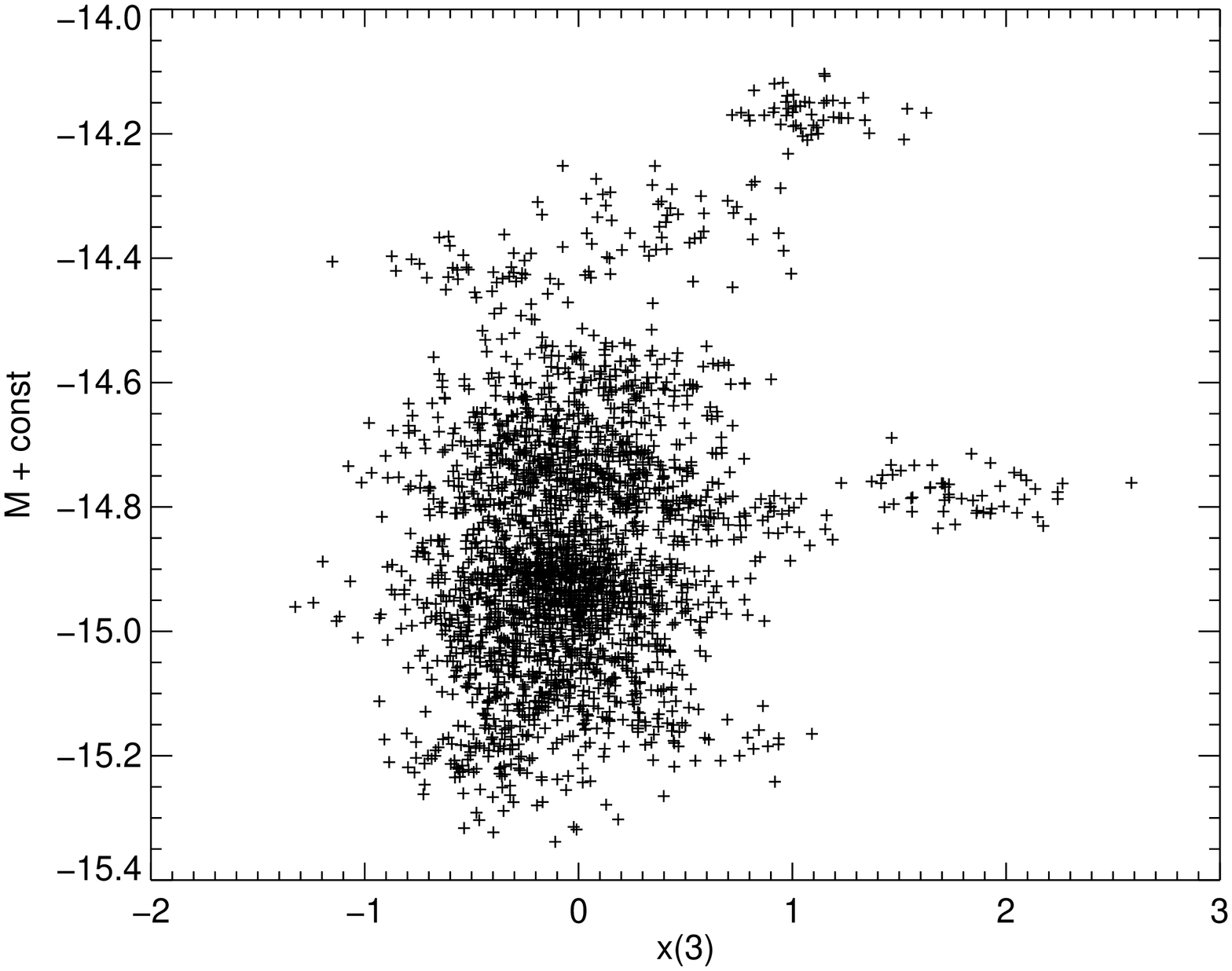} 
  \caption{Training-set absolute magnitudes (with constant offset) as a function of the first four light-curve coordinates in the PCA basis
  for one run of the $z=0.25$-filters calibrating the $i$ band.  Each point represents one supernova realization.
     \label{pcamag:fig}}
\end{figure}

The light-curve and shape coordinates in the PCA basis are correlated with true absolute magnitude
(modulo a constant offset for the arbitrary distance at which the SNe are placed, see \S\ref{data:sec}), as seen in Figure~\ref{pcamag:fig}.

In this section we
relate light-curve and color parameters to absolute magnitudes with a linear
and a Gaussian process model.
The linear model makes an explicit choice for the relationship between light-curve parameters
and absolute magnitude that can be well-constrained by our small dataset.
The Gaussian process methodology lets the data drive the model.
While having the advantage of not requiring
an explicit model, results can be sensitive to statistical fluctuations in the training set or large interpolations
in sparsely populated regions.  An analogous contrast can be made between
SALT \citep{2005A&A...443..781G} who used parameterized functions,
and MLCS \citep{1996ApJ...473...88R} who used data averages to construct light curve models.

\subsubsubsection{Linear Model}
The absolute magnitude at phase $t_0$ and band $\lambda_0$ is
a linear function of the light-curve parameters
 $M_{\left(t_0, \lambda_0\right)} (\mathbf{x};p_{\bar{M}}) = \bar{M}_0+\sum_{j=0}^{n_{max}-1} p_j x(j)$, where $ \bar{M}_0$ and $p_j$ are fit parameters.
Most of the over 100 supernova parameters capture little of the population
variance: as is justified in \S\ref{validation:sec}, $n_{max}=4$ is adopted in this article.

\begin{figure}[htbp] 
   \centering
   \includegraphics[width=5.2in]{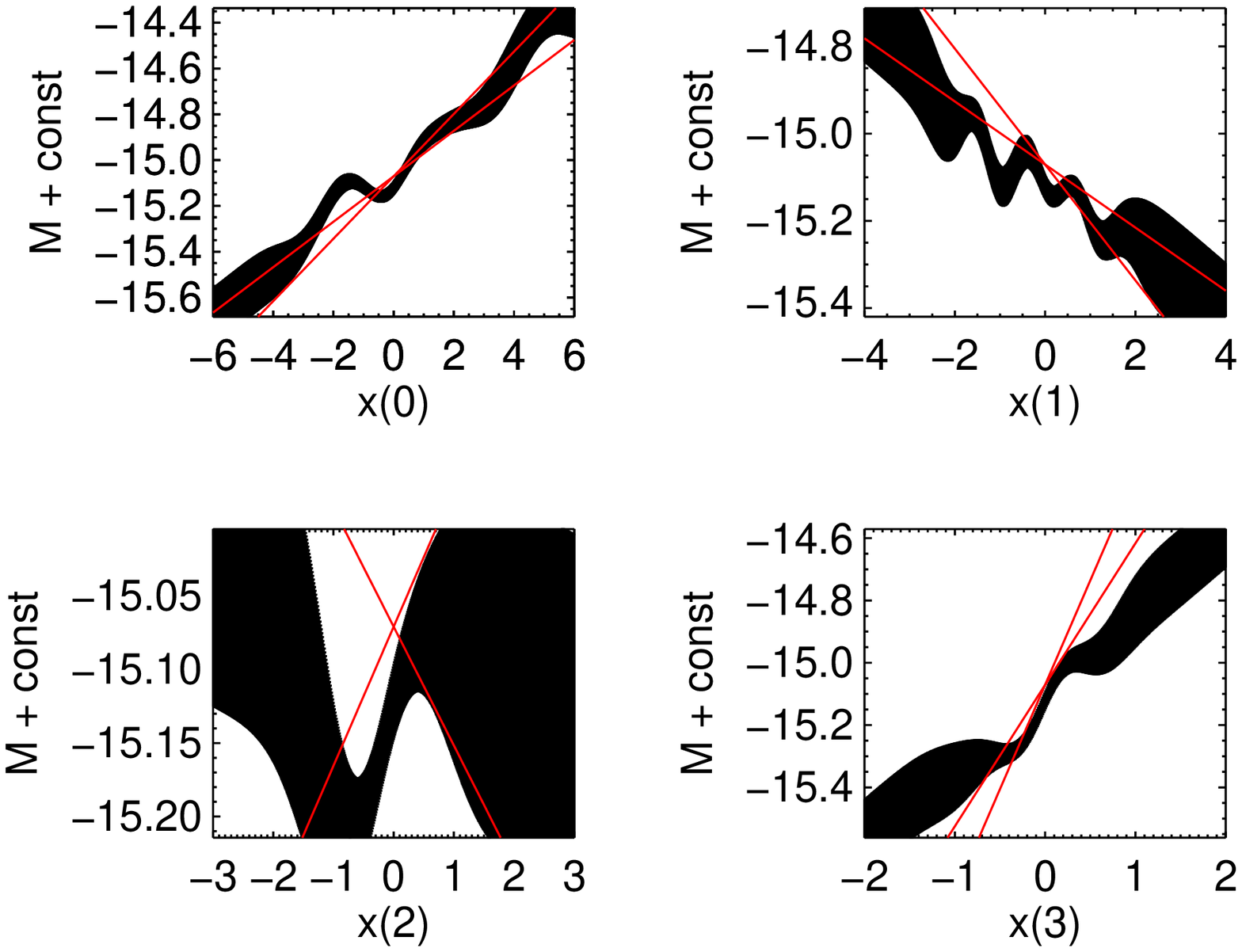} 
  \caption{In red the 1-$\sigma$ range of 
  the inferred absolute magnitude  found for the linear fit 
  and in black the 1-$\sigma$ bands from the Gaussian process model
  as a function of parameter value while holding the coordinates
for all other dimensions at zero,
for one run of the $z=0.00$-filters calibrating the $g$ band.
Note the different ordinate ranges of the plots.
\label{scalzo_lin_g:fig}}
\end{figure}

Figure~\ref{scalzo_lin_g:fig} shows
 the linear-model solution for the  $z=0.00$-filter set calibrating the $g$ band, which closely corresponds
to the bands for which SALT2 is directly trained.  As in SALT2, there are
clear correlations between the first two parameters, which
in turn are correlated with color and $x_1$  respectively
(as shown in Figure~\ref{pca_salt2:fig}).  There is no correlation for the third parameter,
but  a correlation with the fourth parameter is apparent.

\subsubsubsection{Gaussian Process Model}

While previous analyses have modeled linear and quadratic correlations between light-curve parameters and absolute magnitude,
it is well known that such simple relationships are not representative of the full range of light-curve shapes \citep{1996AJ....112.2391H}.  Here
we
infer the absolute magnitude through regression in the high-dimensional PCA parameter space.

The fiducial magnitude at phase $t_0$ and band $\lambda_0$, $M_{\left(t_0, \lambda_0\right)}$, is modeled  as a Gaussian process dependent on light-curve shapes and colors expressed by the parameters $\mathbf{x}$,
\begin{equation}
M_{\left(t_0, \lambda_0\right)} \sim GP\left(\bar{M}(\mathbf{x};\bar{M}_0),k_M(\mathbf{x},\mathbf{x'};\mathbf{l}_{k_M},\sigma_{k_M})+n_M(\mathbf{x},\mathbf{x'};\sigma_{n_M})\right).
\label{Lgp:eqn}
\end{equation}
The mean function 
 $\bar{M}(\mathbf{x};p_{\bar{M}}) = \bar{M}_0+\sum_{j=0}^{n_{max}-1}p_j x(j)$ 
is linear and has free parameters  $\bar{M}_0$ and $p_j$.
The kernel function is again a square exponential
\begin{equation}
k_M(\mathbf{x},\mathbf{x'};\mathbf{l}_{k_M},\sigma_{k_M}) = \sigma_{k_M}^2 \exp{
 \left[
 -\sum_{j=1}^n \left(\frac{x(j)-x'(j)}{l_{k_M}(j)}\right)^2
 \right]
 },
 \label{kernel:eqn}
\end{equation}
where $n$ is the number of elements in $\mathbf{x}$, i.e.\ the number of eigenvectors capturing 95\% of the population variance.
There is an independent per-point nugget, $n_M(\mathbf{x},\mathbf{x'};\sigma_{n_M})=
\sigma_{n_M}^2 \delta_{\mathbf{x},\mathbf{x'}}$. 

Training-set supernovae are used to find the best-fit values for $\bar{M}_0$, $\mathbf{l}_{k_M}$, $p_j$, $\sigma_{k_M}$, and $\sigma_{n_M}$.  The values of $\sigma_{k_M}$,
the strength of magnitude correlations, 
are found to be around 0.18 mag.  
The residual absolute magnitude dispersion after training is
given by the $\sigma_{n_M}$'s, which range from 0.04--0.07 mag except for the $z=0.75$ bands that have a range from 0.07--0.11 mag.

\begin{figure}[htbp] 
   \centering
   \includegraphics[width=3in,angle=270]{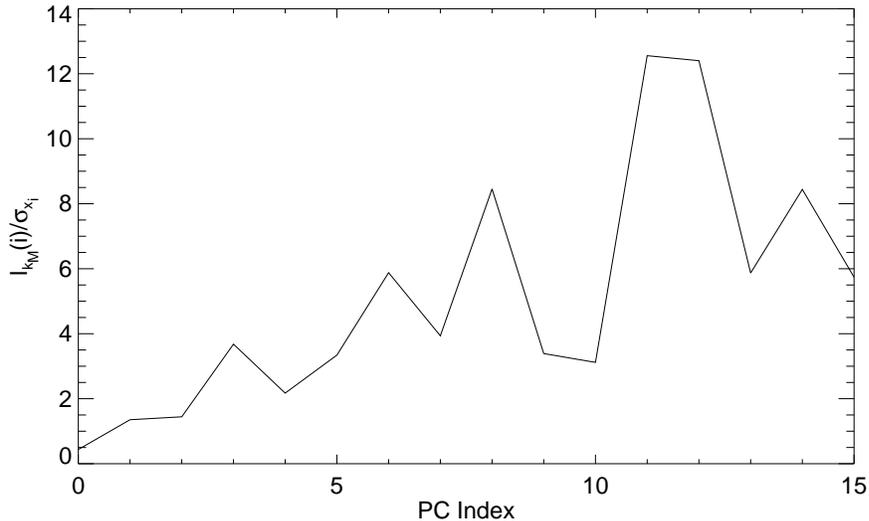} 
  \caption{Characteristic correlation scale in the kernel for each parameter, $l_{k_M}(i)/\sigma_{x_i}$, where $\sigma_{x_i}$ is the standard deviation
of the $i$'th parameter over the training set. \label{pcastrength:fig}}
\end{figure}

The hyperparameters for the length scale of the correlation $\mathbf{l}_{k_M}$ alone do not provide much physical insight.
A more interesting statistic is the ratio of the correlation scale and a typical value of the parameter in the training set; 
the metric represents a typical scale for each parameter in the
exponent in  Eqn.~\ref{kernel:eqn}, which in turn gives the relative contribution
to the correlation strength.  Parameters with low values in Fig.~\ref{pcastrength:fig}
contribute more to the correlation.
Figure~\ref{pcastrength:fig} shows the best-fit $l_{k_M}(j)/\sigma_{x_j}$, where $\sigma_{x_j}$ is the standard deviation
of the $j$'th coefficient of all realizations in the training set of a representative run.

The relationship between each parameter and absolute magnitude is illustrated for a slice in parameter space for
the same run
shown for the linear model; Figure~\ref{scalzo_lin_g:fig} shows the most-likely absolute magnitude as a function of parameter value while holding the coordinates
for all other dimensions at zero.  In this example the coordinate range
spanned by the supernova set decreases with the dimension order: $-6<x(0)<4$, $-2<x(1)<2$,
$-1.8<x(2)<1.8$, and $-1<x(3)<1$. 
The correlation between parameters and absolute magnitude is non-linear.

The Gaussian process and linear solutions for the first two parameters
are in general agreement, consistent with the SALT2 model. 
$X(2)$
does not have a monotonic trend in the region
and extends a small range of influence on luminosity.
The fourth
parameter has agreement in $X(3)$ within the range
where there is training-set data.

To summarize this  step of the analysis, supernova absolute magnitudes are taken to be a function of the light-curve shapes
and colors.
In one analysis a linear fit is performed.
In the other the functional form is not explicit, instead its stochastic behavior is modeled;
training-set supernovae provide several measurements of this function and we use these to train
a Gaussian process description of their correlation.
The trained Gaussian process provides a prescription for
interpolating an ``{\it inferred}'' absolute magnitude from the light-curve shape and color.
It should be emphasized that the training procedure is not optimized to minimize the dispersion in the residuals
of absolute magnitudes but rather to maximize the likelihood of the data for the Gaussian process model.

\subsection{Validation}
\label{validation:sec}
The validation-set supernovae are used to test how well true absolute magnitudes can be predicted from the inferred absolute magnitudes
derived from the trained model
in the procedure given in \S\ref{training:sec}.
We perform {\it cross-validation} by dividing the supernovae into distinct training and validation sets with 25\% assigned to the latter: the results are not expected to be as good
as if all supernovae are used for training, but in this approach there is no question of the independence of the magnitude
dispersion of the validation set from the data used in training.
We measure the stability of our results by
repeating the same analysis on validation sets containing different supernovae; as 25\% of the objects enter the validation set the analysis
is run four times with independent validation but partially intersecting training sets.
Over four runs each supernova enters a validation set once and only once.
Comparing the four almost-independent
cross-validations diagnoses
whether the number and fraction
of training supernovae are a fair representation of the full population.

\subsubsection{Pruned Supernova Sample}
\begin{figure}[htbp] 
   \centering{}
   \includegraphics[width=3in,angle=270]{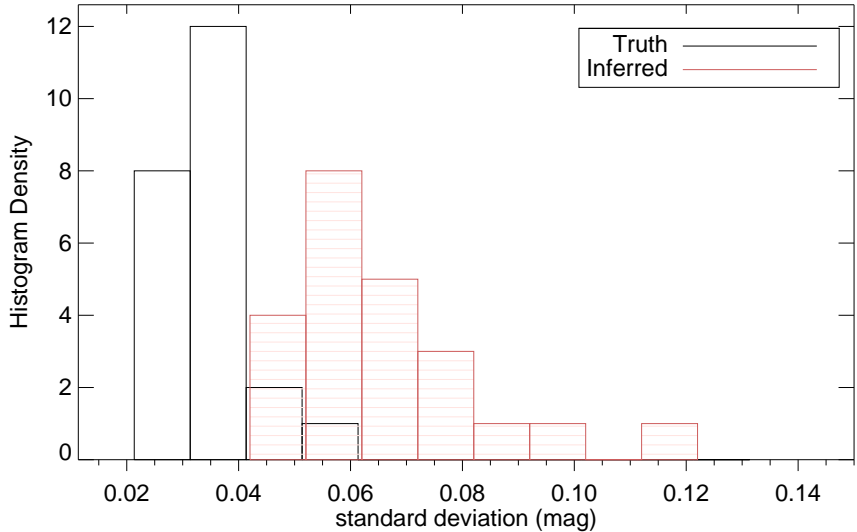} 
  \caption{Histograms of uncertainties in the Gaussian process prediction variance including measurement error
  in the true (unfilled) and inferred (hatched) absolute magnitudes per supernova,
  for a run calibrating the $i$ band of the $z=0.25$ filter set. \label{uncertainties:fig}}
\end{figure}

The SNfactory data do not have perfect temporal sampling or signal-to-noise.
As a selection criterion, we study those validation supernovae with synthetic photometry
in at least eight distinct phases in each of the bands.  This leaves 13--23 validation supernovae per run except in
the $z=0.75$ filter set, which has 7--11 SNe per run.
Measurement uncertainties
contribute to uncertainties in the interpolated absolute magnitude and light curves.  
To illustrate,  the distribution of light curves realized from the Gaussian process predicted
mean and covariance shown in
Figures~\ref{lc:fig}, \ref{bad1lc:fig},  and \ref{bad2lc:fig} indicate the size of the uncertainties.
Figure~\ref{uncertainties:fig} contains a histogram of the standard
deviations of the measured true absolute magnitude at $B$-peak for the $i$-band of the $z=0.25$ filter set,
here the absolute magnitude uncertainties per supernova range from 0.02--0.06 mag.

As a representative example of the light curve regression, we present results for one random
supernova from one of the runs with $z=0.25$ filter set.
Figure~\ref{lc:fig} shows the synthetic photometry, the predicted mean of the underlying light curves in solid, and ten random realizations in dashed lines.

\begin{figure}[htbp] 
   \centering
   \includegraphics[width=4in,angle=270]{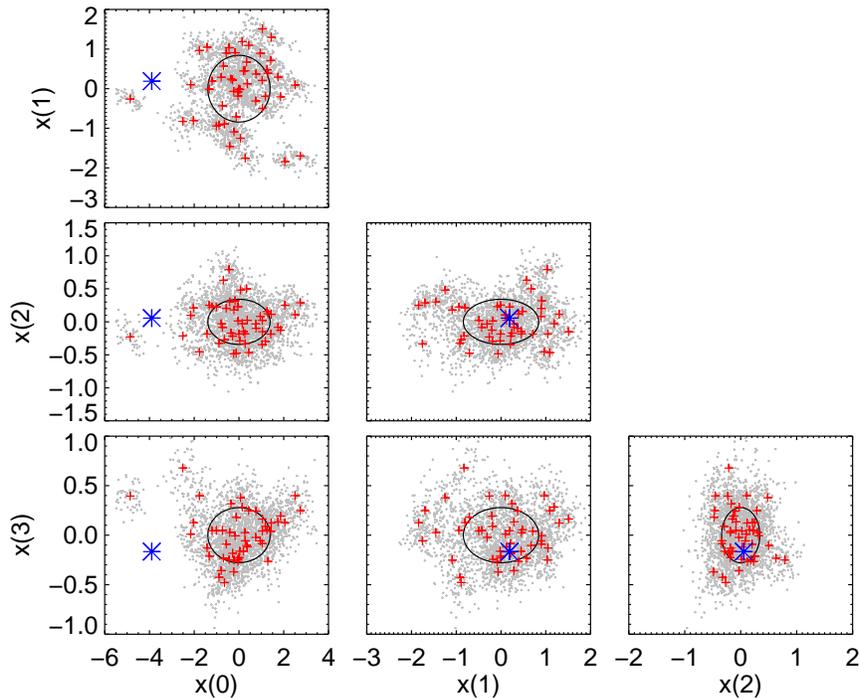} 
  \caption{Distribution of the first four PCA coefficients for  one of the $z=0.00$ filter-set runs used in the training of the
  analysis of SN~\#19.   The symbols are the same as in Figure~\ref{pcacoeff:fig}.  In addition, blue stars show the PCA coefficients
  of the predicted mean of SN~\#19.
\label{pcacoeffsn2007le:fig}}
\end{figure}

We identify two odd supernovae, SN~\#13 and SN~\#19, that are culled from the sample.
 SN~\#19 data and predicted light curves for the $z=0$ filter set
are shown in Figure~\ref{bad1lc:fig}. This supernova does not have the wavelength coverage to make the $z=0.25$ sample and
its PCA coefficients fail the distance cut for $z=0.75$.  It is an extremely slow decliner with a $g$-band $\Delta m_{15}=0.7$
derived from the Gaussian process prediction.
The PCA coefficients from the $z=0$ case are shown among those of the training set in Figure~\ref{pcacoeffsn2007le:fig};
it is clear that SN~\#19 does not have a corresponding analog in the training set.  With $z=0.0055$, this supernova
has the lowest CMB-frame redshift of the full sample of which $>96\%$ has  $z>0.015$;
a peculiar velocity of 300 km s$^{-1}$ would contribute a 0.40 mag residual.
SN~\#13  data and predicted light curves for the $z=0.25$ filter set
are shown in Figure~\ref{bad2lc:fig}.  The discrepancy is accounted for by one spectrum near peak that is fainter
than expected; indeed this discrepancy disappears in subsequent versions of the spectral reduction.

Validation supernovae may not have an analog in the training set.   We approximate the PCA coefficient distribution
as Gaussian, described by the standard deviations $\sigma_{x_i}$ of each dimension.
A validation supernova is included in the analysis
if it lies within the core of the distribution
\begin{equation}
\sum_j{\left(\frac{x_j}{\sigma_{x_j}}\right)}^2 < \chi^2(p,n_{PCA}),
\label{cut:eqn}
\end{equation}
where $\chi^2(p,n)$ is the value where the cumulative $\chi^2$-distribution has $p$-value $0.05$ 
and $n_{PCA}$
is the number of principal components that capture 95\% of the variance.  This
cut criterion was made after the distributions were constructed and is therefore not blind.
However, 0.05 is selected as a nice round number that should capture 95\% of the population and ensuing results are insensitive to departures
from this value.
This cut typically removes 0--2 objects from the sample.

\subsubsection{True and Inferred Magnitudes and their Differences}

The covariance in the regression is accounted for in
our analysis by using
fifty light-curve realizations of each supernovae. 
Each of the fifty multi-band light-curve realizations of each supernova is temporally sampled from $-10$ to 35 days using the best-fit Gaussian process
model of the training set.  
The light curves
are normalized by the true $M_{\left(t_0,\lambda_0\right)}$ and expressed in the coordinate system defined by the PCA eigenvectors.
The PCA coefficients are used to infer $M_{\left(t_0,\lambda_0\right)}$.  
The standard deviation of its fifty realizations provides the measurement uncertainty of $M_{\left(t_0,\lambda_0\right)}$
of a supernova, this value depends on the quality of the measured light curve and ranges.
The means and standard deviations of the 50 realizations of true and inferred absolute magnitudes  from the validation
procedure
(modulo a constant offset) are given in Tables~\ref{sample0:tab}
\ref{sample025:tab}, \ref{sample050:tab}, and \ref{sample075:tab} for the Gaussian-process-model.  (In the tables, omitted
supernovae
were used for light-curve training but did not pass the selection criteria for the peak-magnitude analysis.
In italics are supernovae that have data before two days pre-maximum with light curves that have at least eight points.
In bold are supernovae used for validation that have data before two-days pre-maximum and at least eight points in all light curves.)
Figure~\ref{uncertainties:fig} shows the histogram of the standard
deviations of the inferred absolute magnitude  at $B$-peak for the $i$-band of the $z=0.25$ filter set:
the inferred absolute magnitude measurement
uncertainties per supernova mostly fall within the range 0.04 -- 0.10 mag.

The accuracy to which true absolute magnitudes are predicted by absolute magnitudes inferred from SNfactory synthetic light curves
is the root-mean-square of the
difference between the true and inferred  $M_{\left(t_0,\lambda_0\right)}$ over all realizations of all supernovae.
The weighted rms  ($wrms$) of the differences of each validation set,
and then the mean and standard deviation of those weighted rms' over the four training/validations ($\overline{wrms}$
and $\sigma(wrms)$ respectively)
are calculated.
The wrms weights the squared residual by the inverse measurement variance visualized in Figure~\ref{uncertainties:fig}
\citep[the expression for the weighted rms is given in][]{2011A&A...526A..81B}.
The mean shows how well the algorithm standardizes the distance indicator over all supernovae.
The standard deviation of the wrms of the four runs indicates the
sensitivity to the sample selection.
The weighted rms' have contributions from measurement errors, peculiar velocities,
and the intrinsic ability that the methodology has to calibrate supernovae, $\sigma_{int}$.
Assuming a peculiar velocity dispersion of  300 km s$^{-1}$, for each run we 
report  the mean and standard deviations
of the fit value the intrinsic dispersion $\sigma_{int}$.

The wrms calculated above represents an estimate of the absolute-magnitude uncertainty when using 75\% of the data training, such that none of the validation dispersions are
calculated with  supernovae
used in  the training.
Our analysis lends to easy calculation of the wrms from {\it $K$-fold cross analysis} \citep[as done by][]{2011A&A...526A..81B} for $K=4$, where the dispersion
is calculated for all supernovae together despite the fact they are based on different training sets; these 
results are given in Tables~\ref{sd_lin:tab} and \ref{sd:tab}.
We provide two numbers for the K-fold analysis; the first using the subset after applying the
sample cut of Eqn.~\ref{cut:eqn} and the second  using the full sample of objects.
The former statistic is calculated anticipating that the supernova parameter space is
not densely sampled and considering that our absolute magnitude model may not
extrapolate well to objects outside the training regime.

For the linear model, the number of light-curve parameters related to absolute magnitude
is set as $n_{max}=4$.  This choice is made as follows:
We calculate $\overline{wrms}$ for the linear model
varying  $n_{max}$ from 1 to 8 for the $i$-band calibration of the $z=0.25$ filter set,
getting 0.097, 0.099, 0.104, 0.091, 0.092, 0.094, 0.093, and 0.095 mag.  There is a step
in the  $\overline{wrms}$ when four parameters are used.
In the Gaussian process model, $n_{max}$ is set to the number of principal components that capture 95\%
of the variance.

Calculated dispersions from the validation are shown in Table~\ref{sd_lin:tab} for the linear fit,
and in Table~\ref{sd:tab} for the Gaussian process model.
The weighted mean of each validation set is  $\lesssim 0.01$ mag.

\begin{table} \footnotesize
\center
\begin{tabular}{cc|ccccccc}
Filter System $z$ &Band  &  $\sigma_0$ & $\overline{wrms}$ & $\sigma(wrms)$& $\overline{\sigma_{int}}$ & $\sigma(\sigma_{int})$ & 4-fold cross-validation & $N_{SN}$ \\ \hline
0.00& $g$&  0.322&  0.146& 0.026&  0.123& 0.049& 0.154 (0.158)&          50\\
0.00& $r$&  0.229&  0.118& 0.014&  0.049& 0.036& 0.120 (0.122)&          49\\
0.00& $i$&  0.265&  0.137& 0.018&  0.099& 0.021& 0.136 (0.140) &          50\\ \hline
0.25& $g$&  0.437&  0.141& 0.039&  0.119& 0.052& 0.155 (0.169)&          44\\
0.25& $r$&  0.301&  0.119& 0.039&  0.068& 0.052& 0.122  (0.147)&          43\\
0.25& $i$&  0.236&  0.091& 0.019&  0.025& 0.030& 0.090 (0.102) &         43\\
0.25& $z$&  0.250&  0.136& 0.032&  0.111& 0.032& 0.130    (0.132)   &    44\\ \hline
0.50& $r$&  0.391&  0.153& 0.033&  0.126& 0.042& 0.156 (0.157)  &       50\\ 
0.50& $i$&  0.301&  0.149& 0.023&  0.123& 0.021& 0.148     (0.149) &   48\\
0.50& $z$&  0.248&  0.135& 0.025&  0.093& 0.027& 0.132   (0.131)    &   49\\ \hline
0.75& $r$&  0.320&  0.140& 0.036&  0.093& 0.062& 0.148    (0.184)   &   35\\
0.75& $i$&  0.237&  0.129& 0.042&  0.088& 0.059& 0.133    (0.184)   &   35\\
0.75& $z$&  0.210&  0.131& 0.035&  0.080& 0.058& 0.133    (0.160)   &   35\\ \hline
\end{tabular}
\caption{Results from the linear model: For each filter system and calibrated band: $\sigma_0$, the raw standard deviation of $M_{(t_0,\lambda_0)}$ for the input sample;
$\overline{wrms}$ and $\sigma(wrms)$, the mean and standard deviation over the four validation samples of the wrms difference between true
and inferred magnitude; the mean intrinsic dispersion and its standard deviation over the four validation samples, the 4-fold cross-validation error estimate with (and without) rejection
of objects with extreme parameter values; $N_{SN}$ is the number of validated supernovae.  \label{sd_lin:tab}}
\end{table}

\begin{table} \footnotesize
\center
\begin{tabular}{cc|ccccccc}
Filter System $z$ &Band  &  $\sigma_0$ & $\overline{wrms}$ & $\sigma(wrms)$& $\overline{\sigma_{int}}$ & $\sigma(\sigma_{int})$ & 4-fold cross-validation & $N_{SN}$ \\ \hline
0.00& $g$&  0.322&  0.129& 0.040&  0.099& 0.044& 0.134 (0.145)&          50\\
0.00& $r$&  0.229&  0.120& 0.041&  0.087& 0.039& 0.123 (0.130)&          49\\
0.00& $i$&  0.265&  0.131& 0.040&  0.100& 0.045& 0.136 (0.146)&          50\\ \hline
0.25& $g$&  0.437&  0.101& 0.017&  0.065& 0.018& 0.102 (0.141)&          44\\
0.25& $r$&  0.301&  0.100& 0.021&  0.068& 0.026& 0.102 (0.139)&          43\\
0.25& $i$&  0.236&  0.093& 0.013&  0.064& 0.017& 0.094 (0.128)&          43\\
0.25& $z$&  0.250&  0.113& 0.025&  0.076& 0.024& 0.113 (0.158)&          44\\ \hline
0.50& $r$&  0.391&  0.122& 0.019&  0.085& 0.033& 0.121 (0.118)&          50\\
0.50& $i$&  0.301&  0.118& 0.013&  0.077& 0.021& 0.116 (0.121)&          48\\
0.50& $z$&  0.248&  0.114& 0.005&  0.073& 0.006& 0.113 (0.112)&          49\\ \hline
0.75& $r$&  0.320&  0.129& 0.046&  0.061& 0.072& 0.136 (0.189)&          35\\
0.75& $i$&  0.237&  0.131& 0.054&  0.068& 0.050& 0.147 (0.212)&          35\\
0.75& $z$&  0.210&  0.120& 0.032&  0.047& 0.039& 0.123 (0.163)&          35\\ \hline
\end{tabular}
\caption{As in Table~\ref{sd_lin:tab} for the Gaussian process model. \label{sd:tab}}
\end{table}

The statistics in \citet{1983Efron}  are calculated for the case of the $i$-band from the $z=0.25$ filter set:
The {\it apparent error rate}  is determined from
the residuals of the training-set supernovae as run through the same processing as the validation supernovae.
The linear model gives 0.089 mag.  The resulting
0.025 mag for the Gaussian process model is an  overoptimistic error estimate:
the best-fit $\sigma_{n_M}=0.060$ mag nugget parameter (introduced after
Eqn.\ \ref{kernel:eqn})
gives an estimate of the absolute magnitude dispersion of the training supernovae.
The {\it bootstrap} estimate is calculated from 50 realizations
of training sets generated using random draws with replacement
from the full set and the $\sim 37$\% unselected supernovae forming the validation set; this error estimate is 
0.107 (0.117) mag for the linear (Gaussian process) model.  The {\it .632 estimator}
is not rigorously motivated but empirically exhibits small biases in the estimator 
of the true prediction error in certain cases, it is a function of the apparent and bootstrap estimators for which we get 0.100 (0.083)  mag for the linear (Gaussian process) model.


\section{Discussion}
Training and validation sets have different variance in their predicted magnitudes.  The discrepancy between
the two is not surprising given the sparsity of supernovae; it indicates
that the current training set does not fairly represent the population of the current validation set.
We anticipate that these values should converge as
the number of training supernovae increases.  The validation wrms may decrease with interpolation from more
local values, while
the training standard deviation may increase as singletons disappear.

The subsample of $\sim 50$ training-set
supernovae used to calibrate absolute magnitudes is too small to cover the full range of intrinsic and extrinsic supernova variability.  
This issue is highlighted in SN~\#19, where highly-discrepant absolute magnitudes are predicted when no analogue
exists in the training set. 
We remark that SN~\#19 (IAU name SN~2007le) does have unique properties noted elsewhere  \citep{2009ApJ...702.1157S}:
it is one of only three supernovae that exhibit time-varying Na I D absorption features and the
only one of those that is not highly reddened, suggesting circumstellar extinction.
Nevertheless, with the loose requirement of Eqn.~\ref{cut:eqn} many other supernovae
manage to be reasonably calibrated.
With a larger sample size, it will be interesting to see if large outliers persist despite having training-set analogues in regions of  parameter space where SNe~Ia
are not well standardized.
The trade between criteria for deciding if a supernova has training-set analogues versus the numbers of supernovae that meet that criteria should be studied in
actual cosmological analyses.

Whether the linear or Gaussian process model results in lower dispersion depends on the filter set being
calibrated and the specific statistic of interest.
Figure~\ref{scalzo_lin_g:fig} shows examples of how either model can go wrong.  The wiggles in
the $X(1)$- dependence of absolute magnitude suggest that
the Gaussian process regression is sensitive to fluctuations in the small training set.
The Gaussian process regression for $X(2)$ is clearly not well described by the linear model.
While both linear and Gaussian process solutions for $X(3)$ agree in the region covered by the
training sample, beyond they extrapolate out with different slopes.
For the data used, neither model for absolute magnitude universally outperforms the other.

We get dispersions of 0.09--0.14 mag (linear fit) and 0.09--0.11 mag (Gaussian process model)
using the four bands of the $z=0.25$ filter set.
Filter sets that use only three bands exhibit higher dispersion, although only modestly so.
Although the input magnitude dispersion strongly depends on the band of the magnitude being calibrated, this dependency mostly disappears
after calibration in the Gaussian process fit. 
Some of the dispersion is due to limitations in the data themselves.
There are light curves with suspected misestimated fluxes or uncertainties, SN~\#13 being a
prime example (Figure~\ref{bad2lc:fig}).  Sometimes
there are significant temporal interpolations being made due to coarse light curve sampling, particularly over the secondary
maximum
in the redder bands as seen in
Figures~\ref{lc:fig}, \ref{bad1lc:fig},  and \ref{bad2lc:fig}.  
A further decrease in the dispersion may be possible as more supernova spectrophotometry with higher signal-to-noise and denser temporal sampling becomes available.

In an attempt to avoid an explicit description of the supernova light-curve model,
we use the relatively simple normalized Hsiao templates for the mean function.
Certainly, Gaussian processes can be used to describe residuals about
the SALT2 model whose shape and color parameters (including date of maximum) are included as variables in the absolute
magnitude model.
Inclusion of a dust-absorption model
in the mean function should reduce the number of training-set supernovae needed to
densely sample the full range of dust absorption.

The new expanded set of light-curve parameters can standardize SNe Ia to a level that appears competitive with other fitters. 
A direct comparison
with the quoted dispersions of 0.12 mag for  SALT2 and 0.14 mag of MLCS2k2 is difficult because there is no sharp delineation between training and
validation supernovae reported for these fitters.
A SALT2 analysis on synthetic photometry from this same dataset yields a 0.147 mag residual on the Hubble diagram.
An alternative reference
is the 0.18--0.22 mag scatter found using these fitters on the CfA3 sample \citep{2009ApJ...700.1097H}.
Assuming the dust parameter $R_V=3.1$, \citet{2011ApJ...731..120M} calculate an apparent error rate of 0.10 mag, a bootstrap error of 0.15 mag, and
a .632 estimator of 0.13 mag based on optical--NIR light curves; for a more flexible
model where
the $R_V$ distribution differs in different $A_V$ intervals they obtain
a  .632 estimator of 0.11 mag.
\citet{2010AJ....139..120F} calibrate absolute magnitudes in the redder optical bands $r$ and $i$  with
training-set errors of
0.13 and 0.15 mag rms scatter and inferred intrinsic dispersions of 0.10 and 0.12 mag respectively.

We compare the results of our methodology using
the $g$-band calibration
of the $z=0$, $gri$ filter set, with SALT2 
in wavelength regions that correspond roughly to Johnson-Cousins $B$, $V$, and $R$.
The combined subset dispersions and for the same set of supernovae
are calculated to allow direct comparison with the preceding analysis.
The results are $\overline{wrms}=0.132 \pm 0.008$ mag
and   4-fold cross-validation dispersion of $0.132$ mag; the SALT2 uncertainties in distance are large
and result in an intrinsic dispersion consistent with zero.
The linear-fit performs somewhat worse in reducing
the total dispersions relative
to this second-generation light-curve analyzer, although the Gaussian process method compares
favorably.

Most of the leverage in absolute magnitude comes from the first four parameters, as shown in Figure~\ref{pcastrength:fig}.
This contrasts with MLCS2k2 and SALT2, which describe supernovae with one free light-curve-shape and one free color parameter.
Unlike those light curve fitters, many parameters are readily available in our analysis to account for color variance due to physical mechanisms
within the supernova explosion and different distributions and properties of absorbing dust.

The intrinsic dispersion is dependent on the blueshift of the filter set, a reasonable result
considering that fluxes from different restframe wavelength regions are measured.  Note that
 \citet[MLCS2k2]{2007ApJ...659..122J} and \citet[SALT2]{2007A&A...466...11G} each provide one
 redshift-independent intrinsic dispersion.  As an example, the $z$-band of the $z=0.5$ filter system has
a dispersion of 0.114 mag and the $i$-band of the $z=0.75$ filter system is 0.131 mag.
Given that the $griz$ filters are not logarithmically distributed it is difficult to make direct inferences as to the
exact cause of the different dispersions: whether local wavelength regions or the full wavelength range is playing
the more important role in determining absolute magnitudes. 
It would be of interest to find correlations between the PCA coefficients of the $z=0.5$  analysis
with the excess residuals of the $z=0.75$ analysis; this would mean that systematic errors that depend on redshift are incurred 
due to the limited resolution and wavelength coverage of a fixed filter system.
No such correlations are apparent with the  available statistics.

In the Gaussian process analysis, we restrict ourselves to using as parameters the coefficients of the leading principal components that account for 95\% of the dispersion.
The reduction of the dimensionality of the parameter space was motivated by computational tractability
and anticipating that it is overkill to use $>100$ parameters to deduce absolute magnitudes.  Indeed, our results do
not change appreciably by using the $\sim 30$ parameters that describe 99\% of the dispersion.  Figure~\ref{pcastrength:fig}
hints that even our restricted parameter set is more than needed for this dataset.    This validates our non-trivial
expectation that the largest variances in light-curve shape and color be correlated with  absolute magnitude. 
Further compression of information may be possible: In our methodology the PCA also accounts for the covariance in the light-curve prediction.  This may be avoided by accounting for measurement noise while performing the PCA, which is possible using expectation maximization techniques \citep{2012arXiv1208.4122B}.
Other projections such as from independent
component analysis and kernel PCA may prove more efficient in encapsulating the information, which could then be used
in a Gaussian process.

For a given supernova, we calculate the inferred peak absolute magnitude for each of the observer bands.  These magnitudes
and their residuals from the truth are highly correlated and so their resulting inferred distances are not independent.  
The same would be true if we had calibrated absolute magnitude at phases other than $B$ maximum.  The methodology
presented optimizes the determination of absolute magnitude in a single band and phase.

In an actual cosmology analysis, a customized light-curve model can be made for each redshift or
on a grid sufficiently fine to make interpolation errors unimportant.  For all redshifts,
the distance moduli (difference
between observed peak magnitude and inferred peak absolute magnitude) are based
on a common set of supernovae and can be directly compared, at least to the same
extent being done with current light-curve fitters.

\section{Future Outlook: Spectrophotometric Data}
\label{spec:sec}
The SNfactory generates spectrophotometry for objects at low redshift while the JDEM/ISWG design for a space telescope
dark energy experiment called for spectrophotometric time series of high-redshift supernovae.  In combination
such data would
bring the potential of providing improved distance measurements with lower statistical and systematic uncertainty
by preserving information lost in broadband photometry.

Specific spectral features have been
correlated with absolute magnitude \citep{2006ApJ...647..513B, 2009A&A...500L..17B, 2009ApJ...699L.139W, 2011ApJ...729...55F,2011A&A...529L...4C, 2012arXiv1202.2130S} and may provide
additional information not contained in broadband light curves that further standardize absolute magnitudes.
Using the approach of \S\ref{band:sec} with each dispersion element serving as a filter, the variance in
resolved features can be correlated
with absolute magnitude.

Supernovae have varying magnitude dispersion depending on wavelength  and phase \citep{2006MNRAS.370..933J}
that can be modeled as a Gaussian process.
Measurements $m_{(1+z)t,(1+z)\lambda}$ for observer phase and wavelength $(1+z)t$ and $(1+z)\lambda$ are drawn as
\begin{equation}
m_{\left((1+z)t,(1+z)\lambda\right)} \sim GP\left(\bar{M}(t,\lambda;p_{\bar{M}})+2.5\log{\left(\mu(z,\mathbf{p}_\Omega)\right)},k_m(t,\lambda,t',\lambda';\mathbf{p}_{k_m} )+ n_m(t,\lambda,t',\lambda';\mathbf{p}_{n_m})\right).
\label{specgp:eqn}
\end{equation}
With spectrophotometric data,  the cosmological parameters $\mathbf{p}_\Omega=\{\Omega_M, \Omega_{DE}, w_0, w_a\}$ that determine the distance
modulus $\mu$ predicted by theory can be fit simultaneously with the supernova parameters.

Such an analysis using SNfactory data alone is of interest.  Although insensitive to the cosmological parameters, the hyperparameters $\mathbf{p}_{k_m}$ in particular provide insight into the interplay between wavelength and phase with the standard candle nature of Type Ia supernovae.  This information aids in the planning of future missions and hardware
and quantifies the systematic uncertainties potentially induced when using degraded data (e.g.\ broad-band photometry) only.

\section{Conclusions}
\label{conclusions:sec}
The regressed values from supernova light curves provide a large number of parameters that may correlate with absolute magnitude.
Gaussian processes provide a convenient and parametrizable method of performing interpolations between
measurements of light-curve shape and color and for using this information to predict absolute magnitudes.

For a proof of concept, we
present an example application showing how spectrophotometry of  low-redshift supernovae can be used to determine distances from broad-band photometry
of high-redshift supernovae, such as will be obtained by DES and LSST.  Using SNfactory data, we find  dispersions in distance determinations
 as high as 0.44 mag corrected down to 0.04 mag within the training set and to
as low as 0.09 mag when applied to a validation set.   Some of this dispersion is attributable
to the data themselves.
Our procedure is highly competitive with SALT2 and MLCS2k2, which give training-set dispersions of 0.12 and 0.14 respectively. 

The spectrophotometry of the SNfactory provides an excellent dataset for exploring supernova studies.
Traditionally, broad-band light curves of $z \sim 0$ supernovae are used to train light-curve fitters.
The spectrophotometric templates allow the calibration of the absolute magnitude to be optimized for any redshift. With spectroscopy,
a light-curve model can be tailor made for every object from an imaging survey.
For planning of future supernova-specific imaging surveys,
our methodology and SNfactory data can be used to determine optimal filter sets that minimize intrinsic dispersion over the targeted redshift range.

There are directions for future development of the method and exploration of the data.
\begin{itemize}
\item There are further improvements possible in the implementation of the methodology.
A kernel other than the square exponential may better describe supernova light curves and their relationship with
absolute magnitude. 
We did not perform an exhaustive search for the absolute maximum likelihoods in the parameter space since
the local maxima found in this analysis were sufficient to establish our proof of concept.

\item When constraining the problem to broad-band photometry, the full wavelength range and resolution available is not exploited.
Analysis of intrinsic supernova properties within SNfactory and simultaneous analysis of low- and high-redshift samples
are planned in future work.

\item There is a correlation between supernova absolute magnitudes and host-galaxy properties after light-curve shape corrections
\citep{2010ApJ...715..743K, 2010MNRAS.406..782S, 2010ApJ...722..566L, 2011ApJ...740...92G,
Childressa}.  The search
for such correlations with the absolute magnitude residuals from our method is ongoing.

\item The observed variance of SNfactory measurements, and indeed of all supernova data,
cannot be entirely attributed to photon noise.  Gaussian processes provide a natural way to model and fit for external
sources of photometric uncertainty.

\end{itemize}

\acknowledgements
We thank Bruce Bassett for fruitful discussions and Dan Birchall
for observing assistance.  We thank the technical
and scientific staffs of the Palomar Observatory, the High Performance
Wireless Radio Network (HPWREN), the National Energy Research Scientific
Computing Center (NERSC), and the University of Hawaii 2.2~m telescope.
We wish to recognize and acknowledge the significant cultural role and
reverence that the summit of Mauna Kea has always had within the
indigenous Hawaiian community.  We are most fortunate to have the
opportunity to conduct observations from this mountain.  This work was
supported by the Director, Office of Science, Office of High Energy
Physics, of the U.S. Department of Energy under Contract
No.~DE-AC02-05CH11231; by a grant from the Gordon \& Betty Moore
Foundation; in France by support from CNRS/IN2P3, CNRS/INSU, and PNC;
and in Germany by the DFG through TRR33 ``The Dark Universe.''  NERSC is
supported by the Director, Office of Science, Office of Advanced
Scientific Computing Research, of the U.S. Department of Energy under
Contract No.~DE-AC02-05CH11231.  HPWREN is funded by National Science
Foundation Grant Number ANI-0087344, and the University of California,
San Diego.

\appendix
\section{Review of Gaussian Processes}
\label{appendix:sec}
In this appendix we introduce Gaussian process and provide expressions for the likelihood and
Normal probability distribution function used in this article.
A broader description of Gaussian processes can be found in \citet{GPML}.

A Gaussian process is denoted as
\begin{equation}
f(\mathbf{x}) \sim GP(m(\mathbf{x}),k(\mathbf{x},\mathbf{x}')).
\end{equation}
The expectation of the stochastic function $f(\mathbf{x})$ is the $m(\mathbf{x})$
with a model covariance between any two points $\mathrm{cov}\left(f(\mathbf{x}_p),f(\mathbf{x}_q)\right)=k(\mathbf{x}_p,\mathbf{x}_q)$.
Both $m$ and $k$ may be parameterized and the parameters of the latter are referred to as hyperparameters since
they describe function scatter rather than the function itself.
For a set of input points $X_\star$, the values of the function are drawn from a Normal distribution
\begin{equation}
\mathbf{f}_\star \sim \mathcal{N}\left(m(X_\star),K(X_\star,X_\star)\right),
\end{equation}
where the covariance matrix $K$ has elements filled with all input pairs of $k(\mathbf{x}_{\star,i},\mathbf{x}_{\star,j})$.

Since the probability distribution function is described by a Gaussian, several statistics are expressed analytically after algebraic
manipulation.
The likelihood that $\mathbf{y}$, a set of $n$ measurements of $f$ at inputs $X$, with measurement covariance $V$  described by a Gaussian process 
is written as
\begin{equation}
\log{p(\mathbf{y}|X)}= -\frac{1}{2} (\mathbf{y}-m(X))^T (K+V)^{-1} (\mathbf{y}-m(X))- \frac{1}{2} \log{|K+V|} 
-\frac{n}{2}\log{2\pi}.
\end{equation}

A set of measurements and function values is drawn from a Normal distribution
\begin{equation}
 \left[
\begin{array}{c}
      \mathbf{y} \\
      \mathbf{f}_\star
    \end{array}
  \right] \sim \mathcal{N}\left(
  \left[
      \begin{array}{c}
       m(X) \\
      m(X_\star)
    \end{array}
    \right]
    ,
    \left[
          \begin{array}{cc}
       K(X,X)+V & K(X,X_\star) \\
      K(X_\star,X) &K(X_\star,X_\star)
    \end{array}
    \right]
  \right).
\end{equation}
The conditional distribution of the function values is Gaussian with expected mean
\begin{equation}
\bar{\mathbf{f}}_\star=m(X)+ K(X_\star,X)\left[K(X,X)+V\right]^{-1}\left(\mathbf{y}-m(X)\right)
\label{gpmean:eqn}
\end{equation}
and covariance
\begin{equation}
\mathrm{cov}\left(\bar{\mathbf{f}}_\star\right)=K(X_\star,X_\star)- K(X_\star,X)\left[K(X,X)+V\right]^{-1}K(X,X_\star).
\label{gpcov:eqn}
\end{equation}

\begin{deluxetable}{ccccccc}
\tablecolumns{7}
\tablewidth{0pc}
\tabletypesize{\tiny}
\tablecaption{True and Gaussian-process-inferred
peak absolute magnitudes from validation using the $z=0$ filter set with
a constant additive offset. In italics are supernovae that have data before two days pre-maximum with light curves that have at least eight points.
In bold are supernovae that have data before two-days pre-maximum and at least eight points in all light curves.\label{sample0:tab}} 
 \tablehead{
\colhead{}&\multicolumn{2}{c}{$g$}&\multicolumn{2}{c}{$r$}&\multicolumn{2}{c}{$i$}\\
\cline{2-3} \cline{4-5}  \cline{6-7}\\ 
\colhead{SN \#}&\colhead{true}&\colhead{inferred}&\colhead{true}&\colhead{inferred}&\colhead{true}&\colhead{inferred}
}
\startdata
    0&$\mathbf{  -15.23\pm   0.04}$ & $\mathbf{  -15.22\pm   0.06}$&$\mathbf{  -14.91\pm   0.05}$ & $\mathbf{  -14.90\pm   0.05}$&$\mathbf{  -14.19\pm   0.05}$ & $\mathbf{  -14.19\pm   0.09}$\\
    1&$\mathbf{  -15.12\pm   0.03}$ & $\mathbf{  -15.16\pm   0.06}$&$\mathbf{  -14.90\pm   0.03}$ & $\mathbf{  -14.93\pm   0.05}$&$\mathbf{  -14.18\pm   0.04}$ & $\mathbf{  -14.18\pm   0.07}$\\
    3&$\mathbf{  -15.25\pm   0.03}$ & $\mathbf{  -15.23\pm   0.07}$&$\mathbf{  -14.94\pm   0.03}$ & $\mathbf{  -14.92\pm   0.06}$&$\mathbf{  -14.25\pm   0.04}$ & $\mathbf{  -14.24\pm   0.06}$\\
    4&$\mathbf{  -15.08\pm   0.03}$ & $\mathbf{  -14.76\pm   0.07}$&$\mathbf{  -14.83\pm   0.04}$ & $\mathbf{  -14.53\pm   0.08}$&$\mathbf{  -14.17\pm   0.04}$ & $\mathbf{  -13.86\pm   0.08}$\\
    6&$\mathbf{  -15.29\pm   0.03}$ & $\mathbf{  -15.20\pm   0.13}$&$\mathbf{  -14.99\pm   0.03}$ & $\mathbf{  -14.88\pm   0.10}$&$\mathbf{  -14.32\pm   0.05}$ & $\mathbf{  -14.22\pm   0.13}$\\
    7&$\mathit{  -15.18\pm   0.03}$ & $\mathit{  -15.27\pm   0.08}$&$\mathit{  -14.90\pm   0.03}$ & $\mathit{  -14.97\pm   0.07}$&$\mathit{  -14.14\pm   0.03}$ & $\mathit{  -14.17\pm   0.07}$\\
    8&$\mathbf{  -15.22\pm   0.04}$ & $\mathbf{  -15.06\pm   0.05}$&$\mathbf{  -14.90\pm   0.04}$ & $\mathbf{  -14.70\pm   0.02}$&$\mathbf{  -14.25\pm   0.06}$ & $\mathbf{  -14.11\pm   0.07}$\\
    9&$\mathbf{  -15.16\pm   0.02}$ & $\mathbf{  -15.23\pm   0.08}$&$\mathbf{  -14.84\pm   0.03}$ & $\mathbf{  -14.93\pm   0.07}$&$\mathbf{  -14.13\pm   0.04}$ & $\mathbf{  -14.19\pm   0.08}$\\
\enddata
\end{deluxetable}

\begin{deluxetable}{ccccccccc}
\tablecolumns{9}
\tablewidth{0pc}
\tabletypesize{\tiny}
\tablecaption{True and inferred peak absolute magnitudes from validation using the  $z=0.25$ filter set. \label{sample025:tab}} 
 \tablehead{
\colhead{}&\multicolumn{2}{c}{$g$}&\multicolumn{2}{c}{$r$}&\multicolumn{2}{c}{$i$}&\multicolumn{2}{c}{$z$}\\
\cline{2-3} \cline{4-5}  \cline{6-7} \cline{8-9}\\ 
\colhead{SN \#}&\colhead{true}&\colhead{inferred}&\colhead{true}&\colhead{inferred}&\colhead{true}&\colhead{inferred}&\colhead{true}&\colhead{inferred}
}
\startdata
    0&$\mathbf{  -15.07\pm   0.07}$ & $\mathbf{  -15.07\pm   0.09}$&$\mathbf{  -15.08\pm   0.04}$ & $\mathbf{  -15.11\pm   0.07}$&$\mathbf{  -14.94\pm   0.04}$ & $\mathbf{  -14.95\pm   0.05}$&$\mathbf{  -14.08\pm   0.05}$ & $\mathbf{  -14.09\pm   0.06}$\\
    1&$\mathbf{  -14.88\pm   0.05}$ & $\mathbf{  -14.85\pm   0.09}$&$\mathbf{  -15.01\pm   0.02}$ & $\mathbf{  -14.94\pm   0.06}$&$\mathbf{  -14.92\pm   0.03}$ & $\mathbf{  -14.90\pm   0.05}$&$\mathbf{  -14.05\pm   0.03}$ & $\mathbf{  -14.05\pm   0.05}$\\
    3&$\mathbf{  -15.11\pm   0.05}$ & $\mathbf{  -15.16\pm   0.08}$&$\mathbf{  -15.12\pm   0.02}$ & $\mathbf{  -15.17\pm   0.03}$&$\mathbf{  -14.98\pm   0.02}$ & $\mathbf{  -15.05\pm   0.05}$&$\mathbf{  -14.11\pm   0.03}$ & $\mathbf{  -14.18\pm   0.07}$\\
    4&$\mathbf{  -14.82\pm   0.07}$ & $\mathbf{  -14.67\pm   0.13}$&$\mathbf{  -14.96\pm   0.03}$ & $\mathbf{  -14.84\pm   0.09}$&$\mathbf{  -14.85\pm   0.03}$ & $\mathbf{  -14.70\pm   0.09}$&$\mathbf{  -14.07\pm   0.04}$ & $\mathbf{  -13.96\pm   0.09}$\\
    7&$\mathit{  -15.05\pm   0.07}$ & $\mathit{  -15.13\pm   0.11}$&$\mathit{  -15.04\pm   0.03}$ & $\mathit{  -15.19\pm   0.07}$&$\mathit{  -14.93\pm   0.02}$ & $\mathit{  -15.05\pm   0.06}$&$\mathit{  -14.02\pm   0.03}$ & $\mathit{  -14.11\pm   0.07}$\\
    8&$\mathit{  -15.02\pm   0.08}$ & $\mathit{  -14.80\pm   0.15}$&$\mathit{  -15.05\pm   0.04}$ & $\mathit{  -14.88\pm   0.06}$&$\mathit{  -14.91\pm   0.04}$ & $\mathit{  -14.69\pm   0.04}$&$\mathit{  -14.14\pm   0.04}$ & $\mathit{  -13.95\pm   0.05}$\\
    9&$\mathbf{  -15.02\pm   0.05}$ & $\mathbf{  -15.10\pm   0.08}$&$\mathbf{  -15.02\pm   0.02}$ & $\mathbf{  -15.10\pm   0.07}$&$\mathbf{  -14.87\pm   0.03}$ & $\mathbf{  -14.96\pm   0.05}$&$\mathbf{  -14.00\pm   0.03}$ & $\mathbf{  -14.07\pm   0.04}$\\
\enddata
\end{deluxetable}

\begin{deluxetable}{ccccccc}
\tablecolumns{7}
\tablewidth{0pc}
\tabletypesize{\tiny}
\tablecaption{True and inferred peak absolute magnitudes  from validation using the  $z=0.50$ filter set. \label{sample050:tab}} 
 \tablehead{
\colhead{}&\multicolumn{2}{c}{$r$}&\multicolumn{2}{c}{$i$}&\multicolumn{2}{c}{$z$}\\
\cline{2-3} \cline{4-5}  \cline{6-7}\\ 
\colhead{SN \#}&\colhead{true}&\colhead{inferred}&\colhead{true}&\colhead{inferred}&\colhead{true}&\colhead{inferred}
}
\startdata
    0&$\mathbf{  -15.27\pm   0.04}$ & $\mathbf{  -15.21\pm   0.08}$&$\mathbf{  -14.98\pm   0.04}$ & $\mathbf{  -14.90\pm   0.06}$&$\mathbf{  -14.75\pm   0.03}$ & $\mathbf{  -14.69\pm   0.05}$\\
    1&$\mathbf{  -15.14\pm   0.03}$ & $\mathbf{  -15.22\pm   0.07}$&$\mathbf{  -14.91\pm   0.02}$ & $\mathbf{  -14.83\pm   0.06}$&$\mathbf{  -14.73\pm   0.03}$ & $\mathbf{  -14.69\pm   0.06}$\\
    3&$\mathbf{  -15.27\pm   0.03}$ & $\mathbf{  -15.27\pm   0.07}$&$\mathbf{  -15.01\pm   0.03}$ & $\mathbf{  -15.01\pm   0.06}$&$\mathbf{  -14.79\pm   0.03}$ & $\mathbf{  -14.79\pm   0.06}$\\
    4&$\mathbf{  -15.11\pm   0.04}$ & $\mathbf{  -14.95\pm   0.12}$&$\mathbf{  -14.85\pm   0.03}$ & $\mathbf{  -14.72\pm   0.09}$&$\mathbf{  -14.63\pm   0.03}$ & $\mathbf{  -14.50\pm   0.09}$\\
    6&$\mathbf{  -15.32\pm   0.04}$ & $\mathbf{  -15.23\pm   0.09}$&$\mathbf{  -15.03\pm   0.04}$ & $\mathbf{  -14.91\pm   0.08}$&$\mathbf{  -14.82\pm   0.03}$ & $\mathbf{  -14.70\pm   0.07}$\\
    7&$\mathit{  -15.22\pm   0.04}$ & $\mathit{  -15.29\pm   0.08}$&$\mathit{  -14.96\pm   0.03}$ & $\mathit{  -15.05\pm   0.06}$&$\mathit{  -14.75\pm   0.02}$ & $\mathit{  -14.81\pm   0.07}$\\
    8&$\mathbf{  -15.23\pm   0.05}$ & $\mathbf{  -15.11\pm   0.08}$&$\mathbf{  -14.94\pm   0.04}$ & $\mathbf{  -14.81\pm   0.06}$&$\mathbf{  -14.72\pm   0.04}$ & $\mathbf{  -14.58\pm   0.04}$\\
    9&$\mathbf{  -15.21\pm   0.02}$ & $\mathbf{  -15.35\pm   0.08}$&$\mathbf{  -14.92\pm   0.03}$ & $\mathbf{  -15.05\pm   0.08}$&$\mathbf{  -14.68\pm   0.02}$ & $\mathbf{  -14.82\pm   0.07}$\\
\enddata
\end{deluxetable}

\begin{deluxetable}{ccccccc}
\tablecolumns{7}
\tablewidth{0pc}
\tabletypesize{\tiny}
\tablecaption{True and inferred peak absolute magnitudes from validation using the  $z=0.75$ filter set. \label{sample075:tab}} 
 \tablehead{
\colhead{}&\multicolumn{2}{c}{$r$}&\multicolumn{2}{c}{$i$}&\multicolumn{2}{c}{$z$}\\
\colhead{SN \#}&\colhead{true}&\colhead{inferred}&\colhead{true}&\colhead{inferred}&\colhead{true}&\colhead{inferred}
}
\startdata
    0&$\mathbf{  -14.83\pm   0.07}$ & $\mathbf{  -14.83\pm   0.10}$&$\mathbf{  -15.18\pm   0.04}$ & $\mathbf{  -15.17\pm   0.06}$&$\mathbf{  -14.73\pm   0.04}$ & $\mathbf{  -14.74\pm   0.05}$\\
    1&$\mathbf{  -14.67\pm   0.05}$ & $\mathbf{  -14.58\pm   0.09}$&$\mathbf{  -15.04\pm   0.02}$ & $\mathbf{  -14.96\pm   0.09}$&$\mathbf{  -14.68\pm   0.02}$ & $\mathbf{  -14.64\pm   0.06}$\\
    3&$\mathbf{  -14.89\pm   0.05}$ & $\mathbf{  -14.89\pm   0.10}$&$\mathbf{  -15.18\pm   0.03}$ & $\mathbf{  -15.20\pm   0.08}$&$\mathbf{  -14.78\pm   0.03}$ & $\mathbf{  -14.78\pm   0.05}$\\
    4&$\mathbf{  -14.56\pm   0.07}$ & $\mathbf{  -14.34\pm   0.13}$&$\mathbf{  -15.03\pm   0.03}$ & $\mathbf{  -14.81\pm   0.11}$&$\mathbf{  -14.60\pm   0.03}$ & $\mathbf{  -14.40\pm   0.10}$\\
    6&$\mathbf{  -14.81\pm   0.07}$ & $\mathbf{  -14.58\pm   0.14}$&$\mathbf{  -15.22\pm   0.03}$ & $\mathbf{  -14.99\pm   0.07}$&$\mathbf{  -14.79\pm   0.03}$ & $\mathbf{  -14.55\pm   0.09}$\\
    7&$\mathit{  -14.89\pm   0.06}$ & $\mathit{  -15.01\pm   0.12}$&$\mathit{  -15.12\pm   0.03}$ & $\mathit{  -15.22\pm   0.09}$&$\mathit{  -14.70\pm   0.03}$ & $\mathit{  -14.82\pm   0.07}$\\
    8&$\mathbf{  -14.82\pm   0.05}$ & $\mathbf{  -14.65\pm   0.07}$&$\mathbf{  -15.16\pm   0.05}$ & $\mathbf{  -15.01\pm   0.10}$&$\mathbf{  -14.73\pm   0.05}$ & $\mathbf{  -14.55\pm   0.05}$\\
    9&$\mathbf{  -14.82\pm   0.05}$ & $\mathbf{  -14.90\pm   0.09}$&$\mathbf{  -15.09\pm   0.02}$ & $\mathbf{  -15.15\pm   0.07}$&$\mathbf{  -14.69\pm   0.02}$ & $\mathbf{  -14.76\pm   0.06}$\\
\enddata
\end{deluxetable}

\end{document}